\documentclass[review]{elsarticle}

\usepackage{lineno}
\modulolinenumbers[5]

\usepackage{slashbox,amsmath}
\usepackage{graphicx}
\usepackage{dcolumn}
\usepackage{bm}
\usepackage{references}
\usepackage{hyperref} 

\usepackage[normalem]{ulem}
\usepackage[usenames]{xcolor}
\usepackage{booktabs}

\newcommand{\be}{\begin{equation}}
\newcommand{\ee}{\end{equation}}
\newcommand{\ba}{\begin{eqnarray}}
\newcommand{\ea}{\end{eqnarray}}
\newcommand{\bd}{\begin{displaymath}}
\newcommand{\ed}{\end{displaymath}}

\newcommand{\mm}{\mathcal{M}}

\def\simge{\mathrel{\rlap{\raise 0.511ex
       \hbox{$>$}}{\lower 0.511ex \hbox{$\sim$}}}}
\def\simle{\mathrel{\rlap{\raise 0.511ex
        \hbox{$<$}}{\lower 0.511ex \hbox{$\sim$}}}}
\makeatletter
\renewcommand\paragraph{\@startsection{paragraph}{4}{\z@}%
            {-2.5ex\@plus -1ex \@minus -.25ex}%
            {1.25ex \@plus .25ex}%
            {\normalfont\normalsize\bfseries}}
\makeatother
\usepackage{ulem} 

\journal{Physics Reports}

\begin{document}

\begin{frontmatter}

\title{The Equation of State of Hot, Dense Matter and Neutron Stars}

\author{James M. Lattimer\fnref{fnjml}}
\address{Department of Physics \& Astronomy, Stony Brook University, Stony Brook, NY 11733, USA}

\author{Madappa Prakash\fnref{fnmp}}
\address{Department of Physics and Astronomy, Ohio University, Athens, OH 45701, USA}
\fntext[fnmp]{prakash@ohio.edu}
\fntext[fnjml]{james.lattimer@stonybrook.edu}

\begin{abstract}

Recent developments in the theory of pure neutron matter and experiments concerning the symmetry energy of nuclear matter, coupled with recent measurements of high-mass neutron stars, now allow for relatively tight constraints on the equation of state of dense matter.  We review how these constraints are formulated and describe the implications they have for neutron stars and core-collapse supernovae.  We also examine thermal properties of dense matter, which are important for supernovae and neutron star mergers, but which cannot be nearly as well constrained at this time by experiment.  In addition, we consider the role of the equation of state in medium-energy heavy-ion collisions.

\end{abstract}



\end{frontmatter}


\clearpage

\tableofcontents

\clearpage


\section{INTRODUCTION}
\label{Sec:Intro}
Gerry was long interested in the equation of state (EOS) of dense matter and
supernovae, and formulated many ideas concerning the mechanism
underlying core-collapse supernovae.  Gerry Brown was not one who would let a crisis or controversy in nuclear physics pass him by.  Whether the dense matter EOS was soft or stiff was a debate he plunged into with great gusto. To enable a core-collapse supernova explode through the prompt-shock and rebound mechanism, Gerry argued \cite{BO85} that the EOS at nuclear densities had to be soft with an incompressibility parameter, $K_s$, of isospin symmetric nuclear matter much smaller ($\sim 110$ MeV) than 
$ 220\pm 30$ MeV inferred from the analysis of giant monopole resonances by Blaizot et al. ~\cite{Blaizot80}.  
Thinking today supports the notion that neutrinos and
multi-dimensional effects are required to ensure successful
explosions.  While he also famously argued for
a relatively soft nuclear EOS to explain why no neutron star has yet
been detected in the remnant of SN 1987A, observations of pulsars today indicate that
the maximum mass is quite large~\cite{Demorest10,Antoniadis13}.    In his later years, author JML would try to provoke him (in retaliation for phone calls, during Notre Dame football games, purporting to discuss physics) with
tales of ever larger neutron star mass measurements.  Ironically, it
now appears that uncertainties in the dense matter EOS
have little to do with uncertainties in the core-collapse supernova
mechanism.

Nevertheless, he does not appear to have been completely wrong about softness in
the EOS.  While experiments~\cite{Youngblood99,Garg04,Colo04} indicate that symmetric
matter has a larger incompressibility than he favored, pure neutron
matter, which is much closer to neutron star matter than is symmetric
matter, seems to be relatively soft near the nuclear saturation
density.  Calculations of the properties of pure neutron matter, as
well as experimental results concerning the symmetry properties of
dense matter gleaned from experiments measuring binding energies,
neutron skin thicknesses and giant resonances, support this
perspective.  But to attain a large maximum mass, the EOS at densities beyond twice the nuclear saturation density must
become very stiff.  Gerry would have been fascinated with this
development.

Gerry was also intimately involved in great debates in the late 1970s
concerning thermal effects in dense matter, and strongly argued, with
Hans Bethe, that the importance of excited states in nuclei had been
underappreciated.  Historical comments concerning his work, which
culminated in the ``BBAL'' ~\cite{Bethe79} paper, are contained in
articles appearing in the recent Nuclear Physics Memorial 
Volume~\cite{Lattimer14,Langanke14}.  In this case, of course, he was
correct.  The role of thermal effects has taken on new emphasis with
the realization that high entropies and temperatures exist in hypermassive
neutron stars, the metastable, differentially rotating hot
configurations which are the aftermath of some neutron star mergers.  Their
lifetimes before gravitational collapse to black holes ensues crucially depend on the
specific heat of the hot matter as well as on neutrino emissivities
which determine cooling rates and timescales for dissipation of
differential rotation.

In the 1980's data about collective flow from 0.5-2 GeV per nucleon  heavy-ion collisions at the Bevalac became available.  Initial theoretical analyses indicated that the EOS at near-nuclear and supra-nuclear densities was 
very stiff with an incompressibility parameter close to 400 MeV.  In many works, even larger values were predicted.  Gerry and author MP struggled a lot to reconcile such large values with the much lower values suggested by the analyses of the giant resonances data and theoretical calculations of the EOS.  MP's first paper with Gerry~\cite{ABBCP} was a tortuous experience insofar as none of the authors was convinced about the resolution of the problem on hand. 
We argued fiercely about how the paper was to be written. After many drafts,  
Gerry relegated text of MP's detailed calculations to the appendices, and replaced the main text with many conflicting ideas. The paper took over a year to get published with many revisions after the referee's comments.
One of those ideas, that the momentum dependent interactions could be at the root of the solution, turned out to be right and has stood the test of time. As always, Gerry was generous to competitors (behind their backs, of course); the note added in proof acknowledges preprints by 
Aichelin et al.,~\cite {Aichelin87} and Gale et el.,~\cite{Gale87} which were submited after the completion of our work.  These works had similar ideas and calculations moving in the right direction, although they were incomplete. 

Not being satisfied with what had transpired, MP suggested to Gerry a collaboration with Subal Das Gupta, who with George Bertsch had devised the means  to describe heavy-ion collisions by solving Boltzmann-type transport equations with Monte Carlo techniques~\cite{Bertsch88}. 
Without hesitation, Gerry facilitated Subal's sabbatical to be spent at Stony Brook, a period that proved extremely productive.  
Keeping in line with Gerry's dearly held principle ``KISS'' (Keep It Simple Stupid), Subal, Kuo \& Prakash~\cite{Prakash88b} wrote a short paper  explaining 
why contact interactions of the Skyrme type (with  a quadratic momentum dependence in the single-particle potential) would produce too much flow compared to data whereas finite-range range interactions (with single-particle potentials saturating at high momenta) would produce less collective flow in accord with heavy-ion data. Gerry saw the point, declared the problem solved, and soon thereafter his interest in this subject dwindled rapidly. 
The devils in the details were to be sorted out by Charles Gale (at that time, Subal's graduate student), Gerd Welke (Gerry's graduate student whom MP co-advised), Subal and Prakash.  In the papers that followed~\cite{Welke88,Gale90}, an effective momentum-dependent interaction was devised to fit  the single-particle potential from the variational calculations of Wiringa \cite{w88} 
as well as  optical model fits to nucleon-nucleus scattering.  We also performed Boltzmann-Uhling-Uhlenbeck calculations which successfully accounted for the transverse flow data~\cite{Danielewicz88}. The mean field interaction needed had an incompressibility parameter 
of 215 MeV as required by analysis of giant resonances, as suggested by the analysis of the giant 
resonances~\cite{Youngblood99,Garg04,Colo04}. 
A sizable fraction of the transverse flow generated in the early phases of the reaction, where equilibrium cannot possibly exist, is sensitive to the initial-state correlation, which the momentum dependence of the interaction tends to preserve.
Independent calculations by Danielewicz~\cite{Danielewicz:00,Danielewicz:02}, who also used a momentum-dependent single-particle potential, have confirmed the need for the momentum dependence to saturate at high momenta to describe heavy-ion data.  
In addition, the cold EOS implied by such a mean field is also consistent with the recently discovered 2 ${\rm M_\odot}$ neutron 
star~\cite{Danielewicz02,cons15a}. Much of the credit for the resolution of the thorny problem mentioned above belongs to Gerry who gave the team he built much impetus. 

Gerry's penchant for effective masses is well reflected in his
prescient paper ``Effective Mass in Nuclei'' written with Gunn and
Gould in 1963~\cite{Brown63}. The abstract of this paper stated
``Calculations in finite nuclei indicate that in the region of bound
particles $\approx -8$ to 0 MeV the velocity dependence of the shell
model is, if present, opposite to that usually assumed. This can be
expressed by saying that the ratio of the effective mass to the real
mass is equal to or greater than unity in this energy range.'' The
last sentence of the paper read ``It seems strange to us that people
making calculations in nuclear matter do not worry about this point,
however, since it remains in direct conflict with our conclusions,
unless understood as a specific effect of the finiteness.'' This was
Gerry at his provocative best in print!  Effective masses are central
to the delineation of thermal effects and are further discussed in Sec.~\ref{Sec:Teffects}.

Gerry's protective attitude toward his wards was unmatched. 
When Gerry became seriously ill, he forwarded his last graduate student, Constantinos Constantinou, to Ohio University.  The authors co-advised Constantinos, who completed his Ph.D. degree from Stony Brook in 2013. Constantinos (now a post-doc at Juelich with Ulf-Meissner, another of Gerry's wards) has worked on thermal effects in dense matter relevant to the astrophysics of compact objects, a subject that was close to Gerry's heart. This article contains a brief review of 
Constantinos's work~\cite{cons15a,APRppr,cons15b}, which would have pleased Gerry immensely especially as Landau's Fermi Liquid theory (another subject close to Gerry's heart) was put to good use.

This review will highlight important recent developments in our
understanding of the EOS of dense, hot matter.  In
section \ref{Sec:Nstars}, the basics of neutron star structure are
developed, and constraints for neutron star radii based on general
relativity, causality and pulsar-timing measurements of neutron star
masses are outlined.  A lot hinges upon the behavior of the nuclear
symmetry energy.  It is now realized that the radius of typical
star, usually taken to be $1.4-1.5M_\odot$, is closely connected to
the pressure of neutron star matter near the nuclear saturation
density $n_s\sim0.16$ fm$^{-3}$.  Neutron star matter near $n_s$ has a
very small proton fraction, so that it is nearly the same as pure
neutron matter, which has been the focus of activity in recent years.
And given that the pressure of neutron star matter can be computed
from symmetric nuclear matter with knowledge of the nuclear symmetry
energy, there is also a direct connection between neutron star
structure and experiments that probe the symmetry energy, namely those
measuring binding energies, neutron skin thicknesses and giant
resonances.  The evidence from nuclear experiments supporting symmetry
energy
results deduced from neutron matter studies is explored in section
\ref{Sec:Exper}.  The refined constraints developed along these two
parallel tracks have definite implications for neutron star structure,
when they are coupled with the existence of the hadronic neutron star
crust.  By parameterizing the high-density EOS, these constraints can
be quantified, which is the subject of section \ref{Sec:pw}.  

All neutron
stars emit photons releasing their thermal energy.  If they were blackbody emitters,
measurements of their fluxes, temperatures and distances would suffice
to measure neutron star radii, a subject discussed in section
\ref{Sec:Observ}.  Even though neutron star atmospheres modify their
spectra so that they are not true blackbodies, it is apparent that
astronomical observations can closely connect to neutron star radii.
Although measurements of neutron star radii from observations are not
yet accurate enough to compete with inferences obtained from nuclear
experiment and theory, several proposals for improvements in this
direction are discussed.  In section \ref{Sec:Teffects}, thermal
effects in dense matter are reviewed.  Uncertainties stemming from the
lack of knowledge of effective masses at high densities, as well as
model dependences arising from the nature of the nuclear interaction
model, are discussed.  In section \ref{Sec:NTeffects}, non-thermal effects of importance to heavy-ion collisions are discussed.   Emphasis is placed on the role of the momentum dependence in the single particle potential.  In section \ref{Sec:Conclusions}, we summarize the main points of this review.

\section{THE EOS AND NEUTRON STAR STRUCTURE}
\label{Sec:Nstars}

The EOS of finite temperature matter is an essential ingredient in the modeling of core-collapse supernovae, neutron stars from their birth to old age, and mergers of compact binary stars.  For use in large-scale computer simulations of these phenomena, the EOS is rendered in tabular forms as functions of the baryon density $n$, temperature $T$, and the net electron fraction $Y_e=n_e/n$, all of which vary over wide ranges as indicated in Table \ref{pfield}. Also shown in this table are the values of the entropy per baryon, $S\equiv S(n,T,Y_e)$, which is generally used to gauge the ambient physical conditions.  Examples of EOS tabulations can be found, e.g., 
in Refs.~\cite{Lattimer91,SLM94,Shen1,sys07,Hempel10,Shen2b,shen2011new,Oconnor,Hempel12,Steiner}.

\begin{table}[h]
\begin{tabular}{cccc} 
\\ \hline
& {\bf Core-collapse}  & {\bf Proto-neutron}  &  {\bf Mergers of compact}    \\ 
             & {\bf supernovae} & {\bf stars} & {\bf binary stars} \\ \hline
$n/n_s$ & $10^{-8}$ - 10 & $10^{-8}$ - 10  & $10^{-8}$ - 10   \\ 
$T({\rm MeV})$ & 0 - 30 & 0 - 50   & 0 - 100 \\  
$Y_e$ & 0.35 - 0.45 & 0.01 - 0.3 & 0.01 - 0.6 \\ 
$S(k_B)$ & 0.5 - 10 &  0 - 10  & 0 - 100 \\ \hline
\end{tabular}
\begin{center}
\caption{Ranges of baryon number density $n$, temperature $T$, net electron fraction $Y_e=n_e/n$, and entropy per baryon $S$ encountered in the indicated astrophysical phenomena. }\vspace*{-1.cm}
\label{pfield}
\end{center}
\end{table}
%

\begin{figure}[h]\vspace*{-.5cm}
\begin{center}
\hspace*{-1.5cm}\includegraphics[width=15cm]{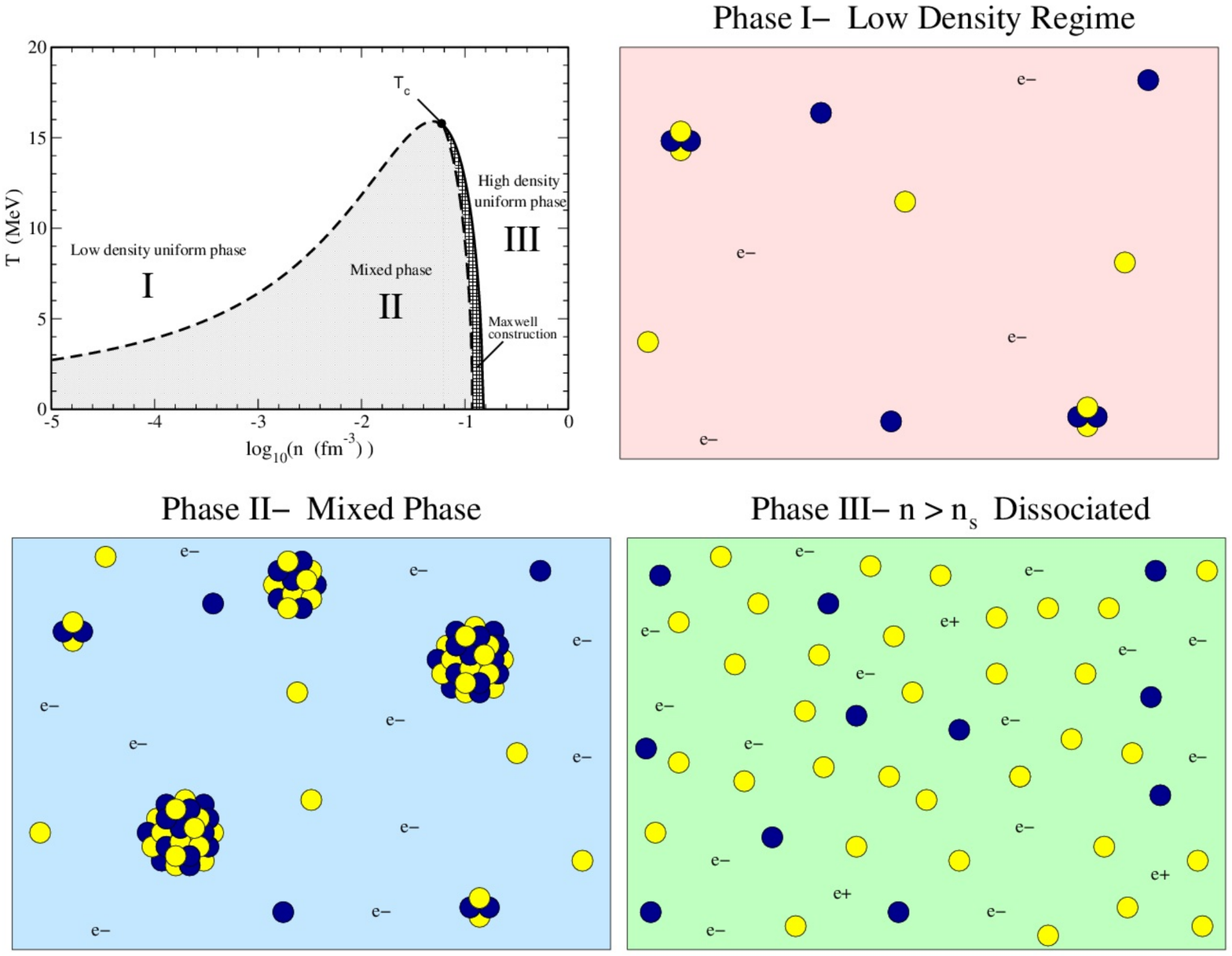}
\end{center}
\vspace*{-1.35cm}
\caption{Phases of sub-nuclear matter.  The upper left figure shows the boundary  separating the three phases illustrated in the other pictures. Figure from Ref. \cite{Carmell:05}.} 
\label{schem}
\end{figure}

At relatively low temperatures, matter below $\sim 0.1~{\rm fm}^{-3}$ comprises an inhomogeneous mixture of unbound nucleons (mostly neutrons), light nuclear clusters ($\alpha$ particles, deuterons, etc.), heavy nuclei, and electrons. At densities less than the neutron-drip density, about $0.0016n_s$, only heavy nuclei and electrons exist.  At densities beyond $0.01n_s$, competition between surface and Coulomb energies deforms nuclei, resulting in the evenutal formation of pasta-like configurations around $n\simeq0.1n_s$.  At $n\sim0.5n_s$, a phase transition from inhomogeneous to homogeneous matter occurs.   In the inhomogeneous phase above a certain temperature the relative proportion of nuclei dwindles, and a state of homogeneous matter forms.
At supra-nuclear densities, probably beyond $2n_s$, Bose condensates, strangeness-bearing baryons, mesons, and quarks may also be present. These phases are illustrated schematically in Fig. \ref{schem}.  

%

The global properties of neutron stars mostly depend on the EOS of matter above the saturation density.  Two of the outstanding problems in nuclear astrophysics concern the
maximum mass for neutron stars and the value of the radius of a
typical $1.4M_\odot$ star.  These problems are coupled: it is shown below that an observed lower limit
for the maximum mass implies a minimum value for this radius.  Recent measurements of neutron star
masses by pulsar timing have led to the discovery of stars with masses
much larger than seemed likely a decade ago, so that this
constraint is no longer academic but practical.

Neutron stars are so compact that general relativity (GR) is essential in determining their structure.  Nevertheless, some useful insights can be obtained by considering the spherical Newtonian structure equations:
\begin{equation}
{dp\over dr}=-{Gm\rho\over r^2},\qquad{dm\over dr}=4\pi\rho r^2,
\label{eq:newt}\end{equation}
where $\rho$ is the mass density and $p$ is the pressure.  Above, $m$
represents the mass interior to the radius $r$.  It is necessary to
supply the EOS $p(\rho)$ to solve these equations.  It
is useful to consider the relations that result from assuming a
polytropic relation $p\simeq K\rho^\gamma$ for the EOS of the star.
$K$ is a constant and $\gamma$ is the constant polytropic exponent.
This can be useful when considering either low-mass or high-mass
neutron stars.  In the former case, the polytropic formula can be
applied to matter with densities below the nuclear saturation density
$n_s$, where matter is dominated by the pressure of relativistic
degenerate electrons and $\gamma\simeq4/3$.  In the latter case, where the
average density is well in excess of $n_s$, a rough approximation is
$\gamma\approx2$.

\subsection{Results from Dimensional Analysis}
\begin{figure}[h]\vspace*{-1.5cm}
\hspace*{-1.5cm}\includegraphics[width=14cm,angle=180]{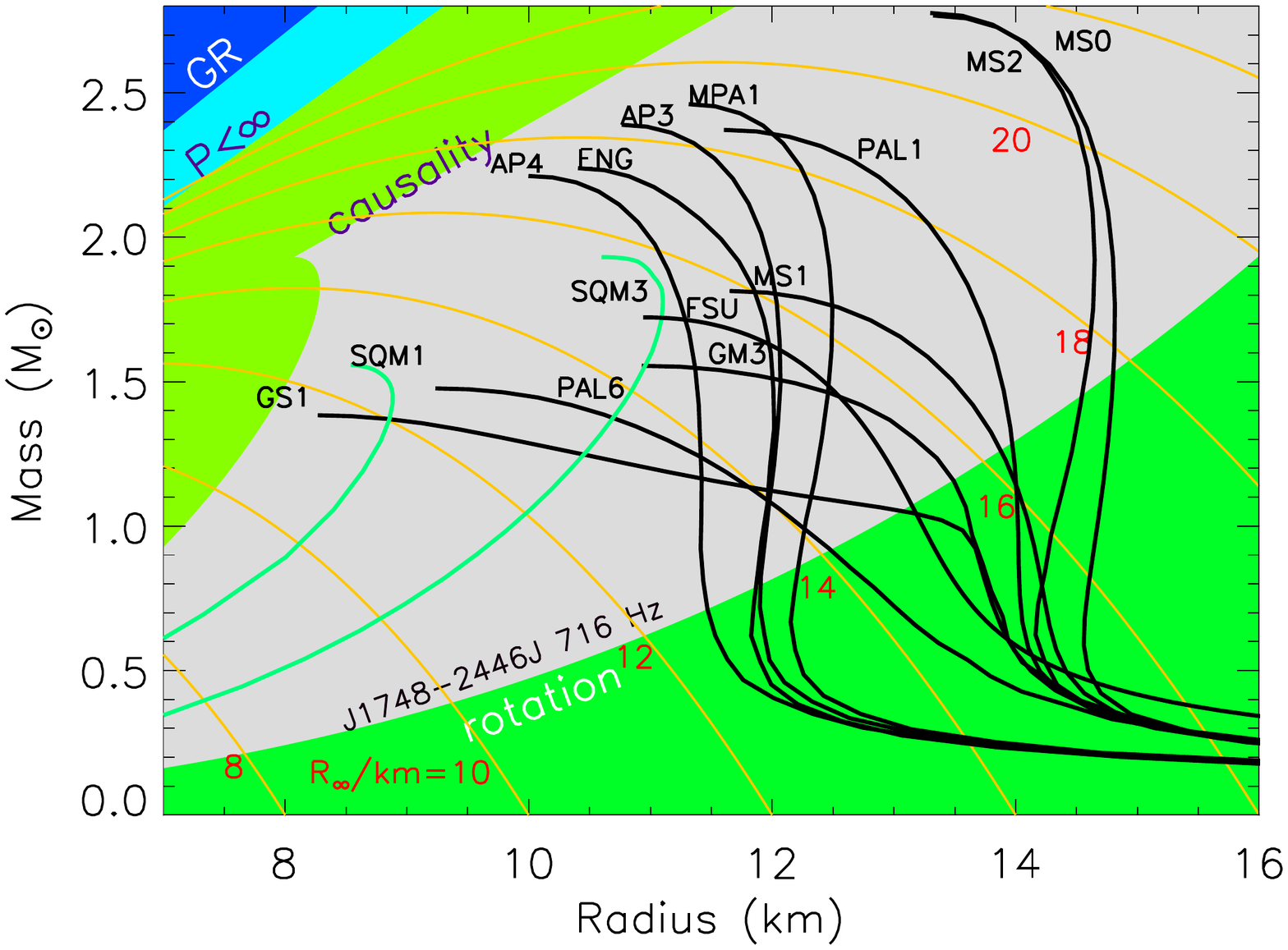}\vspace*{-1.cm}
\caption{Mass-radius curves for a variety of popular EOSs (see Ref.~\cite{Lattimer01} for descriptions).  The green shaded region in the upper left is causally-excluded; the green shaded region in the lower-right is excluded by the most rapidly spinning pulsar.  Black curves are hadronic EOSs; green curves are for strange quark matter configurations.  Lines of fixed $R_\infty=R/\sqrt{1-2\beta}$ are indicated as orange curves.}
\label{fig:mr}
\end{figure}

For a polytrope, dimensional analysis of the structure equations leads to
the mass-radius relation
\begin{equation}
M=-\xi_1^2\theta_1^\prime(4\pi)^{-1/(\gamma-1)}\left({K\gamma\over G(\gamma-1)}\right)^{1/(2-\gamma)}\left({R\over\xi_1}\right)^{(4-3\gamma)/(2-\gamma)},
\label{eq:poly}\end{equation}
where $\xi_1$ and $\theta_1^\prime$ are constants that depend on
$\gamma$.  We therefore find for low-density stars that $M\propto
K^{3/2}R^0$, or the total mass $M$ is independent of the stellar
radius $R$.  This is in fact true, as the mass approaches the
so-called minimum neutron star mass, about $0.1M_\odot$ as the
compactness $\beta={GM/Rc^2}\rightarrow0$.  On the other hand, for
high-density stars, we find $R\propto K^{1/2}M^0$, or the radius is
independent of the mass.  Once again, this is approximately true, at
least until the compactness $\beta$ becomes large (i.e.,
$\beta\simge1/6$) and GR can no longer be ignored.  In
both cases, the value of $K$ is quantitatively important in
determining either the limiting mass or radius.  These features are apparent 
in the $M-R$ curves displayed in Fig. \ref{fig:mr}.

The close connection between neutron star radii and $K$ was exploited by Lattimer \& Prakash~\cite{Lattimer01} who found phenomenological
correlations between the pressure of neutron star matter at selected densities, $p_\beta(n)$ in units of MeV fm$^{-3}$, and the radius of a typical $1.4M_\odot$ star\footnote{This displayed correlations, and one standard deviation errors,
are revisions~\cite{Lattimer13} that incorporate only those EOSs that can support observed neutron star masses.}
\begin{align}
R_{1.4}=(9.52\pm0.49)[p_\beta(n_s)]^{1/4}{\rm~km};\, R_{1.4}&=(5.68\pm0.14)[p_\beta(2n_s)]^{1/4}{\rm~km}.
\label{eq:radius}
\end{align}
The spread of $R_{1.4}$ observed in Fig. (\ref{fig:mr}), 
between 9 and 15 km, reflects uncertainties in the pressure.  
Fig. (\ref{fig:mr}) shows, however, that the spread of $R_{1.4}$ becomes
smaller (11--15 km) if EOSs are required to support 
masses of $2.0M_\odot$ rather than $1.4M_\odot$.  
We will show that the radius uncertainty is further reduced 
due to additional theoretical and experimental constraints.
 
The introduction of GR results in strong gravity for large compactness
(small values of $\beta$).  The structure equations become
\begin{equation}
{dP\over dr}=-{G(mc^2+4\pi r^3p)(\epsilon+p)\over rc^4(r-2Gm/c^2)};\qquad {dm\over dr}=4\pi{\epsilon\over c^2} r^2,
\label{eq:rel}\end{equation}
where $\epsilon$ is the total mass-energy density, defined by
$\epsilon=n(m_bc^2+E)$, with $m_b$ the baryon mass and $E$ the
internal energy per baryon. In the GR case, there exists a maximum
mass irrespective of whether or not causality is
violated.  Imposition of causality, of course, lowers the maximum mass. 

An important distinction between GR and the Newtonian case is 
the appearance of the mass density $\rho=nm_bc^2$ in the latter 
instead of the total energy density $\epsilon$.
As a result, only if $\gamma=d\ln p/d\ln\rho\le4/3$ in the
high-density limit will a maximum mass exist in the Newtonian
case\footnote{Eq. (\ref{eq:poly}) shows that
  $M\propto\rho_c^{(3\gamma-4)/2}$, so if $\gamma>4/3$, $M$ increases
  without bound with increasing central density $\rho_c$.}.
Enforcing causality in the Newtonian sense therefore results in a maximum
mass, since $\gamma\le1$ in this case in the high-density limit.
For neutron stars this is an artificial situation, however, since enforcing causality in the relativistic sense $dp/d\epsilon=1$
implies that $d\ln p/d\ln\rho=\gamma=2$ for any realistic EOS at high densities, and no maximum mass will therefore exist.
The existence of the Chandrasekhar mass for a Newtonian white dwarf is a consequence of restricting densities to values below
where baryonic pressure becomes substantial so that $\gamma=4/3$.  

A simple demonstration of the GR situation occurs in the
incompressible limit, where the density in the star is constant
irrespective of the pressure.  This is effectively the case
$\gamma\rightarrow\infty$, and Eq. (\ref{eq:newt}) shows that
$M\propto K^0R^3$ in the Newtonian limit and there is no maximum mass.  Since the
central density in the GR case is $\epsilon_c=3Mc^2/(4\pi R^3)$, the
explicit solution of Eq. (\ref{eq:rel}) is
\begin{equation}
p(r)=\epsilon_c\left[{\sqrt{1-2\beta(r/R)^2}-\sqrt{1-2\beta}\over3\sqrt{1-2\beta}-\sqrt{1-2\beta(r/R)^2}}\right].
\label{eq:inc}\end{equation}
The denominator vanishes at the star's center for $\beta=4/9$, which
represents the maximum possible compactness\footnote{Note the
  condition $p_c\le\epsilon_c$, required by causality, gives an even
  more restrictive limit $\beta\le3/8$.}.  These limits to $\beta$
hold for any EOS, not just the incompressible case (see Sec. 11.6 in
Ref. \cite{Weinberg72}).  The $\beta=4/9$ limit is indicated together
with the black hole limit $\beta=1/2$ in Fig. \ref{fig:mr}.
 
The denominator term $r-2Gm/c^2$ has a large effect in
modifying the $M-R$ curves, but it is not the reason a maximum mass exists.  If it is ignored, 
the incompressible fluid has the explicit solution
\begin{equation}
p(r)=\epsilon_c\left({\exp[2\beta(1-r^2/R^2)/3]-1\over3-\exp[2\beta(1-r^2/R^2)/3]}\right).
\label{eq:inc1}\end{equation}
The denominator vanishes at the center for $\beta=(3/2)\ln3\simeq1.65$, so the maximum compactness is about 3.7 times larger than when including this term.\footnote{With the condition $p_c\le\epsilon_c$, the more restrictive limit $\beta=(3/2)\ln2\simeq1.04$ is obtained.}
 
\subsection{The {\it Maximally-Compact} EOS: The Roles of Causality and the Maximum Mass\label{sec:mceos}}

An incompressible fluid has an infinite sound speed squared, $c_s^2/c^2=s=\partial p/\partial\epsilon$, and is therefore unrealistic.  Imposition of causality, namely that $s\le1$,  leads to
a larger minimum radius, $R\simge2.823GM/c^2$~\cite{Haensel99}.  Koranda, Stergioulas \& Friedman~\cite{Koranda97} conjectured that the most compact
configurations are produced when the low-density EOS is very soft (i.e., low pressure) and the high-density EOS is very stiff.   The {\it maximally-compact} EOS
is therefore
\begin{align}
p=\epsilon-\epsilon_0\quad{\rm when}\quad\epsilon\ge\epsilon_0;\qquad\qquad
p=0\quad{\rm when}\quad\epsilon<\epsilon_0.
\end{align}
In this case, the structure equations can be cast into a scale-free form by substitution of the variables
\begin{equation}
 w=\epsilon/\epsilon_0,\qquad x=r\sqrt{G\epsilon_0}/c^2,\qquad y=m\sqrt{G^3\epsilon_0}/c^4.
\label{eq:dim}\end{equation}
There is a single parameter $\epsilon_0$, aside from the constants $G$ and $c$, and the dimensionless TOV equations become
\begin{align}
{dw\over dx}=-{(y+4\pi x^3w)(2w-1)\over x(x-2y)},\qquad {dy\over dx}=4\pi x^2w,
\end{align}
with boundary values $y=0$ and $dw/dx=0$ when $x=0$.  The surface is where the pressure vanishes, or $w=1$.  At the surface,
$x$ and $y$ attain the values $x_1=x(w=1)$ and $y_1=y(w=1)$, respectively.  The
central value of $w$, $w_0$, defines a family of solutions.  The {\it maximally compact} solution occurs when
$y_1$ is maximized.  This occurs when $w_{0,c}=3.034$, for which $x_{1,c}=0.2405$ and $y_{1,c}=0.08522$ \cite{Lattimer11}.  This gives the causally limited compactness as $\beta_c=y_{1,c}/x_{1,c}=0.3543=1/2.823$, which is independent of $\epsilon_0$.

The maximum neutron star mass is scaled by $\epsilon_0$:
\begin{equation}
M_{max}={y_{1,c}c^4\over\sqrt{G^3\epsilon_0}}\simeq4.09\sqrt{\epsilon_s\over\epsilon_0}M_\odot,
\end{equation}
where $\epsilon_s\simeq150$ MeV fm$^{-3}$ is the energy density at
$n_s$.  This is essentially the result first found by Rhoades \&
Ruffini~\cite{Rhoades74}.  Since nuclear structure shows that a phase
transition does not occur below $n_s$, the maximum neutron star mass
must be less than about $4.1M_\odot$.  In addition, the largest
accurately measured neutron star mass,
$M_{max}\simeq1.97M_\odot$~\cite{Demorest10,Antoniadis13}, which must be less than
the true maximum mass, gives an upper limit to $\epsilon_0$,
$\epsilon_{0,max}$.  Upper limits to the central energy density and
pressure then follow:
\begin{align}
\epsilon_{c,max}&=w_{0,c}\epsilon_{0,max}\simeq50.8\epsilon_s\left({M_\odot\over M_{max}}\right)^2,\nonumber\\
 p_{c,max}&=(w_{0,c}-1)\epsilon_{0,max}\simeq34.1\epsilon_s\left({M_\odot\over M_{max}}\right)^2.
\label{eq:pe}\end{align}
The discovery of a star with $M>1.97M_\odot$ would decrease these limits.

\begin{figure}[h]\vspace*{-1.5cm}
\center\includegraphics[width=12cm,angle=180]{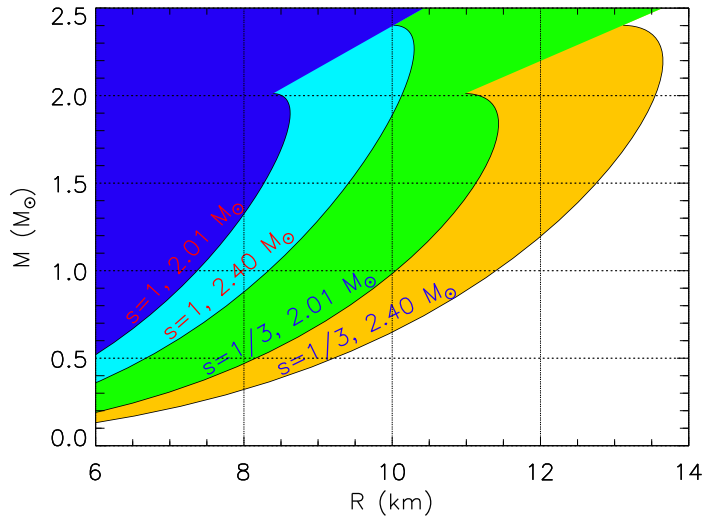}\vspace*{-1.3cm}
\caption{Mass-radius diagram showing regions excluded by causality and the {\it maximally compact} conjecture.  Boundaries are indicated for two values of $M_{max}$ ($2.01M_\odot, 2.4M_\odot$) and for two values of sound speed squared $s$ (1, 1/3).}
\label{fig:maxr}
\end{figure}
 
\begin{figure}[h]\vspace*{-.5cm}
\center\hspace*{-1.5cm}\includegraphics[width=12cm,angle=180]{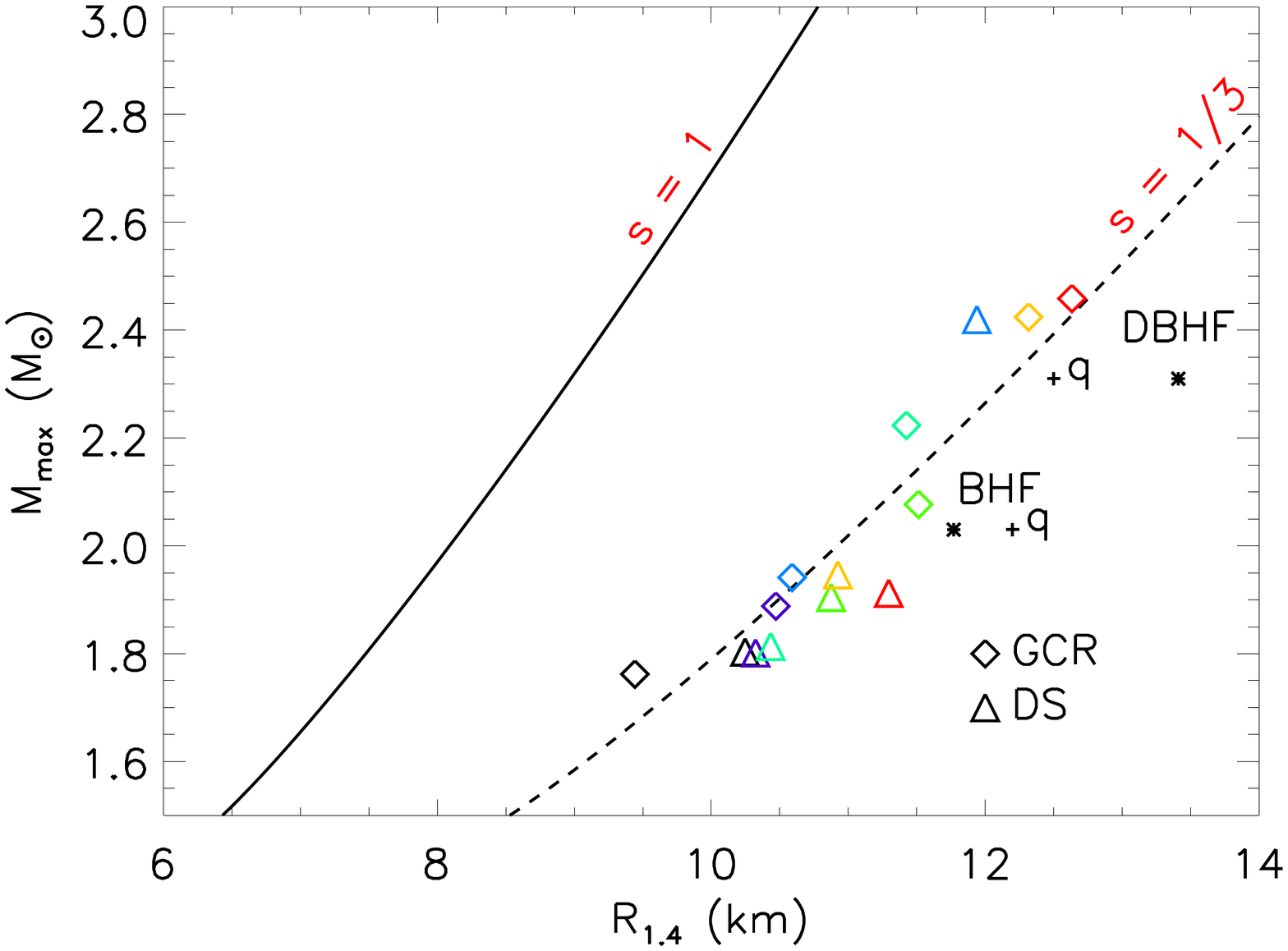}\vspace*{-1cm}
\caption{The minimum radii of $1.4M_\odot$ stars, as a function of $M_{max}$, permitted by causality and the maximum compactness conjecture.  Results are shown for $s=1/3$ and $s=1$.
The symbols show the maximum massses and values of $R_{1/4}$ obtained from TOV integrations after extrapolating different neutron matter calculations from Refs.~\cite{Gandolfi12,Drischler14} using Eq. (\ref{eq:nmext}) to arbitrarily high densities.}
\label{fig:maxr14}
\end{figure}

Another result immediately follows: the minimum radius for a neutron star of mass $M\le M_{max}$ is determined by the coupled equations \cite{Lattimer12}
\begin{equation}
M=M_{max}{y_c\over y_{c,1}},\qquad R={GM_{max}\over c^2}{x_c\over y_{c,1}}={GM\over c^2}{x_c\over y_c}.
\end{equation}
where $x_c(y_c)$ is the {\it maximally compact} solution.  For a given $M$,
$y_c$ is found, and then $R$ is found using $x_c(y_c)$.  For masses
larger than $M_{max}$, the radius limit is $R>2.823GM/c^2$.  This
result is indicated in Fig. \ref{fig:mr} assuming
$M_{max}=1.97M_\odot$ and in Fig. \ref{fig:maxr} for
$M_{max}=2.01M_\odot$ and $2.4M_\odot$.  Fig. \ref{fig:mr} shows that a
number of EOSs are ruled out by the observation that
$M_{max}\simge2M_\odot$ stars.  The green shaded region is ruled out
by observations of these stars.  Should more massive neutron stars be
found, progressivly larger $M-R$ regions become inaccessible.
Fig. \ref{fig:maxr14} shows the limiting radii for a typical
$1.4M_\odot$ star as a function of $M_{max}$.  Also displayed is the
equivalent boundary if the constraint $s=1/3$ is alternately employed.
A phenomenolgical result noted by Ref. \cite{Lattimer12} is that models of neutron
stars containing deconfined quark matter or cores with a mixed
quark-hadron phase instead have radii limited by the {\it maximally compact}
EOS assuming $s=1/3$, i.e., $p=(\epsilon-\epsilon_0)/3$ for
$\epsilon>\epsilon_0$.  This constraint changes the eigenvalues
[$w_{0,c}=4.826$, $x_{1,c}=0.1910$ and $y_{1,c}=0.05169$] and
dramatically increases the limiting radii, as shown in
Fig. \ref{fig:maxr}.  This limit, as shown below, might be closer to
the the radii of realistic EOSs based on experimental studies than is
the $s=1$ boundary, but there are no physical grounds for taking it as
a limit.  It is useful to display the limiting radii for $1.4M_\odot$
stars as a function of $M_{max}$, and this is done in
Fig. \ref{fig:maxr14}.

The limiting radii provided by the $s=1$ assumption, while interesting, assumed $p=0$ at low densities and can be improved with constraints
on the low-density EOS.  Neutron stars are believed to have crusts, and experimental and theoretical information can be brought to bear on the properties of matter near $n_s$.  In the following, we make use of these constraints to examine refined, model-independent limits to neutron star radii.
\subsection{The Neutron Star Crust and the Low-Density EOS\label{Sec:crust}}

There are many observations that indicate most, if not all, neutron
stars have hadronic crusts, {\it i.e.}, they have a surface region with
densities less than approximately $10^{14}$ g cm$^{-3}$ composed
largely of nuclei, neutrons and electrons in beta equilibrium
\cite{Baym71,Kobyakov14}.  Observations of pulsar glitches \cite{Link99}, neutron star cooling
following transient heating events \cite{Brown98}, and the general consistency of
observed thermal emissions from cooling neutron stars \cite{plps04}
favor the existence of a hadronic crust.\footnote{There exist, however, alternative interpretations of these observations involving
a thin ($M\simeq10^{-5}M_\odot$), electrostatically-supported, hadronic shell overlaying a high-density pure strange quark star (also known as
a self-bound star) \cite{Blaschke01,Sedrakian15}.}  The pressure in the crust is largely due
to relativistic, degenerate electrons with at most a 5\% contribution
from nuclei and neutrons.  Since the nuclei are in pressure
equilibrium with the neutrons, they individually contribute almost no
pressure since their internal baryon density is close to the nuclear
saturation density $n_s\simeq0.16$ fm$^{-3}$ or
$\rho_s\simeq2.7\times10^{14}$ g cm$^{-3}$ where uniformly dense
symmetric matter has zero pressure.  The major contribution of baryons
to the pressure is from the collective Coulomb pressure due to the
nuclear lattice, and is therefore largely independent of uncertainties
in the nuclear matter EOS.  

\begin{figure}[h]\vspace*{-1.25cm}
\includegraphics[width=10cm,angle=180]{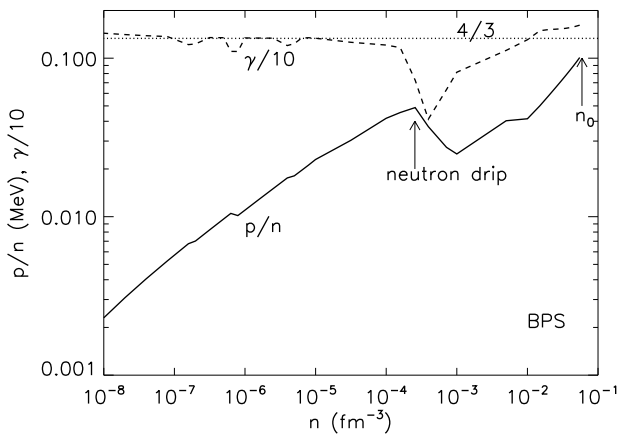}\vspace*{-.5cm}
\caption{The BPS~\cite{Baym71} EOS.  The solid line is the pressure, and the dashed line is the polytropic exponent.  The value corresponding to $\gamma=4/3$ is shown as the dotted line.  The neutron drip density and the core-crust transition density $n_0$ are shown.}
\label{fig:pbps}
\end{figure}

Although the average value of $\gamma$ in the crust, as expected, is about 4/3, it
drops significantly following the onset of neutron drip at a density
$4.3\cdot10^{11}$ g cm$^{-3}$, due to the introduction of a new degree
of freedom.  It then recovers, and even exceeds 4/3 at densities just
below $n_s$.  Fig. \ref{fig:pbps} shows the pressure and polytropic
index as a function of density within the crust for the BPS EOS
\cite{Baym71}.  Estimates from neutron matter calculations \cite{Hebeler13} indicate that the transition from
crustal material to uniform nuclear matter occurs in the range
$0.4\simle n_0/n_s\simle0.5$.  We will assume the crust-core transition density
is $n_0=n_s/2.7$, and use the BPS \cite{Baym71} crustal EOS, for which
the pressure, energy density and internal energy per baryon  at $n_0$ are
\begin{equation}
 p_0=0.243{\rm~MeV~fm}^{-3}, \qquad\epsilon_0=56.39{\rm~MeV~fm}^{-3},\qquad E_0=12.13{\rm~MeV}.  
\label{eq:bps}\end{equation}
Results are relatively insensitive to the exact location of the core-crust boundary.

\subsection{The Intermediate EOS and Pure Neutron Matter\label{Sec:nm}}

Matter in neutron stars at densities between $n_0$
  and at least $2n_s$ is most likely a structureless fluid of nucleons which is 
  extremely neutron rich due to the condition of beta
  equilibrium.  This condition is equivalent to minimization of the
  total energy per baryon with respect to the charge fraction
  $x=n_p/n$ where $n_n$ and $n_p$ are the neutron and proton baryon
  densities, respectively, and $n=n_n+n_p$.  The difference between
  the energy of pure neutron matter and symmetric matter (with equal
  numbers of neutrons and protons) is called the nuclear symmetry
  energy $S(n)$, and the energies of intermediate proton fractions can be
  approximated with a quadratic interpolation between these extremes:
\begin{equation}
E(n,x)\simeq E_{1/2}(n)+S(n)(1-2x)^2,
\label{eq:e}\end{equation}
where $E_{1/2}(n)$ is the energy per baryon of symmetric matter.  We define $E_{1/2}(n_s)=-B$, where $B\simeq16$ MeV is the bulk binding energy per baryon of symmetric matter at the saturation density.  We make the distinction here between $S(n)$ and $S_2(n)$ defined by
\begin{equation}
S_2(n)=(1/8)(\partial^2 E(n,x)/\partial x^2)_{x=1/2}.
\label{eq:s2n}\end{equation}
In many treatments, $S_2$ is also called the symmetry energy, but only if the quadratic approximation is
valid will $S(n)=S_2(n)$.  A convenient estimate for the symmetry energy near $n_s$ is
\begin{equation}
S(n)=S_{v}(n/n_s)^\gamma; 
\label{eq:s}\end{equation}
$S_v=S_2(n_s)$.  Nuclear experimental information and
neutron matter calculations indicate that
$26{\rm~MeV}\simle S_{v}\simle34{\rm~MeV}$ and
$0.3\simle\gamma\simle0.7$, with the value of $\gamma$ positively correlated with $S_v$.  The pressure corresponding to Eqs. (\ref{eq:e}) and (\ref{eq:s}) is
\begin{equation}
p(n,x)=n^2{\partial E(n,x)\over\partial n}\simeq p_{1/2}(n)+S_{v}\gamma n_s\left({n\over n_s}\right)^{\gamma+1}(1-2x)^2,
\label{eq:p}\end{equation}
where $p_{1/2}(n)$ is the pressure of symmetric matter.  Note that, by definition, $p_{1/2}(n_s)=0$; to leading order, near $n_s$, the symmetric matter pressure increases linearly with density, $p_{1/2}(n)\simeq(K_s/9)(n-n_s)$.

Matter in neutron stars is in beta equilibrium with $\mu_n-\mu_p=\mu_e$, which follows from minimizing the total baryon and electron energies with respect to the proton fraction $x$:
\begin{equation}
{\partial[E(n,x)+E_e(n,x)]/\partial x}=0.
\label{eq:ex}\end{equation}
$E_e=(3/4)\hbar cx(3\pi^2nx)^{1/3}$ is the electron energy per baryon assuming relativistic degeneracy.  For the symmetry energy ansatz Eq. (\ref{eq:s}), applicable to uniform nucleonic matter, this is equivalent to
\begin{equation}
4S_{v}\left({n/ n_s}\right)^\gamma(1-2x)=\hbar c(3\pi^2nx)^{1/3}.
\label{eq:be}\end{equation}
This can be solved as a cubic equation for $x$ at a specific density, but since $x$ is small, has the approximate solution
\begin{equation}
x_\beta\simeq{64S_{v}^3(n/n_s)^{3\gamma}\over3\pi^2n_s(\hbar c)^3+384S_{v}^3(n/n_s)^{3\gamma}}
\label{eq:be1}\end{equation}
which has the value $x_\beta\simeq0.040$ when $n=n_s$ and $S_v=31$ MeV.\footnote{In reality, muons should be included because $\mu_e=\hbar c(3\pi^2n_sx_\beta)^{1/3}\simeq113$ MeV $>m_\mu c^2\simeq105$ MeV.  This omission has little effect on our discussion.}  The pressure of pure neutron matter using Eq. (\ref{eq:s}), at $n_s$, is $p(n_s)=\gamma n_sS_{v}.$  In beta equilibrium, to lowest orders, this is modified:
\begin{equation}
p_\beta(n_s)\simeq\gamma n_sS_v\left[1-\left({4S_{v}\over\hbar c}\right)^3{4\gamma-1\over3\pi^2n_s\gamma}\right].
\label{eq:be2}\end{equation}
The correction term in Eq. (\ref{eq:be2}) is of order 1.4\%, and can be ignored to good approximation.  At higher densities, the proton fraction and the correction term generally increase due to the increasing symmetry energy.  There is also a contribution from $p_{1/2}(n)$.  However, for densities up to $2n_s$ the neutron star matter pressure is essentially equivalent to pure neutron matter pressure.  \\

\begin{table}[h]
\begin{tabular}{|c||c|c|c|c||c|c|c|c|c|c|c|c|c}\hline
 Model & $a$ & $\alpha$ & $b$ & $\beta$ & $S_v$ & $L$ &  $p_1$ & $\gamma_1$ \\ 
& MeV && MeV && MeV & MeV  & MeV fm$^{-3}$& \\ \hline
GCR 0 & 12.7 & 0.49 & 1.78 & 2.26 & 30.5 & 31.3 & 7.272 & 2.113\\
GCR 1 & 12.7 & 0.48 & 3.45 & 2.12 & 32.1 & 30.8 & 10.402 & 2.335\\
GCR 2 & 12.8 & 0.488 & 3.19 & 2.20 & 32.0 & 40.6 &10.537 & 2.343\\
GCR 3 & 13.0 & 0.475 & 3.21 & 2.47 & 32.0 & 44.0 & 13.274 & 2.487\\
GCR 4 & 12.6 & 0.475 & 5.16 & 2.12 & 33.7 & 51.5 & 14.304 & 2.533\\
GCR 5 & 13.0 & 0.50 & 4.71 & 2.49 & 33.8 & 56.2 & 18.678 & 2.700\\
GCR 6 & 13.4 & 0.514 & 5.62 & 2.436 & 35.1 & 63.6 &  20.933 & 2.770\\ \hline
DHS 0 & 10.94 & 0.459 & 4.106 & 1.977 & 31.1 & 39.4  & 8.125 & 2.182 \\
DHS 1 & 11.00 & 0.460 & 4.425 & 1.947 & 31.4 & 41.0  & 8.453 & 2.206\\
DHS 2 & 11.95 & 0.495 & 3.493 & 2.632 & 31.4 & 45.3 & 13.760 & 2.509\\
DHS 3 & 11.02 & 0.460 & 4.683 & 1.935 & 31.7 & 42.4  & 8.768 & 2.229\\
DHS 4 & 10.95 & 0.454 & 5.158 & 1.972 & 32.1 & 45.4  & 9.676 & 2.290\\
DHS 5 & 10.34 & 0.429 & 4.954 & 2.024 & 31.3 & 43.4  & 9.180 & 2.258\\
DHS 6 & 10.29 & 0.433 & 7.227 & 1.842 & 33.5 & 53.3  & 11.241 & 2.384\\ \hline
\end{tabular} 
\caption{Neutron matter calculations fit to the energy parameterization of Eq. (\ref{eq:nmext}).  GCR are models from 
Ref.~\cite{Gandolfi12}; DHS are models from Drischler and Schwenk (2015, unpublished).}
 \label{tab:nm}
\end{table}

  Recent calculations of the properties of pure neutron matter have
  produced estimates of the pressure-energy density relation up to
  about $2n_s$.  Ref.~\cite{Gandolfi12} showed that the neutron matter energy for densities less than about $2n_s$ was adequately approximated by the double power law
\begin{equation}
E(n,0)\simeq a(n/n_s)^\alpha+b(n/n_s)^\beta
\label{eq:nmext}\end{equation}
where $a, b, \alpha$ and $\beta$ are parameters.
Table \ref{tab:nm} displays parameter values found by Ref.~\cite{Gandolfi12} for quantum Monte Carlo neutron matter calculations.
We have also displayed parameter values that fit the neutron matter
results of unpublished calculations of Drischler \& Schwenk up to densities $\simeq1.5n_s$.

Naively, we can extend neutron matter calculations to arbitrarily higher densities using Eq. (\ref{eq:nmext}).  
Doing so, and using the TOV equations to produce $M-R$ curves, we find many of the neutron matter calculations in Table \ref{tab:nm} to be too soft to support the observed value of $M_{max}$ (Fig. \ref{fig:maxr14}).  It appears that the EOS must become substantially stiffer at densities not far above $n_s$ in order that observed neutron star masses can be explained.  Nevertheless, it is interesting to observe that $M_{max}$ and $R_{1.4}$ for these extrapolations are near the $s=1/3$ boundary for the {\it maximally compact} EOS.  
 
\section{Nuclear Structure and the Nuclear Symmetry Energy}
To confirm the understanding of the intermediate EOS, we turn to experimental data for nuclei which explore the nuclear symmetry energy.
Experimental information concerning the
symmetry energy is usually encoded in the parameters $S_v$ and $L$,
defined as
\begin{align}
S_v&\equiv{1\over8}\left({\partial^2E(n,x)\over\partial x^2}\right)_{n_s,1/2}\simeq S(n_s)\nonumber\\
L&\equiv{3\over8}\left({\partial^3E(n,x)\over\partial n\partial x^2}\right)_{n_s,1/2}.
\label{eq:sl}
\end{align}
For the symmetry energy of Eq. (\ref{eq:s}), one finds $L=3\gamma S_v$ so that $0.9S_v\simle L\simle2.1S_v$.  For the energy
formula Eq. (\ref{eq:nmext}), we find that $S_v=B+a+b$ and $L=3(a\alpha+b\beta)$.
  For each set of neutron matter calculations, the corresponding values of $S_v$, $L$ have been tabulated in Table \ref{tab:nm}, which reveals that 30.5 MeV $\le S_v\le35.1$ MeV and 30.8 MeV $\le L\le63.6$ MeV.  As anticipated, these parameters are positively correlated.  In the quadratic approximation for the isospin dependence of the nucleon energies, $L=3p(n_s,0)/n_s$. 

\subsection{Hydrodynamic Model of the Nuclear Symmetry Energy}
A variety of experiments reveal information about the nuclear symmetry energy.  None of these experiments are capable, at the present time, of individually pinpointing the parameters $S_v$ and $L$.  Rather, each experiment reveals that the parameters are correlated.   The most important concern nuclear binding energies, the neutron skin thickness in neutron-rich nuclei, and giant dipole resonances.  Each type of measurement provides a different correlation.  By combining different experiments, and examining different correlations, one can restrict ranges for the symmetry parameters.   We begin the discussion by exploring an analytic nuclear model that reveals the origin of these correlations.

Lipparini \& Stringari~\cite{Lipparini89} explored a model with a simplified nuclear Hamiltonian energy density
\begin{eqnarray}\label{ham}
{\cal H}&=&{\cal H}_B(n,\alpha)+{\cal Q}(n)(dn/dr)^2, \cr 
{\cal H}_B(n,\alpha)&=&{\cal H}_B(n,0)+v_{sym}(n)\alpha^2
\end{eqnarray}
where the uniform matter contribution is ${\cal H}_B(n,\alpha)$,
${\cal Q}(n)$ controls the gradient contributions, $v_{sym}=S_2/n$, and $\alpha=n-2n_p=n_n-n_p$ is the
isovector density. 
In nuclei, the charge repulsion among the protons redistributes
neutrons and protons and reduces the neutron skin thickness. To take
this into account, and to extend the model of Ref. \cite{Lipparini89}, we 
include a Coulomb contribution:
\begin{equation}
{\cal H}={\cal H}_B(n,\alpha)+{\cal H}_C(n,\alpha)+{\cal Q}(n)(dn/dr)^2,
\label{eq:ham1}\end{equation} 
where, in spherical symmetry, ${\cal H}_C=n_pV_C/2$ and the Coulomb potential is
\begin{equation}\label{coul}
V_C(r)={e^2\over r}\int_0^r n_p(r^\prime)~d^3r^\prime+\int_r^\infty {e^2\over r^\prime} n_p(r^\prime)~d^3r^\prime.
\end{equation}
If the protons are uniformly distributed for $r<R$,
\begin{equation}\label{coul1}
V_C={Ze^2\over R}\left({3\over2}-{r^2\over2R^2}\right)
\end{equation}
for $r<R$ and $V_C=Ze^2/r$ for $r>R$. We have found that a reasonable
approximation for a Woods-Saxon proton distribution, and one that
keeps the model analytic, is provided by assuming Eq.
(\ref{coul1}) to apply for all $r$. Furthermore, the Coulomb potential
and the total Coulomb energy when the Coulomb potential is
self-consistently determined are adequately described by the same
approximation. Where the discrepancy between this approximation and
the real potential is large, the proton density is small.

We now optimize the total nuclear energy with respect to the densities $n$ and $\alpha$
subject to the constraints
\begin{equation}\label{const}
A=\int\rho ~d^3r,\qquad N-Z=\int\alpha ~d^3r,
\end{equation}
 producing the chemical potentials $\mu$ and $\bar\mu$:
\begin{equation}\label{ham1}
{\delta\over\delta n}[{\cal H}-\mu n]=0,\qquad {\delta\over\delta\alpha}[{\cal H}-\bar\mu\alpha]=0.
\end{equation}
These lead to
\begin{eqnarray}\label{ham2}
\mu&=&{\partial[{\cal H}_B+{\cal H}_C]\over\partial n}-2{d\over dr}\left[{\cal Q}n^\prime\right]+{\partial{\cal Q}\over\partial n}\left(n^\prime\right)^2,\cr
\bar\mu&=&{\partial[{\cal H}_B+{\cal H}_C]\over\partial\alpha}.
\end{eqnarray}

\subsubsection{The isoscalar density $n$}
Multiplying the first of Eq.~(\ref{ham1}) by $dn/dr$
and the second by $d\alpha/dr$, their sum can be integrated:
\begin{equation}\label{ham3}
{\cal Q}(n)(dn/dr)^2={\cal H}_B(n,\alpha)+{\cal H}_C(n,\alpha)-\mu n-\bar\mu\alpha,
\end{equation}
for which the boundary condition $\mu n_o+\bar\mu\alpha_o={\cal
  H}_B(n_o,\alpha_o)$ at the center. 

We make the quadratic approximation for the energy density of uniform symmetric
matter:
\begin{equation}\label{bulk}
{\cal H}_B(n,0)=n\left[-B+{K_s\over18}(1-u)^2\right]
\end{equation}
with bulk binding energy $B$, and
$u=n/n_s$. In the case
that $\alpha_o\simeq0$, one has $n_o=n_s=0.16$ MeV fm$^{-3}$.
For laboratory nuclei, $\alpha_o$ is small. We are primarily interested in the behavior of $\alpha$, and will assume that the total density, to lowest order, can be found by assuming $\alpha\simeq0$ in Eq. (\ref{ham3}). We also
approximate ${\cal Q}(n)=Q/n$, with $Q$ constant, which leads to an equation for the
isoscalar density as a function of position:
\begin{equation}\label{bulk1}
{du\over dz}=-u(1-u),\qquad a=3\sqrt{2Q\over K_s},
\end{equation}
where $u=n/n_o$ and $z=r/a$, where the surface thickness
parameter is $a$. This has the solution of a Fermi function, or
Woods-Saxon distribution (which was assumed by Ref.~\cite{Lipparini89}),
\begin{equation}\label{fermi}
u={1\over1+e^{z-y}}.
\end{equation}
Here $y$ is a constant of integration, determined from the first
of the constraints Eq. (\ref{const}):
\begin{eqnarray}
A&=&\int n~d^3r=4\pi n_o a^3F_2(y),\cr
\hspace*{-.5cm}F_i(y)&=&\int_0^\infty\!\!\!\!{z^i~dz\over1+e^{z-y}}\simeq{y^{i+1}\over i+1}\left[1+{i(i+1)\over6}\left({\pi\over y}\right)^2+\cdots\right].
\label{eq:a}\end{eqnarray}
Here $F_i$ is the usual Fermi integral, and the right-most
approximation holds for $y>>1$ and $i\ne-1$ if we ignore an exponentially
small term. This is justified, since one finds [ Eq. (\ref{eq:a})] that $y\simeq
R/a\simeq13$ for $^{208}$Pb ($a$ is evaluated
below).

The parameter $K_s\simeq240$ MeV from experiment\cite{Youngblood99,Garg04,Colo04}, and the value of $Q$
follows from the observed value\cite{Myers69} of the 90-10 surface thickness:
\begin{equation}\label{bulk2}
t_{90-10}=a\int_{0.1}^{0.9}{du\over du/dz}=4a\ln(3)\simeq2.3{\rm~fm},
\end{equation}
giving $a=0.523$ fm and
\begin{equation}\label{bulk3}
Q={K_s\over18}\left({t_{90-10}\over4\ln(3)}\right)^2\simeq3.65{\rm~MeV~fm}^2.
\end{equation}
As a check, the liquid droplet surface tension parameter is the
semi-infinite, symmetric matter, surface thermodynamic potential per
unit area:
\begin{eqnarray}\label{bulk4}
\hspace*{-.5cm}\sigma_o&=&\int[{\cal H}-\mu n]~dz=2Q\int_0^\infty{1\over n}\left({dn\over dz}\right)^2~dz\cr
&=&{2Qn_o\over a}\int_0^1(1-u)~du
={Qn_o\over a}\simeq1.17{\rm~MeV~fm}^{-2}.
\end{eqnarray}
This gives a value $E_s=4\pi r_o^2\sigma_o\simeq19.2$ MeV, with $n_o=4\pi r_o^3/e$, for the
 surface energy parameter in the liquid droplet model,
 very close to the accepted value \cite{Myers66,Myers69}.
Therefore, the simple approximations Eq. (\ref{bulk}) and ${\cal Q}=Q/n$ fit the
most important observed properties of the symmetric matter surface, its tension and thickness, as well as the matter's observed 
incompressibility.

\subsubsection{The isovector density $\alpha$}
Using the second of the variations Eq. (\ref{ham2}) leads to
\begin{equation}
\alpha={\bar\mu+\partial{\cal H}_C/\partial\alpha\over2v_{sym}}={\bar\mu+V_C/2\over v_{sym}},
\label{eq:alpha}\end{equation}
where we note that $n_pV_C$ is proportional to $n_p^2=(n-\alpha)^2/4$
and we treated $n_pV_C$ to be primarily a function of $\alpha$ and not
$n$.  This can be justified on the grounds that while the Coulomb
potential acts directly on the protons only, the neutrons adjust
due to the strong symmetry interaction.

Using the second of the constraints (Eq. (\ref{const})), we obtain
\begin{equation}\label{ham4}
N-Z=\int{\bar\mu+V_C/2\over2v_{sym}}~d^3r\equiv{\bar\mu\over2}H_0+{G\over4},
\end{equation}
where we define $H_i$ and $G$ as
\begin{equation}
H_i=\int\left({r\over R}\right)^i{d^3r\over v_{sym}},\qquad G=\int {V_c\over v_{sym}}~d^3r={Ze^2\over2R}(3H_0-H_2).
\label{eq:HG}\end{equation}  
It then follows that
\begin{equation}\label{ham5}
\alpha={N-Z-(G-V_CH_0)/4\over v_{sym}H_0}.
\end{equation}

\subsubsection{Nuclear structure}
The total symmetry and Coulomb energy is
\begin{align}
E_{sym}+E_C=&\int v_{sym}\alpha^2~d^3r+{1\over4}\int(n-\alpha)V_C~d^3r\nonumber\\
=&{(N-Z)^2\over H_0}-{(N-Z)G\over4H_0}+{1\over4}\int nV_Cd^3r\label{eq:esym}\\
=&{(N-Z)^2\over H_0}+{3\over5}{Z^2e^2\over R}+{Ze^2\over8R}(N-Z)\left({H_2\over H_0}-{3\over5}\right).\nonumber
\end{align}
We used $\int nr^2~d^3r\simeq3AR^2/5$ which follows from Eq. (\ref{eq:a}) to lowest order.
\footnote{The total symmetry and Coulomb energy is independent of including polarization effects on $\alpha$, i.e., treating ${\cal H_C}$ to be a function of $\alpha$.}

The dipole static polarizability, $\alpha_D$, is found by performing the constrained variation\cite{Lipparini89}
\begin{equation}\label{dipole}
{\delta\over\delta\alpha}\left(\int{\cal H}~d^3r-\epsilon\int z\alpha~ d^3r\right)=0,
\end{equation}
with $\epsilon$ a small parameter. Defining $\alpha_d$ as the function
$\alpha(r)$ which solves Eq.~\ref{dipole}, the dipole polarizability
is
\begin{equation}
\alpha_D={1\over2\epsilon}\int z\alpha_d~ d^3r.
\end{equation}
The solutions for $\alpha_d$ and the dipole
polarizability are
\begin{equation}\label{dipole2}
\alpha_d={\epsilon z+V_C\over2v_{sym}},\qquad \alpha_D={1\over4\epsilon}\int z~{\epsilon z+V_C\over v_{sym}}~d^3r={R^2H_2\over12},
\end{equation}
where $z^2=r^2/3$ within the integral.  \footnote{The second term in the middle expression for $\alpha_D$ in Eq. (\ref{dipole2}) vanishes because of symmetry, so that this result is also independent of including polarization effects on $\alpha$.}

From Eqs. (\ref{ham5}) and (\ref{eq:HG}), the central isovector density is
\begin{equation}\label{skin2}
\alpha_o={N-Z-G/4+V_{C,o}H_0/4\over v_{sym,o}H_0}={n_o\over S_vH_0}\left[N-Z+{Ze^2\over8R}H_2\right],
\end{equation}
where $v_{sym,o}=S_v/n_o$ and $V_{C,o}$ are central values. 
Polarization thus results in an increase in asymmetry near the nuclear center.

The neutron skin thickness $R_n-R_p$, the difference between the mean
radii of neutrons and protons, is defined by
\begin{equation}\label{skin3}
{4\pi\over3}\left(R_n^3-R_p^3\right)=\int\left({n_n\over n_{no}}-{n_p\over n_{po}}\right)~d^3r
\end{equation}
where $n_{no}$ and $n_{po}$ are the central values of the neutron and
proton densities. $R_n$ and $R_p$ represent the ``squared-off'' radii.
We find
\begin{equation}
{n_n\over n_{no}}-{n_p\over n_{po}}={n_o\over2n_{no}n_{po}}\left(\alpha-n{\alpha_o\over n_o}\right)
\label{eq:skin1}\end{equation}
and
\begin{equation}
{4\pi\over3}\left(R_n^3-R_p^3\right)={n_oA\over2n_{no}n_{po}}\left(I-{\alpha_o\over n_o}\right)
\label{eq:skin2}\end{equation}
Treating $R_n-R_p<<R$ and keeping the lowest-order
term:
\begin{align}\label{skin4}
{R_n-R_p\over R}\simeq{2\over3}{I-\alpha_o/n_o\over1-\alpha_o^2/n_o^2}
={2\over3}\left[I\left(1-{A\over S_vH_0}\right)-{Ze^2\over8RS_v}{H_2\over H_0}\right],
\end{align}
where $I=(N-Z)/A$.  The denominator term can be neglected since $I$ and $\alpha_o/n_o$ are of the same small magnitude.
Polarization effects obviously decrease the neutron skin thickness.\footnote{In the case of a symmetric nucleus, there is, in fact, a {\it proton} skin.}

Experimentally, however, it is preferable to measure the differences
of the mean-square neutron and proton radii,
which, if the densities are uniform up to $R_n$ or $R_p$, is $\sqrt{3/5}(R_n-R_p)$.  These radii, in the hydrodynamical model, are
\begin{eqnarray}\label{skin8}
r^2_{n,p}&=&{1\over (N,Z)}\int n_{n,p}r^2~d^3r={1\over2(N,Z)}\int(n\pm\alpha)r^2~d^3r\cr
&=&{R^2\over1\pm I}\left[{3\over5}\pm\left({IH_2\over H_0}+{Ze^2\over8RA}\left({H_2^2\over H_0}-H_4\right)\right)\right],
\end{eqnarray}
where the upper (lower) sign refers to $n(p)$.  We then find
\begin{equation}\label{skin10}
{r_{np}\over R}\simeq\sqrt{3\over5}\left[I\left({5H_2\over3H_0}-1\right)+{5Ze^2\over24RA}\left({H_2^2\over H_0}-H_4\right)\right]\left(1-I^2\right)^{-1/2}.
\end{equation}

\subsubsection{Solutions for arbitrary $v_{sym}$}

Lattimer \& Steiner~\cite{Lattimer14a} noted that if $v_{sym}$ is expanded according to
\begin{equation}
{S_v\over n_ov_{sym}(u)}=\sum_{j=1}^Jb_ju^j,\qquad 0<u<1
\label{vexp}\end{equation}
and the density has the Fermi shape of Eq. (\ref{fermi}), the $H_i$ integrals are analytic:
\begin{equation}
H_i=\int\left({r\over R}\right)^i{d^3r\over v_{sym}}={4\pi n_oa^{3+i}\over S_vR^i}\left[F_{i+2}(y)-(2+i){\cal T}F_{i+1}(y)+\cdots\right],
\label{eq:h}\end{equation}
where
\begin{equation}
{\cal T}=b_2+3b_3/2+11b_4/6+25b_5/12+137b_6/60+\cdots
\label{eq:t}\end{equation}
Note that $\sum_{j=1}^Jb_j=1$ ensures $v_{sym}(u=1)=S_v/n_o$.  In principle, any $v_{sym}$ can be fit if $J$ is large enough, but as a simple example, consider a three term fit to the conventional expression of the symmetry energy
\begin{equation}
S_2(u)\simeq S_v+{L\over3}(u-1)+{K_{sym}\over18}(u-1)^2+\cdots;
\label{eq:s2}\end{equation}
if $K_{sym}=18(L/3-S_v)\simeq-210$ MeV, then $S_2(0)$ approximately vanishes.  Fitting
the energy and its first two derivatives at $u=1$
 give\footnote{The following solutions differ from those in Ref.~\cite{Lattimer14a}.}
\begin{align}
b_1&=1+{L\over3S_v}+\left({L\over3S_v}\right)^2-{K_{sym}\over18S_v}\simeq2.37,\nonumber\\
b_2&={K_{sym}\over9S_v}-{L\over3S_v}-2\left({L\over3S_v}\right)^2\simeq-2.14,\nonumber\\
b_3&=\left({L\over3S_v}\right)^2-{K_{sym}\over18S_v}\simeq0.76,\label{eq:b}\\
{\cal T}&=-{L\over3S_v}-{1\over2}\left({L\over3S_v}\right)^2+{K_{sym}\over36S_v}\nonumber\\
&=-{1\over2}\left[1+{L\over3S_v}+\left({L\over3S_v}\right)^2\right]\simeq-0.992.\nonumber
\end{align}
In the last expression for ${\cal T}$ we used $K_{sym}=18(L/3-S_v)$ following from assuming 
$S_2(0)=0$.\footnote{This fit to Eq. (\ref{eq:s2}) is not very good for small $u$ since $[nv_{sym}(n)]_{n\to0}=S_v/b_1$ which should, in fact, vanish; a more realistic fit with larger $J$ gives more negative values of ${\cal T}$.}

It is now possible to make a connection with the liquid droplet
model~\cite{Myers66,Myers69}.  Without polarization corrections, the
total symmetry energy of a nucleus in the liquid droplet model is
\begin{equation}
E_{sym}={S_vAI^2\over1+S_sA^{-1/3}/S_v},
\label{eq:syme}\end{equation} 
where $S_s$ is the surface symmetry
parameter.  In the hydrodynamic model, this energy is $A^2I^2/H_0$
according to Eq. (\ref{eq:esym}).  Using
Eq. (\ref{eq:h}) to leading orders,
\begin{equation}
H_0={A\over S_v}\left(1+{S_s\over S_vA^{1/3}}\right)= {A\over S_v}\left(1-{3{\cal T}a\over r_oA^{1/3}}\right),
\label{eq:h0}\end{equation}
showing that $S_s=-3{\cal T}S_va/r_o$.  Therefore, once $v_{sym}$ is
specified, the liquid droplet parameter $S_s$ can be calculated and
the total symmetry and Coulomb energy, the dipole polarizability, and
the neutron skin thickness can be found.  For the above example,
$S_s/S_v\simeq1.36$.  It also must follow that
\begin{equation}
H_i={A\over S_v}\left({3\over3+i}+{S_s\over S_vA^{1/3}}\right).
\end{equation}
We can now form final expressions for the dipole polarizability and the neutron skin thickness:
\begin{align}
\alpha_D&\simeq{AR^2\over20S_v}\left(1+{5\over3}{S_s\over S_vA^{1/3}}\right),\label{eq:pol}\\
r_{np}&\simeq\sqrt{3\over5(1-I^2)}{2r_o\over3}\left(1+{S_s\over S_vA^{1/3}}\right)^{-1}
\left[I{S_s\over S_v}-{3Ze^2\over140r_oS_v}\left(1+{10\over3}{S_s\over S_vA^{1/3}}\right)\right].
\label{eq:skin}\end{align}

\subsubsection{Predicted experimental correlations}

Nuclear experimental data concerning a single aspect of nuclear
structure cannot pinpoint the symmetry parameters $S_v$ and $L$
uniquely.  We first consider the correlation implied by nuclear binding 
energies~\cite{Lattimer92,Lattimer96}.  In the liquid 
droplet model~\cite{Weizsacker35,Myers66,Myers69}, the total nuclear energy is 
\begin{equation}
E(N,Z)=-BA+E_sA^{2/3}+E_{sym}+E_{C}+E_{pair}+E_{shell}.
\label{eq:ld}\end{equation}
Ignoring $E_{pair}$ and $E_{shell}$, the pairing and shell corrections which 
are globally small, the free parameters of this expression are $B$, 
the bulk binding energy per nucleon, $E_s$, the surface energy parameter, $r_o$, and $S_v$ and $S_s$ 
if we take Eq.~(\ref{eq:syme}) to represent the symmetry energy $E_{sym}$.  By examining energy differences between nuclei with the same $A$ but different $N$ and $Z$, one can isolate $E_{sym}$ for many nuclear pairs.    In reality, the total Coulomb and symmetry energies contain a term proportional to $I$ as seen in Eq. (\ref{eq:esym}), but this term is generally small and we ignore it for this illustration.  One can thus optimize  the parameters $S_s$ and $S_v$ independently of the other liquid droplet parameters.

However, even though the binding energies of thousands of nuclei are
precisely known, the range of values of $A^{1/3}$ for nuclei more
massive than about $A=20$, large enough for the model to be
applicable, is not large, varying from 3 to 6.  Therefore, only a
correlation between $S_s$ and $S_v$ can be accurately determined.  The
correlation is very nearly linear between $S_s/S_v$ and $S_v$ and can
be expressed as a confidence ellipse in the $S_s/S_v$--$S_v$ plane.
One seeks to minize the quantity
\begin{equation}
\chi^2=\sum_i^N(E_{exp,i}-E_{sym,i})^2/(N\sigma^2),
\label{eq:chi}\end{equation}
where $E_{exp,i}$ and $E_{sym,i}$ (i.e., Eq. (\ref{eq:syme})) are the experimental and predicted
energies of the $i$th nuclear pair, respectively, $N$ is the total number of
nuclear pairs, and $\sigma$ is a nominal error, usually taken to be about 1 to 2
MeV.  A $\chi^2$ contour 1 unit above the minimum, which is the confidence
ellipse, represents the $1\sigma$ confidence level.  The properties of
this ellipse are determined by the second derivatives of $\chi^2$ at
the minimum,\footnote{We neglected terms containing $(E_{exp,i}-E_{sym,i})$ in Eq. (\ref{eq:chivv}) as
they are relatively small near the minimum.} 
\begin{align}
\chi_{vv}=&{\partial^2\chi^2\over\partial S_v^2}\simeq
{2\over{\cal N}\sigma^2}\sum_i^NI_i^4A_i^2\left(1+{S_s\over S_vA_i^{-1/3}}\right)^{-2},\nonumber\\
\chi_{vs}=&{\partial^2\chi^2\over\partial S_v\partial(S_s/S_v)}\simeq
{2\over{\cal N}\sigma^2}\sum_i^NI_i^4A_i^{5/3}S_v\left(1+{S_s\over S_vA_i^{-1/3}}\right)^{-3},\label{eq:chivv}\\
\chi_{ss}=&{\partial^2\chi^2\over\partial(S_s/S_v)^2}\simeq
{2\over{\cal N}\sigma^2}\sum_i^NI_i^4A_i^{4/3}S_v^2\left(1+{S_s\over S_vA_i^{-1/3}}\right)^{-4}.\nonumber
\end{align}
The orientation of the
confidence ellipse with respect to the $S_s/S_v$ axis is
$\theta=(1/2)\tan^{-1}[2\chi_{vs}/(\chi_{vv}-\chi_{ss})]$, and the
error widths are $\sigma_v=\sqrt{\chi^{-1}_{vv}}$ and
$\sigma_s=\sqrt{\chi^{-1}_{ss}}$.  The correlation coefficent is
$\chi_{vs}/\sqrt{\chi_{vv}\chi_{ss}}$.  Because the summations in Eq. (\ref{eq:chivv})
tend to be dominated by the more abundant heavier nuclei, we find 
\begin{equation}
{\chi_{vv}\over\chi_{vs}}\simeq\sqrt{\chi_{vv}\over\chi_{
ss}}\simeq\left<A^{1/3}\right>+{S_s\over S_v}
\label{eq:chirat}\end{equation} 
and so the slope of this correlation in $S_v-S_s/S_v$ space is about 7.\footnote{Had we instead written the symmetry
energy as $I_i^2A_i(S_v-S_sA^{-1/3})$ as in the simple liquid drop model, 
the second derivatives would have been independent of the location of the
minimum in $S_s-S_v$ space and the ellipse properties would have depended
only upon the chosen values of $A_i$ and $I_i$.  The slope of the correlation would
then be $<A^{1/3}>\simeq6$.}
It is apparent that the correlation between $S_s$ and
$S_v$ is dependent on neither  the  model for the nucleus nor the nuclear force properties.

The inferred correlation between $L$ and $S_v$ is dependent upon assumptions about the nuclear interaction.
For the model of Eqs. (\ref{eq:s2}) and (\ref{eq:b}),   $S_s/S_v$ depends only on $L/S_v$, so that
\begin{equation}
{d(S_s/S_v)\over d(L/S_v)}={dS_s/dS_v-S_s/S_v\over dL/dS_v-L/S_v}\simeq(a/2r_o)[1+2L/(3S_v)]\approx0.5,
\label{eq:srat}\end{equation}
and one obtains the slope $dL/dS_v\approx12$ for the energy correlation in $S_v-L$ space.

In contrast to binding energies, which are available for over a thousand nuclei, measurements of giant resonances and skin thicknesses are more limited.  Nevertheless, one can obtain correlations between $S_s$ and $S_v$ by considering individual measurements.
In the case of the dipole polarizability, it is obvious from Eq. (\ref{eq:pol}) that the resulting correlation in $S_v-S_s/S_v$ space will have a
slope approximately 3/5 that of the energy correlation.  Without polarization corrections, the slope of the neutron skin thickness correlation would
be zero, since Eq. (\ref{eq:skin}) shows $r_{np}$ would depend only on $S_s/S_v$.  The slope of this correlation with polarization corrections included is actually negative, as can
most easily be seen by considering the equations
\begin{equation}
d(S_s/S_v) = a~ d L + b~ dS_v,\qquad dr_{np}=\alpha~ d(S_s/S_v) + \beta~ dS_v.
\end{equation}
Holding $r_{np}$ fixed implies that the correlation slope in $S_v-S_s/S_v$ space would be negative if $\beta/\alpha>0$; the slope in $S_v-L$ space 
would be negative if $\beta/\alpha>-b$.  From Eq. (\ref{eq:skin}) we observe that, for $^{208}$Pb,
\begin{equation}
{\beta\over\alpha}={3Ze^2\over140r_oS_v}{(1+(10/3)(S_s/S_v)A^{-1/3})(1+(S_s/S_v)A^{-1/3})\over I-(7/3)(S_s/S_v)A^{-1/3}}\simeq0.025
\label{eq:ba}\end{equation}
so $S_s/S_v$ and $S_v$ are anticorrelated irrespective of the nuclear force model.  On the other hand,
\begin{equation}
b=-{L\over S_v^2}{d(S_s/S_v)\over d(L/S_v)},
\end{equation}
which is sensitive to the interaction.  For the simple interaction of Eq. (\ref{eq:b}), $b\simeq-0.03$ and in this case $r_{np}$ has a positive correlation in $S_v-L$ space with a slope $dL/dS_v\simeq0.3$ which is, nevertheless, nearly flat.  Without polarization effects, we would have found $dL/dS_v\simeq1.8$.   Hartree Fock calculations~\cite{Chen10,Lattimer13} with
a variety of realistic interactions indicated that $dL/dS_v$ ranges from -3 to -5.  Irrespective of the exact value, it is clear that this correlation is nearly orthogonal to those from energies or dipole polarizabilities and is therefore extremely valuable in determining values for $S_v$ and $L$.

\subsection{Symmetry Parameter Constraints from Experiment}
\label{Sec:Exper}

\begin{figure}[h]\vspace*{-0.6cm}
\center\includegraphics[width=8cm]{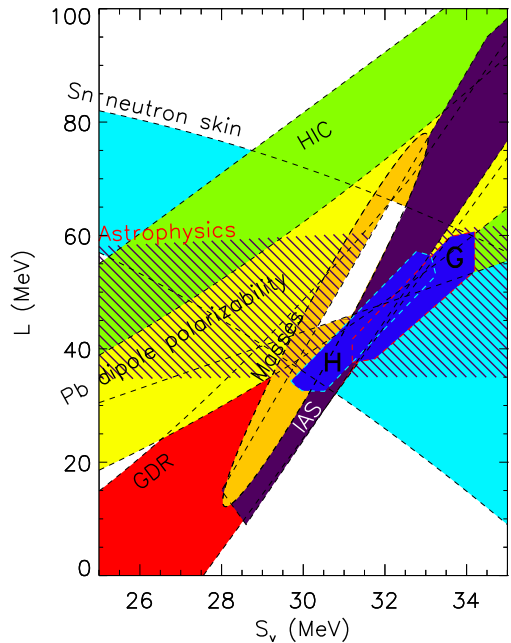}
\vspace*{-1.2cm}\caption{Experimental constraints for symmetry energy parameters, adapted and revised from Ref.~\cite{Lattimer13}.  See the text for further discussion and references to the experimental data and interpretation.  G and H refer to the neutron matter studies of Gandolfi et al.~\cite{Gandolfi12} and Hebeler et al.~\cite{Hebeler10}, respectively. \label{fig:cor}}
\end{figure}

Measurements of all three types of observables,
which have different correlations, can restrict the ranges of the
symmetry parameters considerably.  The experimental situation has been summarized in Refs.~\cite{Lattimer12,Lattimer13} and 
is displayed in Fig. \ref{fig:cor}.

The binding energy correlation (labelled ``Masses'') is taken from
Hartree-Fock calulations with the UNEDF0 density
functional~\cite{Kortelainen10}, in which the nominal fitting error was arbitrarily
chosen to be $\sigma=1$ MeV.\footnote{Ref.~\cite{Kortelainen10} had chosen $\sigma=2$ MeV, but this value seems to be an overestimate, as negative values for $L$ would exist within the $1\sigma$ confidence ellipse.  Negative values of $L$ imply negative neutron matter pressures at $n_s$.}  Importantly, and demonstrating the robustness of this result, the shape and
orientation of the ellipse are the same as predicted by the liquid droplet model, $dL/dS_v\approx12$, once the dependence of $S_s$ on $S_v$ and $L$ is taken into account. 

The constraints for the neutron skin thickness of $^{208}$Pb used in Fig. \ref{fig:cor} are taken
from a study by Chen et al. \cite{Chen10}, who
converted the experimental results~\cite{Ray79,Krasznahorkay94,Krasznahorkay99,Trzeinska01,Klimkiewicz07,Terashima08} for Sn isotopes into an equivalent value for $^{208}$Pb:
$r_{np}\simeq(0.175\pm0.020)$ fm.  Assuming $S_v=31$ MeV, the liquid droplet pediction, Eq. (\ref{eq:skin}), is $S_s/S_v=3.0\pm0.4$.  However, it is known that the liquid droplet model underestimates $r_{np}$ because the neutron and proton distributions have different diffusenesses (i.e., different values of the surface thickness $a$ in Eq. (\ref{bulk1}))\cite{Warda12}.  For Pb$^{208}$ the underestimate can be as large as $0.055$ fm.  When this is taken into account, the predicted value of $S_s/S_v= 2.0\pm0.3$  assuming $S_v=31$ MeV.\footnote{Danielewicz \& Lee~\cite{Danielewicz14} compiled a different set of $^{208}$Pb neutron skin thickness
measurements~\cite{Ray79,Friedman12,Clark03,Zenhiro10,Starodubsky94} to establish a
 weighted average $r_{np}=(0.179\pm0.023)$ fm, just 0.004 fm larger than
determined by Ref.~\cite{Chen10}.  Ref.~\cite{Trzeinska01} gives $0.16\pm0.02$ and a more recent study~\cite{Krasznahorkay13} found $r_{np}=0.161\pm0.042$ fm for $^{208}$Pb.  The latter yields $S_s/S_v=1.7\pm0.6$ when diffuseness corrections are applied.}  Performing a series of Skyrme
Hartree-Fock calculations of $^{208}$Pb, in which values of $S_v$ and
$L$ were systematically varied, Ref.~\cite{Chen10} established that
\begin{eqnarray}\label{skinpb}
{r_{np}\over{\rm fm}}&\simeq&-0.094669+{7.2028S_v\over{\rm GeV}}+{2.3107L\over{\rm GeV}}\cr
&-&{8.8453S_v^2\over{\rm GeV}^2}-{47.837S_vL\over{\rm GeV}^2}+{4.003L^2\over{\rm GeV}^2}.
\end{eqnarray}
This formula, with the aforementioned value for $r_{np}$, establishes the
correlation slope $dL/d S_v\simeq-4.6$, assuming $S_v=31$ MeV and $L=45$
MeV.  More work is necessary to check the model dependence of this result, although it is in essential agreement with the liquid droplet prediction.

The constraint for the electric dipole polarizability
$\alpha_D$ of $^{208}$Pb is taken from data produced by Tamii et
al.~\cite{Tamii11}: $\alpha_D\simeq(20.1\pm0.6)$ fm$^3$. The liquid droplet prediction, Eq. (\ref{eq:pol}), is $S_s/S_v=1.09\pm0.14$ assuming $S_v=31$ MeV, which is slightly inconsistent
 with the neutron skin prediction.   Roca-Maza et
al.~\cite{Roca-Maza13} showed from studies with a series of
relativistic and non-relativistic interactions that $\alpha_D, S_v$ and $r_{np}$ for $^{208}$Pb can be constrained by
\begin{equation}\label{ajr}
\alpha_DS_v\simeq(325\pm14)+(1799\pm70)(r_{np}/{\rm fm}){\rm~Mev~fm}^3.
\end{equation}
By use of Equation (\ref{skinpb}), this is converted onto the $S_v-L$
correlation shown in 
Fig. \ref{fig:cor}.\footnote{The slope of this correlation is significantly different
than shown in~\cite{Lattimer13}, which relied on the analysis in
Ref.~\cite{Reinhard10} that erroneously concluded $\alpha_D\propto
r_{np}$.}

There is also experimental data for which the liquid droplet model has not established predictions.  This includes the centroid energy of the giant dipole resonance
for $^{208}$Pb~\cite{Trippa08}, isospin diffusion in heavy ion collisions~\cite{Tsang09}, and energies of excitations due to isobaric analog states~\cite{Danielewicz14}. For more details of
the constraints they imply, see Ref.~\cite{Lattimer13} and Fig. \ref{fig:cor}.
The white region displayed in Fig. \ref{fig:cor}
represents the consensus agreement of the six experimental constraints
we have discussed, giving a range 44 MeV $<L<66$ 
MeV.\footnote{In comparison to the consensus region found in 
Ref.~\cite{Lattimer13}, the region displayed in Fig. \ref{fig:cor} is
slightly smaller because of the incorporation of the additional
constraint from isobaric analog states.} Since the model dependencies
of these constraints have not been thoroughly explored, the size of
this consensus region may well be underestimated. Treating the
white region as a 68\% confidence interval, Eqs. (\ref{eq:sl}) and (\ref{eq:radius}) predict
$R_{1.4}\simeq(12.1\pm1.1)$ km to 90\%
confidence. As we will see, this range is
quite compatible with several astrophysical observations.

The conclusion from experimental studies is that the ranges of values for $S_v$ and $L$ are consistent with those established from neutron matter calculations.  
In the next section, we explore parameterizations for the high-density EOS that can be constrained by causality, $M_{max}$ and these permitted symmetry parameters.

\section{The High-Density EOS and Mass and Radius Constraints}
\label{Sec:pw}
For densities above the core-crust transition density, we will utilize
separate polytropic regions with continuous pressures at the boundaries.  Each region has the EOS $p_i=K_in^\gamma_i$.\footnote{Alternatively, one could use  a  parameterization employing energy densities rather than baryon densities, and using $p_i=K_i(\epsilon/\epsilon_{i-1})^\gamma_i$.  It would produce no substantial difference to our conclusions.}    This is motivated by the three piecewise polytrope scheme explored by
Read et al. \cite{Read09}.  They found that a scheme with only three parameters could match, to about 4\% rms error, the pressure-energy density relations for a large set of candidate EOSs up to densities below the central densities of $1.4M_\odot$ stars.  With an additional parameter,  more exotic EOSs with kaons or hyperons could also be matched.  In this scheme the starting density of the first polytrope is the core-crust transition density $n_0$ and the starting densities of the upper two polytropes are fixed at $1.85n_s$ and $3.7n_s$, respectively.  The four parameters are the three polytropic indices and the pressure at the starting density of the second polytrope.  A more general scheme would have 7 parameters,  three parameters
corresponding to boundary density points above the core-crust transition and four corresponding to the
polytropic exponents in each region.

In the models we will consider, the boundaries are
denoted $n_i$ with $i=0-3$, and the adiabatic exponents between
the densities $n_i$ and $n_{i-1}$, for $i=1-3$, are $\gamma_i$, such that the pressure in that region is $p=p_{i-1}(n/n_{i-2})^{\gamma_i}$.  In addition, for densities $n>n_3$ matter is treated as a continuation of region 3, so that $p=p_2(n/n_2)^{\gamma_3}$.\footnote{In the scheme of Read et al.~\cite{Read09}, the density boundary $n_3$ is used to define $\gamma_3$, see Eq. (\ref{eq:ge}), but is otherwise irrelevant because $\gamma_3$ applies for all densities $n>n_2$.  Note that our notation is not identical to that of Ref.~\cite{Read09}.}  We assume values at the
core-crust transition are given by Eq. (\ref{eq:bps}) with  $n_0=n_s/2.7$ and that the EOS in region 1 matches neutron matter calculations. 

\subsection{A First Model: Fixed Boundaries\label{Sec:fixed}}
 In our first model, we choose $n_1=1.85n_s$ and $n_2=3.74n_s$ to match the optimum transition densities found by Ref. \cite{Read09}.  The three remaining
free parameters are the polytropic exponents $\gamma_i$ for $i=1-3$.  We choose as parameters, instead, the boundary pressures $p_i$ for $i=1-3$, following Ref. \cite{Ozel15}, by taking $n_3=7.4n_s$.  The first boundary, $n_1$, is close to the upper range of validity of neutron matter calculations, so values for $p_1$ were evaluated for each neutron matter EOS using Eq. (\ref{eq:nmext}); results are given in Table \ref{tab:nm}.  We restrict the allowed $p_1$ range to lie within these results. except that we will slightly expand the upper and lower limits.  The polytropic exponents and the energy densities
within each region are given, assuming continuity of both
energies and pressures at the boundary points $n_i$, by
\begin{align}
n(p)&=n_{i-1}\left({p\over p_{i-1}}\right)^{1/\gamma_i},\qquad
\gamma_i={\ln(p_i/p_{i-1})\over \ln(n_i/n_{i-1})}\\
\epsilon(p)&=\left(\epsilon_{i-1}-{p_{i-1}\over\gamma_i-1}\right)\left({p\over p_{i-1}}\right)^{1/\gamma_i}+{p\over\gamma_i-1},\\
\label{eq:ge}\end{align}
where $\epsilon_{i-1}=n_{i-1}(m_n+E_{i-1})$ and $E_{i-1}$ is energy density
and energy per baryon at the point $n_{i-1}$.  For $p>p_3$, the same expressions are used as for $p_2<p<p_3$.

\begin{figure}[h]\vspace*{-1cm}
  \includegraphics[width=13cm,angle=180]{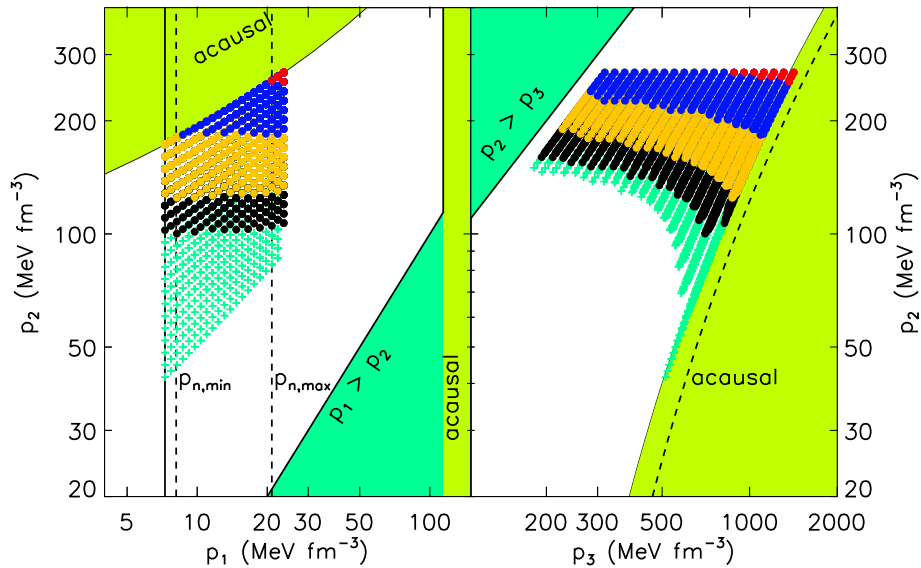}\vspace*{-0.8cm}
  \caption{Permitted ranges of $p_1, p_2$ and $p_3$ from causality and stability. Values ruled out by
    causality (hydrodynamic instability) are indicated by the
    light-(dark-)green shading. Realistic minimum and
    maximum values of $p_1$ are shown as vertical dashed lines, and
    the value $p_{1,\min}=7.56$ MeV fm$^{-3}$ is the vertical solid
    line. Solid dots show parameter values permitting maximum masses,
    respectively, of $1.97~M_\odot$ (black), $2.10~M_\odot$ (orange),
    $2.30~M_\odot$ (blue) and $2.50~M_\odot$ (red). Teal crosses show
    parameter values that lead to acausal configurations or those having
    $M_{max}<1.97~M_\odot$.}
  \label{fig:p1p2p3}
\end{figure}

In this scheme the sound speed
increases monotonically with density, so that
causality within each region requires
\begin{equation}
  \frac{c_{s,i}^2}{c^2}=\left(\frac{\partial p}{\partial\epsilon}
  \right)_i=\frac{\gamma_{i,\max}p_{i,\max}}{\epsilon_{i,\max}+p_{i,\max}}
  \le 1,
  \label{eq:caus}
\end{equation}
which is an implicit equation for $p_{i,\max}$ after using
Eq. (\ref{eq:ge}).  In general, $p_{1,\max}$ is so much larger than
the realistic range for $p_1$ established from neutron matter
studies\footnote{$p_{1,\max}\simeq113.9$ MeV fm$^{-3}$ from causality
  considerations, see Fig. \ref{fig:p1p2p3}} that this causality
condition is not important to our studies.  $p_{3,\max}$ depends upon
$p_2$ and $\epsilon_2$, and therefore also upon $p_1$. However, its
dependence upon $p_1$ is weak, as shown in Fig. \ref{fig:p1p2p3}. The
central density of the star is, in many cases, larger than $n_3$, in
which case the limiting value $p_{3,\max}$ is smaller than the limit
given by Eq. (\ref{eq:caus}).  The actual limit must be found
numerically from TOV integrations of the star's structure requiring
that the maximum sound speed at the center of the maximum mass star be
smaller than $c$.

Minimum values for $p_1$, $p_2$ and/or $p_3$ also exist in order to
satisfy hydrodynamic stability, which requires that $p_i>p_{i-1}$.
Parameter ranges allowed by hydrodynamical stability and causality are
portrayed in Fig. \ref{fig:p1p2p3} as the white regions.\footnote{We note that the causality boundaries differ from those shown in Ref.~\cite{Ozel15}.} For $p_2$ and
$p_3$, more restrictive minima can result if one requires that the
maximum mass exceeds the largest well-measured neutron star mass,
which we take to be $M_{max}=1.97~M_\odot$, the $1\sigma$ lower limit
to the measured mass of PSR J0548+0432~\cite{Antoniadis13}. (Note that
there is no minimum value for $p_1$ based on this condition, due to
the presence of the polytropic regions 2 and 3.) These limits
must be found numerically from TOV integrations, which indicate
that the effective lower limit to $p_2$ is approximately 100 MeV
fm$^{-3}$ for virtually all realistic choices of $p_1$
(Fig. \ref{fig:p1p2p3}).

The result of each TOV integration with a different EOS (i.e., different combinations of $p_1, p_2$ and $p_3$) is indicated
by a symbol in Fig.~\ref{fig:p1p2p3} (many parameter
combinations yield nearly identical configurations and cannot be
distinguished). It is clear that causal configurations capable of supporting
$M_{max}=1.97~M_\odot$ must have $p_2\simge100$ MeV fm$^{-3}$, and if
$M_{max}=2.1~M_\odot$, $p_2\simge125$ MeV fm$^{-3}$. On the other hand,
the specific value of $p_3$ plays relatively little role as long as
$p_2<p_3<p_{3,\max}$ leads to causal configurations of the required
mass.

\begin{figure}[h]\vspace*{-1cm}
  \includegraphics[width=13cm,angle=180]{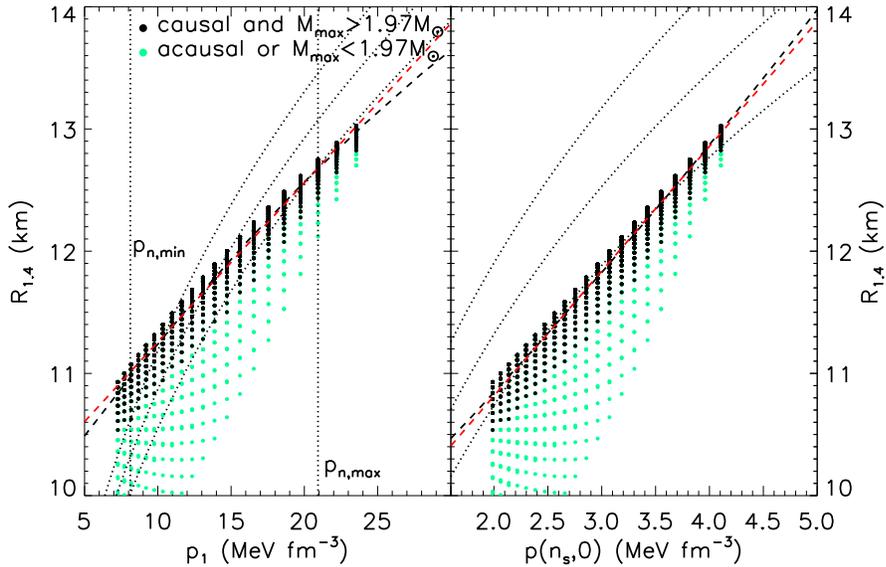}\vspace*{-0.8cm}
  \caption{The correlations between radii of $1.4~M_\odot$ stars,
    $R_{1.4}$, and $p_1$ (left panel) and $p(n_s)$ (right panel). Parameters producing causal configurations
    capable of supporting $1.97~M_\odot$ are indicated as black circles;
    all others are indicated by teal circles. The solid (dashed)
    lines indicate quadratic (linear) fits to the black
    circles.   The dotted curves
    show the correlations inferred from Eq. (\ref{eq:radius})~\cite{Lattimer13} with
    $1\sigma$ errors.}
  \label{fig:rad}
\end{figure}

As expected, neutron star radii will be most sensitive to the
parameter $p_1$.  Values of $R_{1.4}$ for different choices of
parameters are shown in Fig. \ref{fig:rad} as functions of either
$p_1$ or $p(n_s)=p_1(p(n_s)/p1)^{\gamma_1}$.  The spread of radius
values for a given $p_1$ or $p(n_s)$ shows the influence of variations
in $p_2$ and $p_3$ and is small.  We compare the obtained correlations
with the formulae earlier obtained by Ref.~\cite{Lattimer13},
Eq. (\ref{eq:radius}) in Fig. \ref{fig:rad}.  The radii obtained here tend to be somewhat
smaller due to the ceiling placed on $p_1$ from neutron matter
calculations.

\begin{figure}[h]\vspace*{-1cm}
  \includegraphics[width=13cm,angle=180]{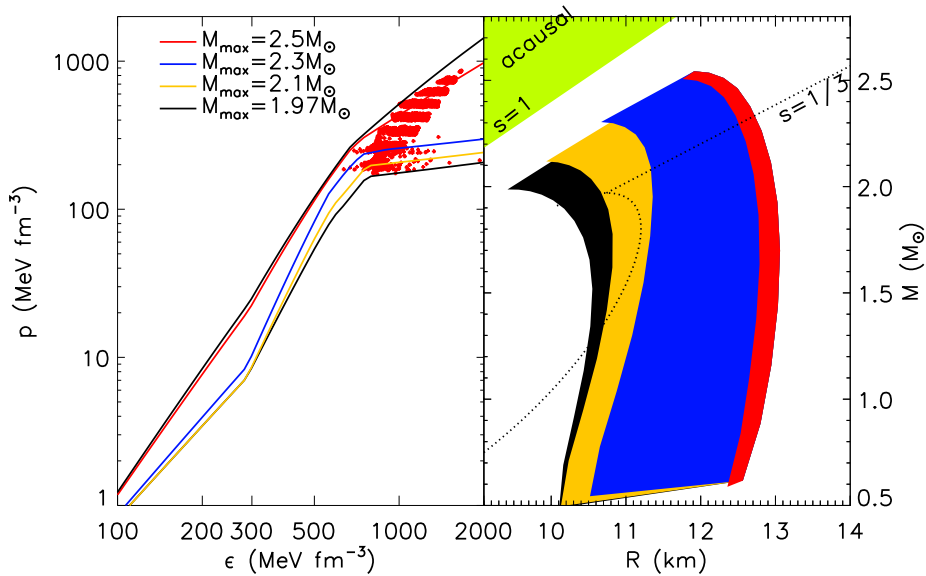}\vspace*{-0.8cm}
  \caption{Left panel:  Allowed pressures as a function of energy density permitted
    by the assumed constraints on the low-density EOS, causality, and
    selected values for $M_{max}$. Red crosses indicate the central
    conditions for surviving EOSs. The black, yellow, blue and red lines
    are for $M_{max}=1.97~M_\odot$, $2.1~M_\odot$, $2.3~M_\odot$, and $2.5~M_\odot$,
    respectively.  Right panel:  Allowed masses and radii for selected values of $M_{max}$.}
  \label{fig:pemr}
\end{figure}
  
The restrictions of causality and large maximum masses severely
restrict the allowed EOSs. The left panel of Fig. \ref{fig:pemr} shows boundaries in the
pressure-energy density plane with different assumptions for $M_{max}$
permitted by causality and the assumed low-density EOS for the crust
and for neutron matter. For $M_{max}=1.97~M_\odot$, the maximum uncertainty
in pressure for a given energy density is no larger than a factor of 3
(which occurs near $n_1$), and is slightly larger than a factor of 2
near the central densities of maximum mass stars. The corresponding
regions in the mass-radius plane that can be populated by EOSs
satisfying the $M_{max}$, causality and the low-density EOS constraints
are also shown in this figure. Collectively, one observes the
importance of neutron star mass measurements: the larger the minimum
value of $M_{max}$, the more restricted the ranges of $p(\epsilon)$
and $R(M)$ and the more accurately the EOS can be predicted.

The range 10.7 km $\le R_{1.4}\le13.1$ km results from the restrictions we imposed from neutron matter calculations and causality.  This is of interest since many recent estimates of neutron star radii from observations have exceeded this range from both sides, as we will observe in Sec. \ref{Sec:Observ}.

\subsection{A Second Model:  Phase Transitions}

Lattimer \& Steiner~\cite{Lattimer14a} noted that phase transitions near
$n_s$ have the effect of allowing both smaller and larger ranges of
$R_{1.4}$.  Similar results have been noted by Hebeler et
al.~\cite{Hebeler13}, who reported 9.7 km $<R_{1.4}<13.9$ km.  However,
the range is very sensitive to the density at which the phase
transition is assumed to exist, and the widest range is observed when
the phase transition begins at $n_s$.  This seems an extreme lower limit,
as there exists no experimental evidence for a phase transition in
nuclei.

We explore the role of phase transitions by allowing the lower and
upper transition densities, $n_1$ and $n_2$, respectively, to vary.
We consider two models for matter with $n\ge n_u$, both described by
$p=s\epsilon-\epsilon_B$ where $s$ is 1 or 1/3 and $\epsilon_B$ is a
constant related to the energy density where the pressure vanishes.
The case $s=1$ specifies a phase transition to the stiffest EOS
allowed by causality, while $s=1/3$ mimics a phase transition to a
deconfined strange quark matter phase where $\epsilon_B=4B/3$
with $B$ the bag constant.  These models are similar to the cases
considered by Alford et al. \cite{Alford15}.

\begin{figure}[h]\vspace*{-1cm}
 \hspace*{-1cm} \includegraphics[width=15cm,height=9cm,angle=180]{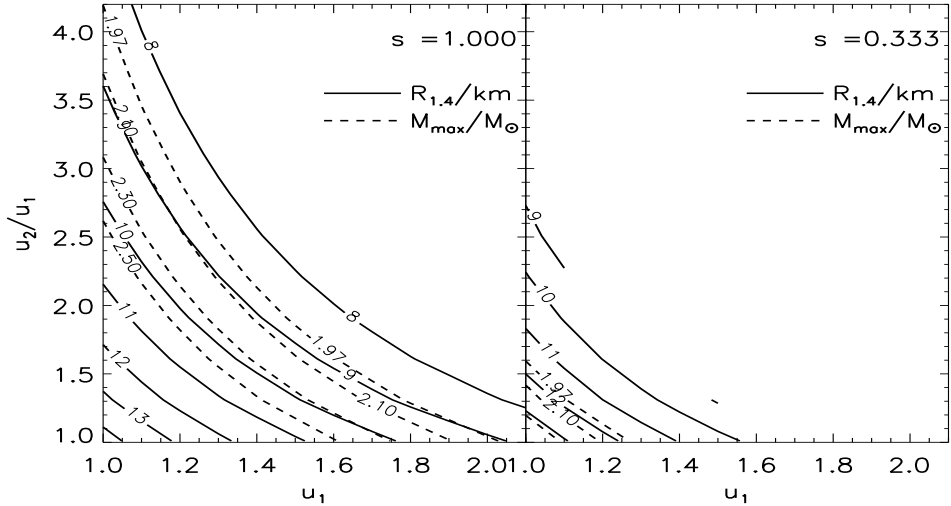}\vspace*{-0.8cm}
  \caption{Radius limits for the case including phase transitions.  Values of $R_{1.4}$ and $M_{max}$ are displayed as functions of $u_1=n_1/n_s$, $u_2=n_2/n_s$ and $s$.  The left (right) panel shows the case $s=1$ ($s=1/3$).}
  \label{fig:u1u2}
\end{figure}

The phase transition (assumed to be of first order here) is accomodated by requiring that $p_\ell=p_u$ and
$\mu_1=\mu_2=(\epsilon_1+p_1)/n_1=(\epsilon_2+p_2)/n_2$ have the same
values on both sides of the transition, $n_1$ being in the hadronic
phase, and $n_2$ being in the high-density phase.  For given values of
$s$ and $\epsilon_B$, this uniquely specifies $n_2$ in terms of $n_1$:
\begin{equation}
n_2={(1+s)p_1(n_1)+\epsilon_B\over s\mu_1(n_1)}.
\label{eq:n2}\end{equation}
The energy density changes by the amount
$\epsilon_2-\epsilon_1=\mu(n_2-n_1)$ across the transition.
Therefore, all possible phase transitions can be parameterized by
$n_1, n_2$ and $s$.  Results are shown in Fig. \ref{fig:u1u2} for the
two cases and are similar to those found by Ref. \cite{Alford15}.  As expected, both the smallest and largest values of
$R_{1.4}$ are found when $n_1=n_s$ and $s=1$.  The largest values,
about 14.3 km, are found for $n_2=n_1$.  The smallest values, about
8.4 km, occur when $n_2\simeq4.2n_1$, and are nearly as small as that
deduced from the {\it maximally compact} EOS for $M_{max}=1.97M_\odot$
($\simeq8.1$ km from Fig. \ref{fig:mr}).  A phase transition is compatible with $M_{max}=1.97M_\odot$ only if
the transition occurs close to $n_s$; in fact, $n_1\simle2n_s$ is required.

In the case that $s=1/3$, one sees from Fig. \ref{fig:u1u2} that 11.7 km $<R_{1.4}<13.8$ km, which has an
upper limit nearly the same as in Hebeler et al.~\cite{Hebeler13} but a
lower limit above what we established above.  Moreover, as noted by Ref. \cite{Alford15}, only a small range
of values of $u_1$ and $u_2$ support a phase transition and $M_{max}\simge1.97M_\odot$.  Using only observations of pulsar masses and estimates of
symmetry properties from neutron matter studies and nuclear
experiments, we can conclude, at present, that while the most extreme
permitted range is 8.4 km $<R_{1.4}<14.3$ km, a realistic range is
about half as broad, being 10.7 km $<R_{1.4}<13.1$ km.  It is
interesting to compare these results with the radius values
estimated from X-ray observations of neutron stars.
 
\section{Summary of Observational Constraints\label{Sec:Observ}}
\subsection{Photospheric Radius Expansion Bursts}

Ref.~\cite{vanParadijs79} proposed using photospheric radius expansion
(PRE) X-ray bursts to obtain simultaneous mass and radius measurements
in 1979, but the method did not lead to interesting constraints until
2006~\cite{Ozel06}. Type I X-ray bursts originate in accreting binaries when sufficient combustible mass
accumulates onto a neutron star surface.  The fuel ignites, with burning spreading quickly across the entire stellar
surface.  A subset of these bursts, PRE bursts are energetic enough to lift the stellar photosphere to great heights.  This requires
luminosities exceeding the so-called Eddington luminosity\footnote{Quantities with the $\infty$ subscript are redshifted, as viewed at the Earth.}
\begin{equation}
L_{Edd,\infty}={4\pi cGM\over\kappa}\sqrt{1-{2GM\over R_{phot}c^2}},
\label{eq:ledd}\end{equation}
where $\kappa$ is the atmosphere's mean opacity and $R_{phot}$ is the
distance of the photosphere from the star's
center\footnote{$R_{phot}>>R$ during the first part of the burst, and
  in principle $R_{phot}=R$ when ``touchdown" occurs and $L_{Edd}$ is
  measured.}.  The bursts are short, lasting a matter of seconds, and
the lifted material quickly loses opacity and falls back to the
stellar surface.  The burst decays as the stellar surface cools.
Measurements of the flux and temperature during this decay permit the
extraction of an angular area for the source
\begin{equation}
A={F_\infty\over\sigma_B T_{c,\infty}^4}={1\over f_c^4}\left({R_\infty\over D}\right)^2,
\label{eq:finfty}\end{equation}
where the flux and color and effective temperatures at Earth
are $F_\infty, T_{c,\infty}$ and $T_{eff,\infty}$, respectively.
$f_c=T_{eff}/T_c$ is the color correction factor which accounts for
the effects of a non-blackbody atmosphere,
$R_\infty=R/\sqrt{1-2GM/(Rc^2)}$ is the apparent radiation radius and
$D$ is the distance.  

\begin{figure}[h]\vspace*{-1.5cm}
 \hspace*{-2cm} \includegraphics[width=15cm,height=11cm,angle=180]{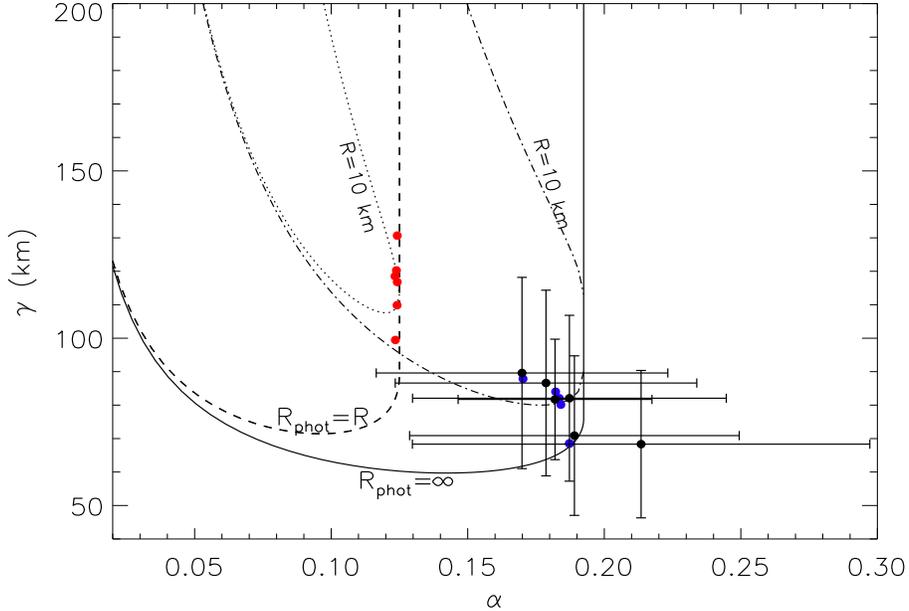}\vspace*{-1cm}
  \caption{Allowed regions in $\alpha-\gamma$ space for real and causal solutions from PRE bursts lie above the dashed (solid) boundary curve  for the $R_{phot}=R$ ($R_{phot}=\infty$) model. The boundaries represent $M$ and $R$ from the {\it maximally compact} solution with $M_{max}=1.97M_\odot$ translated to $\alpha$ and $\gamma$ using Eq. (\ref{eq:br}) [Eq. (\ref{eq:br1})].  Data (black filled circles) and $1\sigma$ errors for PRE burst sources are taken from Ref.~\cite{Ozel15}.  The dotted (dot-dashed) curve is the contour $R=10$ km in the $R_{phot}=R$ ($R_{phot}=\infty$) model.  The red (blue) filled circles show the maximum of the Monte Carlo distributions for the  $R_{phot}=R$ ($R_{phot}=\infty$) model.}
  \label{fig:ag}
\end{figure}

Quantities extracted from observations include $A$, $F_{Edd,\infty}=L_{Edd,\infty}/(4\pi D^2)$, and $D$ (see Ref.~\cite{Ozel15} for a summary and references). Quantities estimated from theory are $\kappa$ and $f_c$. One can form two parameters from the observables:
\begin{align}
\alpha&\equiv{F_{Edd,\infty}\over\sqrt{A}}{\kappa D\over f_c^2c^3}=\beta(1-2\beta),\qquad\gamma\equiv{A\over F_{Edd,\infty}}{f_c^4c^3\over\kappa}={R\over\beta(1-2\beta)^{3/2}}
\label{eq:fa}\end{align}
The quantity $\gamma$ is independent of $D$, and $R_\infty=\alpha\gamma$ is independent of $\kappa$ and $F_{Edd,\infty}$. Note that $R\le\sqrt{27/3125}~\gamma$ since $\beta(1-2\beta)^{3/2}$ has a maximum when $\beta=1/5$. The solutions for mass and radius become
\begin{equation}
\beta={1\over4}\pm{\sqrt{1-8\alpha}\over4},\qquad R_\infty=\alpha\gamma.
\label{eq:br}\end{equation}
$\alpha$ should be less than 1/8 for real solutions to exist.
In practice, however, observed values of $\alpha$ are
greater than 1/8 to more than $1\sigma$ (Fig. \ref{fig:ag}).  If one does Monte Carlo
sampling within the error ranges of the observables, the resulting values of $\alpha$ and $\gamma$
mostly lead to complex solutions or acausal solutions (according to the {\it maximally compact} solution with $M_{max}=1.97M_\odot$).  If one restricts acceptances
 to valid trials, $\alpha\approx1/8$ becomes strongly
preferred, leading to $\beta\approx1/4$.  Preferred values for $\gamma$ lie
near their $1\sigma$ upper limits.
Values for $R_\infty\approx\gamma/8$ then range from 11 to 14 km, so that values for
$R\approx R_\infty\sqrt{1/2}$ range from 8 to 10 km~\cite{Steiner10}.  That only extremely small fractions (i.e., about 4\%) of trials are valid indicates serious problems with the model.  The weighted means of the radii and masses of the sources are $10.6\pm0.7$ km and $1.63\pm0.15 M_\odot$.\footnote{This radius is about 1 km larger than in previous studies with this method, cf. 
Refs.~\cite{Ozel09,Ozel10}, due to the failure to impose causality on solutions.} 

One possible solution to the small acceptance, suggested by Steiner et al. \cite{Steiner10}, is that the Eddington flux is measured while the photosphere is still well above the surface, so the redshift factor in Eq. (\ref{eq:ledd}) is absent.  In this case, one finds 
\begin{equation}
\alpha=\beta\sqrt{1-2\beta},\qquad\gamma={R\over\beta(1-2\beta)},
\label{eq:br1}\end{equation}
whose real solution requires $\alpha\le3^{-3/2}\simeq0.192$.  All PRE burst sources shown in Fig. \ref{fig:ag} satisfy this to within $1\sigma$.  As a result, a much larger trial acceptance rate ensues, about 65\%, after ensuring causality with $M_{max}=1.97M_\odot$.  Note now that $R\le\gamma/8$, a factor 1.34 times larger than for the previous model.  The weighted means of the radii and masses of the sources become ($11.2\pm0.7$) km and ($1.59\pm0.13) M_\odot$.  The mean radius agrees to within $1\sigma$ with those predicted from nuclear experiment and theory, and the mean mass agrees with the observed pulsar mass distribution.  Nevertheless, the actual radius at touchdown is but one example of systematic uncertainties that arise in this simple description of PRE bursts. Ref.~\cite{Suleimanov11} used longer bursts and a more complete model
of the neutron star atmosphere to obtain a radius greater than 14 km
for the single source studied.  The possible obscuration of a large fraction of the burst luminosity by the accretion disc was suggested as an important systematic uncertainty as well.

Another approach is to limit the study to those bursts in which one can avoid the uncertainty of the location of the  photospheric radius, i.e., choosing only ``passive'' bursts, and also using information from the whole cooling tail rather than just near touchdown and at late times.  N\"attila et al. \cite{Nattila15} recently performed a study of three sources with this {\it cooling tail method} and showed that the parameter $F_{Edd,\infty}$  could be determined to at least 3\% accuracy and $A^\prime=D/R_\infty=(f_c^4A)^{-1/2}$ to about 1\%.  One finds that the parameters $\alpha$ and $\gamma$ from Eq. (\ref{eq:fa}) for the three sources each have nearly the same central values: $\alpha\simeq0.117$ and $\gamma\simeq128$ km to about 2\%.  The major uncertainty in $\alpha$ is due to the roughly 10\% distance uncertainty, while $\gamma$ has less than 4\% uncertainty. Since $\alpha$ is smaller than 1/8 to about $1\sigma$, most solutions from parameters taken within the error distributions will be real.  From Eq. (\ref{eq:br}), $\beta\simeq0.185$ or 0.315 and $R_\infty\simeq15$ km (all are central values).  The centroids of the two solutions thus have $M=1.5M_\odot$ and $R=11.9$ km, and $M=1.94M_\odot$ and $R=9.1$ km, respectively.  The second solution can be rejected on the basis of its incompatibility with the symmetry energy of nuclear matter; it is also extremely close to the acausal boundary in $M-R$ space with $M_{max}=1.97M_\odot$ (Fig. \ref{fig:maxr}). In any case, the small uncertainty in $\gamma$ and its value requires that $R<12$ km for any value of $\alpha$. 

\subsection{Quiescent Low-Mass X-Ray Binaries}

Another popular method of measuring neutron star radii using
observations of thermal emission utilizes accreting neutron
stars in binary systems, known as quiescent low-mass X-ray binaries
(qLMXBs). The earliest such observations led to blackbody radius estimates of
less than 1 km, much smaller than that predicted from theoretical
models~\cite{vanParadijs87}. For accreting neutron stars, the
photosphere is expected to consist of hydrogen \cite{bildsten92}, and
Ref.~\cite{Rutledge99} showed that models of thermal emission from
pure hydrogen photospheres \cite{rajagopal96,zavlin96} give inferred
radii of the correct order of magnitude.  Continued observations and a consistent treatment of the surface
gravity~\cite{Heinke06,Webb07} led to refined
estimates of radii between 8 and 16 km for sources in globular clusters where $D$ is reasonably well known.  However, by themselves, this result
cannot rule out a significant number of  EOS
models~\cite{Lattimer01}.

Guillot et al.~\cite{Guillot13} combined new qLMXB observations and a ``common
radius" model (motivated by the generic vertical shape of $M-R$
curves, cf. Fig. \ref{fig:mr}) to obtain a rather small
preferred neutron star radius with a corresponding small error, $9.0\pm1.4$  km.
However, when individually analyzed, the sources predicted
radii in a wide range (7 to 20 km). Ref.~\cite{Steiner13} concluded
that two systematic uncertainties were responsible for this result:
(i) some of the smaller radius neutron stars may have helium rather than
hydrogen atmospheres, and (ii) the estimated amount of X-ray absorbing matter between the sources and Earth might be wrong.
Ref.~\cite{Servillat12}
showed that a helium (rather than hydrogen) atmosphere changes the
inferred radius range for the neutron star in M28 from 6 to 11.5 km to
7 to 17 km.
Heinke et al.~\cite{Heinke14} confirmed that the absorbing mass
inferred for the qLMXB with the largest inferred radius was indeed overestimated, leading to a
systematic radius reduction. Ref.~\cite{Heinke14} also showed that the
choice of the galactic abundance model was an important systematic affecting all sources.

A consequence of the common radius model is the tendency to underpredict the radius.  Measurements of qLMXB spectra are 
better able to constrain $R_\infty$ than the surface gravity $g=GM/[R^2\sqrt{1-2\beta}]$ of a source, since the main global observables are $F_\infty$, $T_{c,\infty}$ and $D$, and for a blackbody
\begin{equation}
F_\infty=\left({R_\infty\over D}\right)^2\sigma_BT_{c,\infty}^4.
\label{eq:f}\end{equation}
Corrections for the atmosphere, most often a low-magnetic field, pure H atmosphere, are readily produced (i.e., through $f_c$).
A variable in $M$ and $R$ orthogonal to $R_\infty$, such as the surface gravity, is poorly constrained by available data for every source.  Therefore, simultaneously refitting all sources under the constraint of a common radius is accommodated most easily by variations in $g$ rather than $R_\infty$.  As is clear from the $R_\infty$ contours seen in Fig. \ref{fig:mr}, sources with widely varying estimated values of $R_\infty$ can satisfy the constraint of a common radius only if that radius is smaller than the smallest $R_\infty$ in the sample.  Sources with large $R_\infty$ can be reconciled by increasing their mass (and $g$).  For the analysis of Guillot et al. \cite{Guillot13}, the mean $R_\infty$ of the individually analyzed sources was 14.0 km; for the joint analysis with a common radius, it became 13.2 km.  The predicted masses of the individually analyzed sources ranged over $0.8M_\odot<M<1.8M_\odot$; for the joint analyses the range increased to $0.7M_\odot<M<2.3M_\odot$.

\subsection{Bayesian Analyses of Observations with Theoretical Priors}
 Steiner et al.~\cite{Steiner10} used Bayesian inference to combine
mass and radius constraints from both qLMXBs and PRE X-ray bursts to
obtain constraints on the EOS.   This work used a nuclear
physics-based parameterization for matter near the nuclear saturation
density and piecewise polytropes at higher densities (in a slightly
different form than that presented in Ref.~\cite{Read09}). 
The Bayesian inference-based method to obtain the EOS parameters is
similar to the method developed in Ref.~\cite{Ozel09}, except that
Ref.~\cite{Steiner10} additionally obtained probability
distributions for the $M-R$ curve by marginalizing over the
posteriors for $R$ as a function of $M$. Also, their direct use
of Bayesian inference allowed the use of
more parameters capable of exploring the uncertainties in the EOS
near saturation densities and led to novel constraints on the density
dependence of the nuclear symmetry energy.

Ref.~\cite{Steiner10} showed that the assumption that the
photosphere at touchdown is coincident with the surface (i.e., $R_{phot}=R$) is not consistent with the data (within the context of the
model being used for PRE X-ray bursts in Refs.~\cite{Steiner10} and
\cite{Ozel10}). This was confirmed with the introduction of Bayes factors in Ref. \cite{Steiner13}.  Bayes factors also showed that helium, not hydrogen, is favored for at least one qLMXB source, 
which allows for larger radius estimates from these sources than found in Ref. \cite{Guillot13}. The final result was a range of $R_{1.4}$ from 10.7 to 12.5 km.

Ref. \cite{Nattila15} performed a Bayesian analysis of the 3 sources they analyzed with the cooling tail method, using the same EOS and causality constraints as in Refs. \cite{Steiner13,Lattimer15} and find, if strong phase transitions are disallowed, that 11.3 km $<R_{1.4}<12.8$ km to 95\% confidence.  Allowing strong phase transitions changes this range to 10.5 km $<R_{1.4}<12.5$ km. All these ranges determined with the Bayesian method are tantalizingly similar to that predicted by neutron matter and nuclear experimental studies. However, reducing their uncertainty is still hostage to our lack of understanding of important systematics in the interpretation of observations.   

\subsection{Other Observations}

We conclude this section with a brief survey of proposed radius measurements.  These include measurements of the moment of inertia of the more massive pulsar in the pulsar binary PSR J0737-3039, pulse-profile observations of pulsars and bursting sources by X-ray telescopes to measure $z$ and $R$, and gravitational wave observations of merging compact objects:  black hole-neutron star (BH-NS) and neutron star-neutron star (NS-NS) binaries.
\subsubsection{Moments of Inertia}
The moment of inertia is readily calculated in GR for the case of a uniformly, slowly rotating star.   It can be expressed as 
\begin{equation}
I={c^2\over G}~{w(R)R^3\over6+2w(R)},
\label{eq:iy}\end{equation}
where $w(0)=0$ and $w(R)$ is the solution, at the surface $R$, of the differential equation
\begin{equation}
{dw\over dr}={4\pi G\over c^2} ~{(\epsilon+p)(4+w)r\over c^2-2Gm}-{w\over r}(3+w),
\label{eq:iw1}\end{equation}
where $m(r)$ is the mass enclosed within radius $r$.

The measurement of the moment of inertia from spin-orbit coupling in a compact binary is especially intriguiging since it is seemingly less subject to uncertain systematic effects than current methods.     Spin-orbit coupling leads to geodetic precession of the angular momentum vector $\vec{\bf L}$ of the orbital plane
around the direction of the total angular momentum $\vec{\bf J}$ of the system.  Since the total angular
momentum $\vec{\bf J}=\vec{\bf L}+\vec{\bf S_A}+\vec{\bf S_B}$ is conserved (at this order), there
are compensating precessions of the spins $\vec{\bf S_A}$ and $\vec{\bf S_B}$ of the two
stars. Since the orbital angular momentum dominates the spin angular
momenta in binaries, the geodetic precession amplitude is
very small, while the associated spin precession amplitudes are
substantial. In addition to geodetic precession, spin-orbit coupling
also manifests itself in apsidal motion (advance of the periastron).

According to Barker \& O'Connell~\cite{Barker75}, the spin and orbital angular momenta evolve as
\begin{align}
{d\vec{S_i}\over dt}&={G(4M_i+3M_{-i})\over2M_ia^3c^2(1-e^2)^{3/2}}\vec{\bf L}\times\vec{\bf S_i},\\
{d\vec{\bf L}^{\rm SO}\over dt}&=\sum_i{G(4M_i+3M_{-i})\over2M_ia^3c^2(1-e^2)^{3/2}}\left(\vec{\bf S_i}-3{\vec{\bf L}\cdot\vec{\bf S_i}\over|\vec{\bf L}|^2}\vec{\bf L}\right),
\label{eq:sil}\end{align}
where the superscript ``SO'' refers to the spin-coupling contribution
only (there are also first- and second-order post-Newtonian terms, unrelated to the spins, that contribute
to this order).  Here $a$ is the semi-major axis, $e$ is the eccentricity and $M_i$ is component $i$'s mass.
Note that if the spins are parallel to $\vec{\bf L}$ there is no precession.  The change in inclination, due to changes in $\vec{\bf L}$, is~\cite{Damour88}
\begin{equation}
{di\over dt}={G\over ac^2}{\pi\over(1-e^2)^{3/2}}\sum_i{I_i(4M_i+3M_{-i})\over M_ia^2P_i}\sin\theta_i\cos\phi_i
\label{eq:di}\end{equation}
where $\theta_i$ is the angle between $\vec{\bf S}_i$, $I_i$ is the moment of inertia of pulsar $i$, and $\vec{\bf L}$ and $\phi_i$ is a projection angle between the line of sight and $\vec{\bf S}_i$.  The amplitude of pulsar's $i$ precession is $\delta_i=(|\vec{\bf S}_i|/|\vec{\bf L}|)\sin\theta_i$.  Also $|\vec{\bf S}_i|=2\pi I_i/ P_i$ where $P_i$ is pulsar $i$'s spin period.  The periodic departure of pulse arrivals has an amplitude
\begin{equation}
\delta t_A={M_{B}\over M}{a\over c}\delta_A\cos i\simeq(1.7\pm1.6)~I_{A,80}~\mu{\rm s},
\end{equation}
where $M$ is the total mass.  We evaluated this for the case of pulsar PSR J0737-3039 and it is clear it is essentially unobservable since the system is
nearly edge-on: $i\approx90^\circ$.  We put the moment of inertia in units of $80M_\odot$ km$^{2}$.

\begin{figure}[h]\vspace*{-1.cm}
\hspace*{-1cm} \includegraphics[width=10.5cm,angle=90]{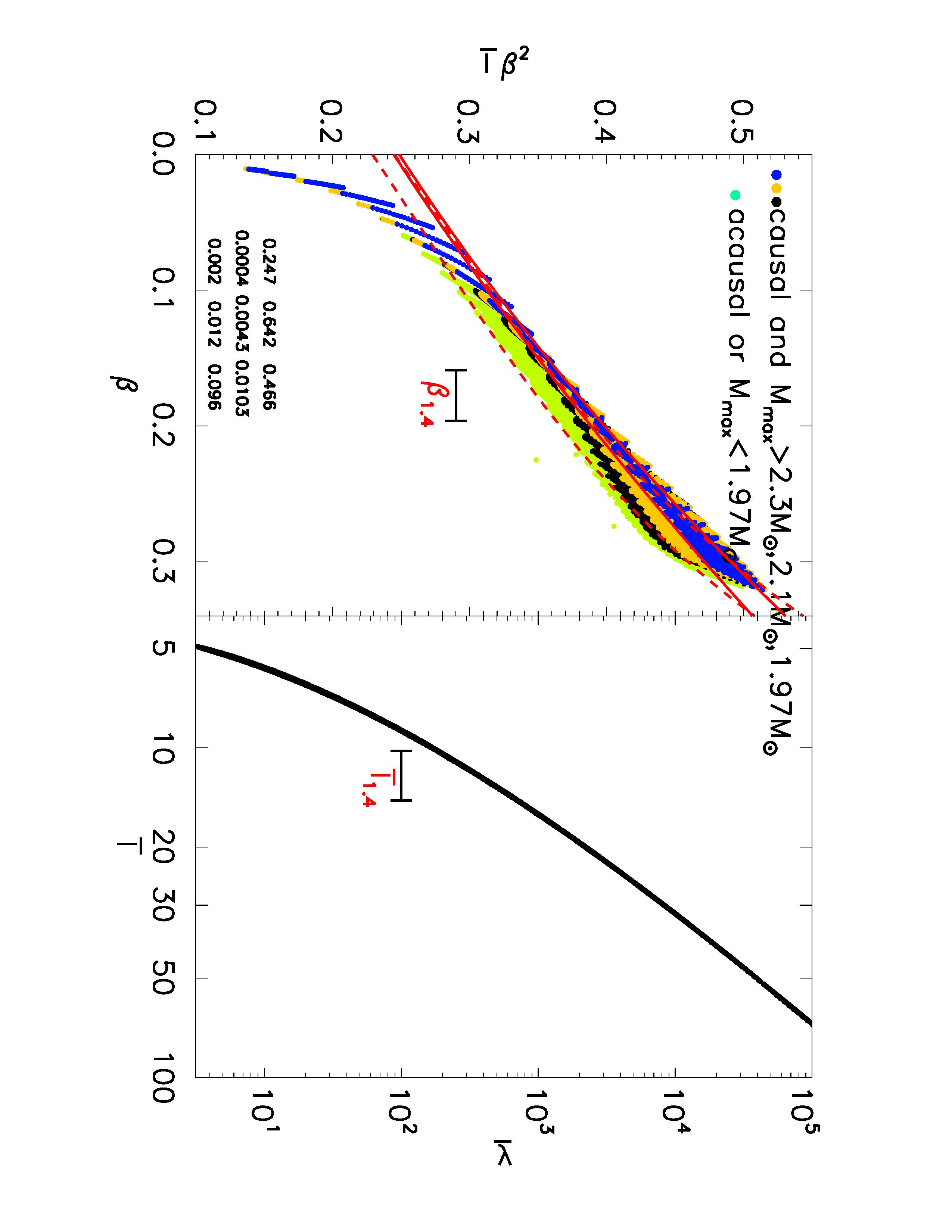}
\vspace*{-1.5cm}
  \caption{Quasi-universal behaviors of the moment of inertia  with respect to $\beta$ ( left panel) and tidal deformability ($\bar\lambda)$, right panel) of neutron stars.  Note that $I\beta^2/M^3\propto I/(MR^2)$, $\bar I=Ic^4/(G^2M^3)$ and $\bar\lambda=\lambda G^4/(Mc^2)^5$.  Each point is a stellar model computed with a piecewise polytropic EOS as described in Sec. \ref{Sec:fixed}.  Models violating causality or not supporting $M_{max}=1.97M_\odot$ are indicated by green points; other colors show causal results for various maximum masses.  The compactness range $\beta_{1.4}$ for $1.4M_\odot$ stars is indicated.  Solid and dashed lines in the left panel indicate Eqs. (\ref{eq:mom2}) and (\ref{eq:mom1}), respectively.
Adapted from Ref. \cite{Lattimer15}.}
  \label{fig:mom}
\end{figure}

On the other hand, the advance of the periastron is proportional to $I_i$ and its observability is maximized when $i=\pi/2$, i.e., for edge-on orbits.  It's observability is likely because it has a magnitude similar to the 2nd post-Newtonian correction to periastron advance:\footnote{$\theta_A\simeq13^\circ$ in PSR J0737-3039, so a small term from Ref.~\cite{Lattimer05} is consequently ignored.}
\begin{align}
{A_p\over A_{2PN}}&\simeq-{8Pc^2\over G(1-e^2)^{1/2}M^2a}\left({189\over1-e^2}-47\right)^{-1}
\sum_i{I_i(4M_i+3M_{-i})\over M_iP_i}
\cos\theta_i\nonumber\\
&\simeq-1.30^{+0.04}_{-0.12}~I_{A,80}.\label{eqn:periastron}
\end{align}
It has been suggested~\cite{Lattimer05} that a measurement accuracy of 10\% is
sufficient to place strong constraints on the EOS.  In a recent unpublished talk in Montreal in 2015, P. Freire argued that
systematics are in our favor in the case of PSR J0737-3039 in that a near cancellation of certain effects is occuring.  The implication is that a measurement accurate to 10\% is perhaps a decade away.  To illustrate its usefulness, we note that 
nearly universal, i.e., EOS-independent, correlations between the moment of inertia and the stellar compactness exist (Fig. \ref{fig:mom}).  Such a quasi-universal relation was quantified by Ref.~\cite{Lattimer05}:
\begin{equation}
I\simeq MR^2\left(0.237\pm0.008\right)\left[1+2.844\beta+18.91\beta^4\right].
\label{eq:mom1}\end{equation}
An improved fit, limited to $\beta\ge0.1$ and based on an array of piecewise polytrope EOSs as described in Sec. \ref{Sec:fixed} with $M_{max}\simge1.97M_\odot$, is \cite{Lattimer15}
\begin{equation}
I\simeq MR^2\left[0.247\pm0.002+\left(0.642\pm0.012\right)\beta+\left(0.466\pm0.096\right)\beta^2\right].
\label{eq:mom2}\end{equation}
The masses in the PSR J0737-3039 system are known to high precision.
The uncertainty in the fitting coefficients is sufficiently small
that a 10\% error in the measurement of $I$ 
will dominate the uncertainty ($\sim6-7$\%) of a radius measurement.

\subsubsection{Pulse-profile observations}

Surface emission from spinning neutron stars  produces a periodic brightness change as hot and cold spots move in and out of the line of sight.  Brightness variations could be due to magnetic fields, by non-uniform thermonuclear burning in an X-ray burster, or anisotropic accretion from a companion.  The strong surface gravity of a neutron star affects the pulse shapes and amplitudes; generally, the more compact the star, the more of its surface is visible to a distant observer, and the smaller the amplitudes.  Fortunately, using pulse profiles to measure compacness does not depend on the star's distance or intervening absorbing matter, unlike measurements from PRE bursts or qLMXBs.  The method is one motivation for the funded NASA's {\it NICER} (Neutron star Internal Composition ExploreR)~\cite{Gendreau12} and the proposed ESO {\it LOFT} (Large Observatory For x-ray Timing)~\cite{Feroci12} missions. 

Early attempts to employ this technique generally studied bolometric emissions~\cite{Pavlov97, Poutanen03, Bogdanov07, Leahy08} which could not untangle degeneracies between $M$ and $R$ and produced estimated values for $\beta$ that, while consistent with neutron stars, lacked enough accuracy to be useful.  It has been suggested, however, that pulse profiles observed in two or more energy bands supply additional constraints that can break these 
degeneracies~\cite{Strohmayer97,Miller09,Psaltis14,Miller15}.  The method makes use of nearly universal relations, similar to the correlations shown in Fig. \ref{fig:mom}, among properties of rotating neutron stars. These include the surface ellipticity, the specific angular momentum $a=Ic/(GM^2P_s)$, where $P_s$ is the spin period, the star's quadrupole moment, $M$, and $R_{eq}$, the equatorial radius~\cite{Morsink07,Yagi13,Baubock13}.  Models~\cite{Psaltis14} show that the pulse profile is nearly independent of the angular size of a spot~\cite{Bogdanov07}, and, using the aforementioned semi-universal relations, the profiles are determined essentially by $M, R_{eq}$, the inclination $i$, and $\theta_s$, the colatitude of the spot.  $M, R_{eq}$ and $P_s$ are sufficient to determine $R$, the non-rotating radius for the same mass $M$, and, hence, will predict the EOS.  The {\it NICER} mission was designed to acquire $M$ and $R$ to about 5\% accuracy from at least one pulsing source, PSR J0437-4715.

\subsubsection{Gravitational wave observations}
The merger of compact binary stars is expected to be the main source of gravitational radiation signals observed with Advanced LIGO and VIRGO~\cite{Abadie10}.  Detection of a regular sinusoidal waveform prior to the merger should allow determination of the so-called chirp mass ${\cal M}=(M_1M_2)^{3/5}/M^{1/5}$ to high accuracy.  Tidal effects during inspiral allow the possibility of breaking the remaining mass degeneracy~\cite{Flanagan08} through the measurement of the tidal deformability $\lambda=2k_2R^5/(3G)$, where $k_2$ is the tidal Love number.
The Love number, like the moment of inertia, can be found from the solution of a first-order differential equation~\cite{Postnikov10}
\begin{align}
{dy\over dr}&=-{y^2\over r}-{y-6\over r-2Gm/c^2}-rQ,\nonumber\\
Q&={4\pi G\over c^4}{(5-y)\epsilon+(9+y)p+(\epsilon+p)/c_s^2\over1-2Gm/c^2}-\left[{2G(m+4\pi pr^3/c^2)\over r(rc^2-2Gm)}\right]^2,\nonumber\\
k_2&={8\over5}\beta^5(1-2\beta)^2\left[2-y_R+2\beta(y_R-1)\right]/{\cal R}\label{eq:yq}\\
{\cal R}&=6\beta(2-y_R+\beta(5y_R-8))+4\beta^3\left[13-11y_R+\beta(3y_R-2)+2\beta^2(1+y_R)\right]\nonumber\\
&+3(1-2\beta)^2\left[2-y_R+2\beta(y_R-1)\right]\ln(1-2\beta),\nonumber
\end{align}
where $y(0)=2$ and $y_R=y(R)$.

The seemingly chaotic signal immediately following tidal disruption, interestingly, seems to have characteristics that depend primarily on the neutron star's tidal deformability.  For black hole-neutron star (BH-NS) mergers, it has been argued that combining pre-merger and post-merger information best fits the tidal deformability, compared to other neutron star 
properties~\cite{Lackey12,Lackey14}.   For a neutron star-neutron star merger, the gravitational wave frequency at the peak of the signal (when tidal disruption occurs) has been found to strongly correlate with $\bar\lambda$~\cite{Read13}.  There are two dominant peaks in the gravitational wave spectrum of the postmerger phase in a neutron star-neutron star merger, which originates from  fundamental quadrupolar oscillation modes, and lie in the range of 2 -- 3.5~kHz.  The lower frequency appears to correlate with the compactness $\beta$ in a quasi-universal relation~\cite{Bernuzzi15}.  The higher frequency correlates with the radius $R_{1.6}$ of $1.6M_\odot$ stars~\cite{Bauswein14}, but the relation is mass dependent.  Takami et al.~\cite{Takami15} instead argue that the higher frequency nicely correlates with the tidal deformability $\bar\lambda$.  Thus, at least for neutron star mergers where masses can be sufficiently well determined, there are three separate possible constraints.  While a single merger may not provide accurate enough information to determine $M$ and $R$ precisely, observations of several mergers may allow strong statistical information to be derived.   This is especially true, given the extremely strong universal correlation between the tidal polarizability and the moment of inertia (Fig. \ref{fig:mom})~\cite{Yagi13}.

\subsubsection{Rotation Periods}

Yet another nearly universal relation holds between the maximum spin frequency of a uniformly rotating neutron star and its average density~\cite{Lattimer04,Haensel09}
\begin{equation}
f_K=1.08\left({M\over M_\odot}\right)^{1/2}\left({10{\rm~km}\over R}\right)^{3/2}{\rm~kHz},
\label{eq:fk}\end{equation}
where $f_K$ indicates the Keperlian frequency at which mass shedding occurs.  The $M$ and $R$ in this formula do not refer to the mass and radius of the spinning configuration but rather to the non-spinning star.  Although this formula was phenomenologically established from full axially-symmetric solutions of Einstein's equations, the dependence on mass and radius is the same as in Newtonian gravity.  Furthermore, the coefficient 1.08 kHz is very close to the prediction of the analytic relativistic Roche model (which assumes the gravitational potential of the spinning star is given by concentrating all its mass at the 
origin)~\cite{Shapiro83,Shapiro89}, $\simeq1.00$ kHz.  It is also interesting that the equatorial radius of the maximally rotating star is about 1.44 times $R$, the non-spinning radius for a star of the same mass $M$, compared to the analytic result of 1.5 from the relativistic Roche model.   The highest frequency pulsar known is PSR J1748-2446ad~\cite{Hessels06} with $f=716$ Hz, and its limiting radius as a function of mass is shown in Fig. \ref{fig:mr}.  Unfortunately, the mass of this pulsar is unknown, so an upper limit to its radius is unable to be established.  However, if most rapidly spinning pulsars, so-called millisecond pulsars, acquired their angular momentum by accretion from a companion, it is not out of the question that a rapidly rotating millisecond pulsar will eventually be found in a binary system with measured masses.  Assuming that $M\simle2M_\odot$, Eq. (\ref{eq:fk}) implies $R\simle16.6$ km.

\subsubsection{Binding Energies}

The neutron star binding energy is the difference in mass between a star's matter if at infinite dispersion (the baryon mass ${\cal N}m_b$) and the gravitational mass $M$:
\begin{equation}
{\rm BE}={\cal N}m_b-M=\int{nm_b\,d^3r\over\sqrt{1-2Gm/(rc^2)}}-\int\epsilon\,d^3r \,.
\label{eq:be3}\end{equation}
Its utility rests on quasi-universal relations connecting the binding energy to the compactness $\beta$ (cf. Ref. \cite{Lattimer01})
\begin{equation}
{{\rm BE}\over Mc^2}\simeq(0.60\pm0.06){\beta\over1-\beta/2}
\label{eq:be4}\end{equation}
 or to the moment of inertia $I$. An improved fit, based on an array of piecewise polytrope EOSs with $M_{max}\simge1.97M_\odot$ and limited to $\beta\simge0.1$, is~\cite{Lattimer15}, 
\begin{equation}
{\rm BE}\simeq\beta Mc^2\left[0.564\pm0.010+(0.521\pm0.077)\beta\right].
\label{eq:be5}\end{equation}
 An accurate determination could in principle be made from the neutrino signal from a supernova, in which the detector serves, essentially, as a calorimeter.  Radius determinations would depend on the accuracy with which the mass of the remnant can be determined.  Measurements of neutrinos from SN 1987A yielded a value in qualitative agreement with a neutron star of $1.2-1.5M_\odot$ and a moderate radius, BE $\approx(2-3)\cdot10^{53}$ ergs.  A contemporary supernova observation of a galactic supernova is likely to yield not a dozen events, as in the case of SN 1987A, but tens of thousands of events and a substantial statistical improvement.  

Another suggested measure of binding energy is related to the lightest pulsar masses measured in neutron star binaries.  Under the assumption that nearly no accretion has accumulated onto a non-recycled pulsar, theoretical estimates for the baryon mass of the remnant based on stellar evolution can be compared to the measured mass~\cite{Podsiadlowski05}.    The lightest neutron stars may be formed by electron-capture supernovae, for which theoretical models imply a well-defined baryon mass ${\cal N}m_b\simeq(1.370\pm0.005)M_\odot$.  As one example, the mass of pulsar B in the system PSR J0737-3039 is $(1.2489\pm0.0007)M_\odot$ \cite{Kramer06} and is thought to have been formed in an electron-capture supernova.  The difference between its present mass and the mass before collapse ($\simeq{\cal N}m_b$), about $(0.121\pm0.005)M_\odot c^2$, represents its BE.  For the case at hand, using Eq. (\ref{eq:be5}),  $\beta\simeq(0.154\pm0.008)$ and $R\simeq(12.0\pm0.6)$ km.  However, if fall-back of matter onto the neutron star after the supernova occurs is significant, i.e., is more than $0.01M_\odot$, the BE is undererestimated and $R$ is overestimated.

In summary, the future appears bright with prospects for further constraining neutron star radii and the dense matter EOS.  We now turn to investigations of the thermal properties of dense matter.

\section{THERMAL EFFECTS IN BULK MATTER}
\label{Sec:Teffects}

In homogeneous bulk matter comprised of nucleons (phase III in Fig. \ref{schem}), the
thermal contributions to the state variables such as the energy, pressure, chemical potentials, specific heats, etc., play important roles in the evolution of compact objects. Except under circumstances when matter is in the degenerate 
($T$ small compared to the Fermi temperature $T_F$) or  the non-degenerate  ($T\gg T_F$) region, a transparent understanding of what governs the thermal effects is not possible because the relevant Fermi integrals  cannot be performed analytically to yield closed-form expressions. Because of the varying concentrations of fermions (neutrons, protons and leptons) and bosons (photons, possible Bose condensates at supra-nuclear densities) encountered, one or the other constituent will likely  lie in different regions of degeneracy. The limiting cases when non-degenerate or degenerate  conditions prevail provide valuable insights besides serving as consistency checks on numerical calculations of the thermal variables, and are discussed below.

\subsection{Non-degenerate Limit}
\label{Sec:Ndeg}
When the fugacities $e^{\mu_i/T}$ are small ($\mu_i$ being the chemical potential of species $i$), non-degenerate (or dilute gas) conditions prevail. In this case, corrections to the ideal gas expressions for the thermal state variables are generally obtained by using the fugacity expansion or the virial expansion approach in which measured phase shifts serve as an input. For contact interactions, non-relativistic Skyrme models being a prototype example, the thermal pressure $P_{th}$ and energy density $\epsilon_{th }$ for a single species of fermions can be written in terms of their ideal gas counterparts (calculated with the Landau effective mass $m^*(n)$ instead of $m_b$) as \cite{Prakash87}
\ba
P_{th}(n,T) &=& P_{th}^{id}(n,T;m^*) \left( 1 - \frac 32 \frac {n}{m^*} \frac {dm^*}{dn} \right) \nonumber \\
\varepsilon_{th}(n,T) &=& \varepsilon_{th}^{id}(n,T;m^*)\,, \qquad
P_{th}^{id}/\varepsilon_{th}^{id} = 2/3 \,, 
\label{ndegth}
\ea
where $m^* = p_{F}(\partial \epsilon(p)/\partial p)_{p_{F}}^{-1}$ with $p_F$ denoting the Fermi momentum and $\epsilon(p)$ the single-particle spectrum. These results are valid for all regions of degeneracy and help to develop simple expressions in the non-degenerate (and degenerate) conditions (see e.g., Ref. \cite{APRppr}).   For non-relativistic models with finite-range interactions, a method involving next-to-leading order steepest descent approach for the calculation of all the state variables has been developed  in Ref. \cite{cons15a}. For relativistic mean-field theoretical (MFT) models,  the  appropriate formulas in the dilute gas limit can be found in 
Ref. \cite{Prakash87}.

\subsection{Degenerate Limit}
\label{Sec:Deg}
For nucleons in the homogeneous phase, Landau's Fermi Liquid Theory (FLT) provides simple analytical expressions that are model independent for the thermal state variables  in the limit of extreme degeneracy ($T/T_F \ll 1$, $T_F$ being the Fermi temperature) \cite{flt}.  In the absence of collective excitations close to the Fermi surface, thermal effects are primarily determined by the nucleon's Landau 
effective mass and its first density derivative which in turn depend on the momentum-dependence of the $T=0$ single-particle energy spectrum.  Recently, analytical formulas valid to next-to-leading order (NLO) effects in $T/T_F$ for all of the thermal state variables have been developed in Ref. \cite{cons15b}, and are summarized below. 

\subsubsection{Non-relativistic models}

In non-relativistic models, the single particle spectrum has the  structure
\be
\epsilon = p^2/2m_b +  \mathcal{U}(n) + R(n,p) \,,
\ee
where $\mathcal{U}(n)$ denotes the contribution that depends only on the density $n$, whereas $R(n,p)$ contains the density- and momentum-dependent part .  The implicitly temperature-dependent Landau effective mass function 
\be
\mm(p) = m_b\left[1+\frac{m_b}{p}\frac{dR(p)}{dp}\right]^{-1} \, 
\label{genm}
\ee
is related to the Landau effective mass $m^*$ via
\be
\mm(p=p_F;T=0) = m^*\,,
\ee
where $p_F$ is the Fermi momentum. 
For a single-species system of spin $1/2$ particles, the leading low-temperature correction to the Fermi-liquid behavior renders the entropy density to be
\be
s \simeq 2anT - \frac{16}{5\pi^2}a^3nT^3(1-L_F) \label{s32} \,,
\ee
where the level-density parameter $a=\pi^2/(4T_F)=\pi^2m^*/(2p_F^2)$, and  
\be
L_F \equiv \frac{7}{12}p_F^2\frac{\mm_F'^2}{m^{*2}} + \frac{7}{12}p_F^2\frac{\mm_F''}{m^*} + \frac{3}{4}p_F\frac{\mm_F'}{m^*} \,,
\label{LF}
\ee
where the primes denote derivatives with respect to momentum $p$. In general,
\be
\left.\frac{d\mm(p)}{dp}\right|_{p_F} = \mm_F' \ne m^{*\prime}=\frac{d\mm(p_F)}{dp_F}
\ee
as $R$ can contain both $p$ and $p_F$ (via $n$).
The quantity $L_F$ arises from nontrivial momentum dependencies in the single-particle potential. 
For free gases (where $R(p)=0$), and for systems having 
only contact interactions where $R(p) \propto p^2$ (such as Skyrme models), $L_F=0$. For a multi-component system, the total entropy density is a sum of the contributions from the 
individual species where, in Eq. (\ref{s32}), the Fermi momentum, the effective mass, and its derivatives all carry a particle-species 
index $i$. NLO results for the entropy per particle $S$, thermal energy $E_{th}$, and thermal pressure $P_{th}$ of a single species system of spin 1/2 particles are:
\begin{eqnarray}
S &\simeq& 2aT -\frac{16}{5\pi^2}a^3T^3(1-L_F)\,, \nonumber\\
E_{th} &\simeq& aT^2 - \frac{12}{5\pi^2}a^3T^4(1-L_F) \,, \label{eth3}\\
P_{th} &\simeq& \frac{2}{3}anQT^2    
      - \frac{8}{5\pi^2}a^3nQT^4\left(1-L_F+\frac{n}{2Q}\frac{dL_F}{dn}\right) \,, \nonumber 
\end{eqnarray}
where
\be
Q = 1-\frac{3n}{2m^*}\frac{dm^*}{dn} \,.
\ee
In simulations of astrophysical phenomena, thermal effects are often incorporated via the thermal index 
$\Gamma_{th} = 1 + P_{th}/\epsilon_{th}$, where $\epsilon_{th}$ is the thermal energy density.  For non-relativistic models, 
\be
\Gamma_{th} = 1+\frac{P_{th}}{nE_{th}} \simeq 1+\frac{2Q}{3}-\frac{4}{5\pi^2}a^2nT^2\frac{dL_F}{dn} \,.
\ee

\subsubsection{Relativistic models}
\label{Sec:MFT}
The single-particle energy spectrum of relativistic mean-field theoretical (MFT) models \cite{Muller96}
has the structure
\be
\epsilon = E^* + U(n)\,, \quad E^* = \sqrt{p^2 + M^{*2}(n,T)}\,. \label{mftmstar}
\ee
The potential $U(n)$ is the result of vector meson exchanges, whereas the Dirac effective mass $M^*$ 
arises from scalar meson interactions. In this case, the NLO thermal variables are \cite{cons15b}
\ba
S &\simeq& 2aT -\frac{16}{5\pi^2}a^3T^3(1-L_F)\,, \label{sarel3}\nonumber\\
E_{th} &\simeq& aT^2 - \frac{12}{5\pi^2}a^3T^4(1-L_F) \,,  \label{ethrel3} \\
P_{th} &\simeq& \frac{1}{3}anT^2(1+q)    
      - \frac{4}{5\pi^2}a^3nT^4\left[1-L_F+q\left(1-\frac{L_F}{3}-\frac{10}{9}\frac{p_F^4}{E_F^{*4}}\right)\right]  \nonumber
\ea
where, using the Fermi temperature $T_F = p_F^2/(2E_F^*)$,
\be
a = \frac{\pi^2}{4T_F} \,,  \quad
q = \frac{M^{*2}}{E_F^{*2}}\left(1-\frac{3n}{M^*}\frac{dM^*}{dn}\right)   \quad {\rm and} \quad
L_F = \frac{11}{12}\frac{p_F^2}{E_F^{*2}}-\frac{5}{12}\frac{p_F^4}{E_F^{*4}} \,.
\ee
Relations applicable to field theoretical calculations beyond MFT are easily derived 
(following the procedure adapted in the non-relativistc case) by adding to Eq. (\ref{mftmstar}) contributions from two-loop contributions \cite{cons15b}. Although such contributions do not lend themselves to analytical forms, straightforward numerical derivatives are easily calculated. 

\subsection{Illustrative Results}

In the results above, the entropy density and specific heats are carried to ${\cal O} (T/T_F)^3$ whereas the energy density and pressure to ${\cal O} (T/T_F)^4$, extending the leading order results of FLT.  These extensions employed  a generalized Landau effective mass function which enables the calculation of the entropy density, and thereafter the other state variables from Maxwell's relations, for a general single-particle spectrum. For models with contact interactions, knowledge of the Landau effective mass suffices, whereas for models with finite-range interactions, momentum derivatives of the Landau effective mass function up to second or third order are required. 

The value of the NLO results lies in the fact that thermal effects in the near-degenerate to degenerate limits can be calculated from the $T=0$ single-particle spectra from microscopic calculations (e.g., non-relativistic and relativistic Brueckner-Hartree-Fock  calculations or effective field theoretical models) which may prove time consuming at finite $T$.  
To illustrate this advantage, models that are widely used in nuclear and neutron star phenomenology were  chosen in Ref. \cite{cons15b} for demonstration purposes. 

\subsubsection{Non-relativistic models}

In the category of non-relativistic potential models,  the finite-range
model, referred to as MDI(A) \cite{Welke88,Das03}, that reproduces the empirical properties of isospin symmetric
and asymmetric bulk nuclear matter \cite{cons15a}, optical model fits to
nucleon-nucleus scattering data \cite{Hama:90}, heavy-ion flow data in the energy
range 0.5-2 GeV/A \cite{Danielewicz02}, and the largest well-measured neutron star mass of
2 $\rm{M}_\odot$ , was contrasted in Ref. \cite{cons15b}
with a conventional zero-range Skyrme model known as SkO$^\prime$ \cite{Reinhard99}.  Both models predict nearly
identical zero-temperature properties (energy density and pressure) at all densities and proton
fractions, and thus the neutron star maximum mass \cite{cons15a}, but differ in their
predictions for heavy-ion flow data \cite{Prakash88b}.

\begin{figure}[h]
\centering
\begin{minipage}[b]{0.495\linewidth}
\centering
\includegraphics[width=6.74cm,height=7cm]{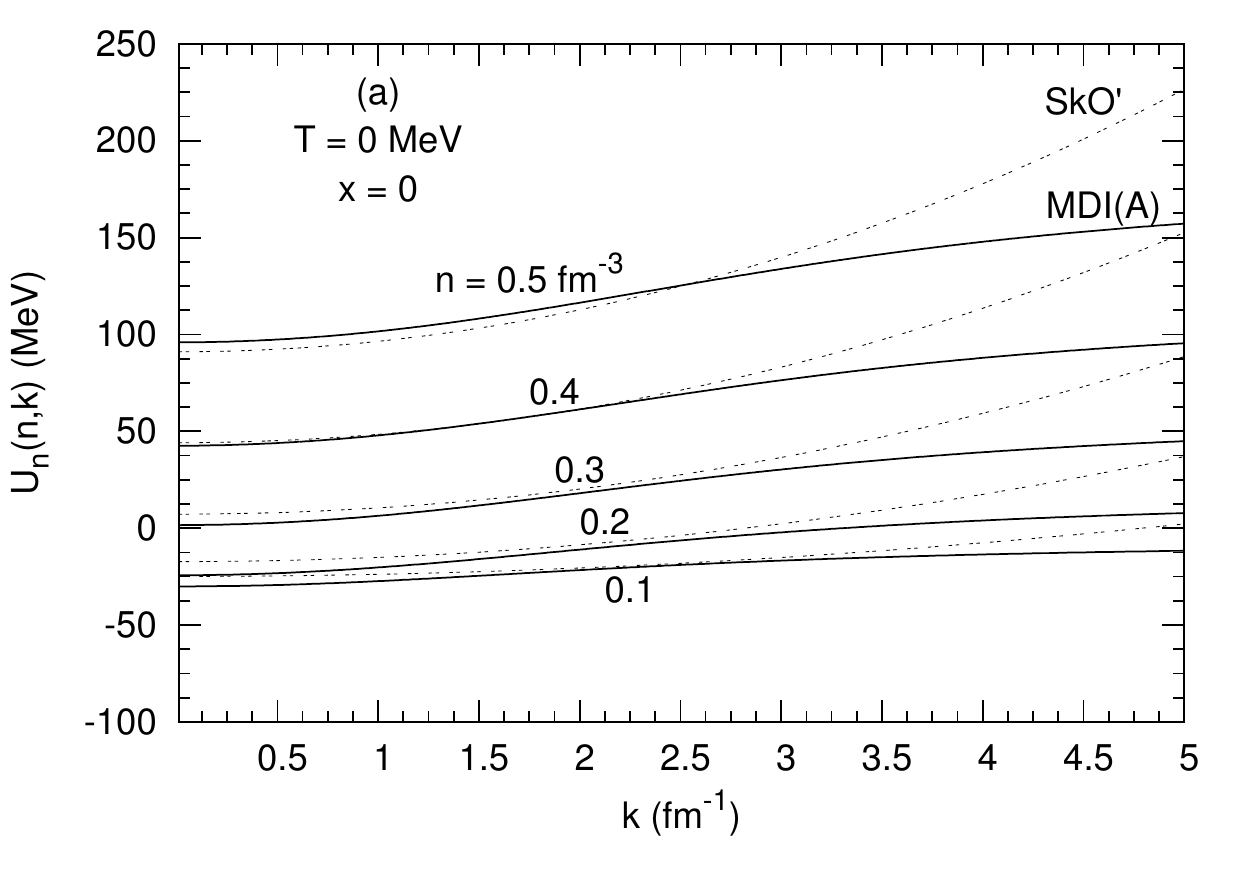}
\end{minipage}
\begin{minipage}[b]{0.495\linewidth}
\centering
\includegraphics[width=6.74cm,height=7cm]{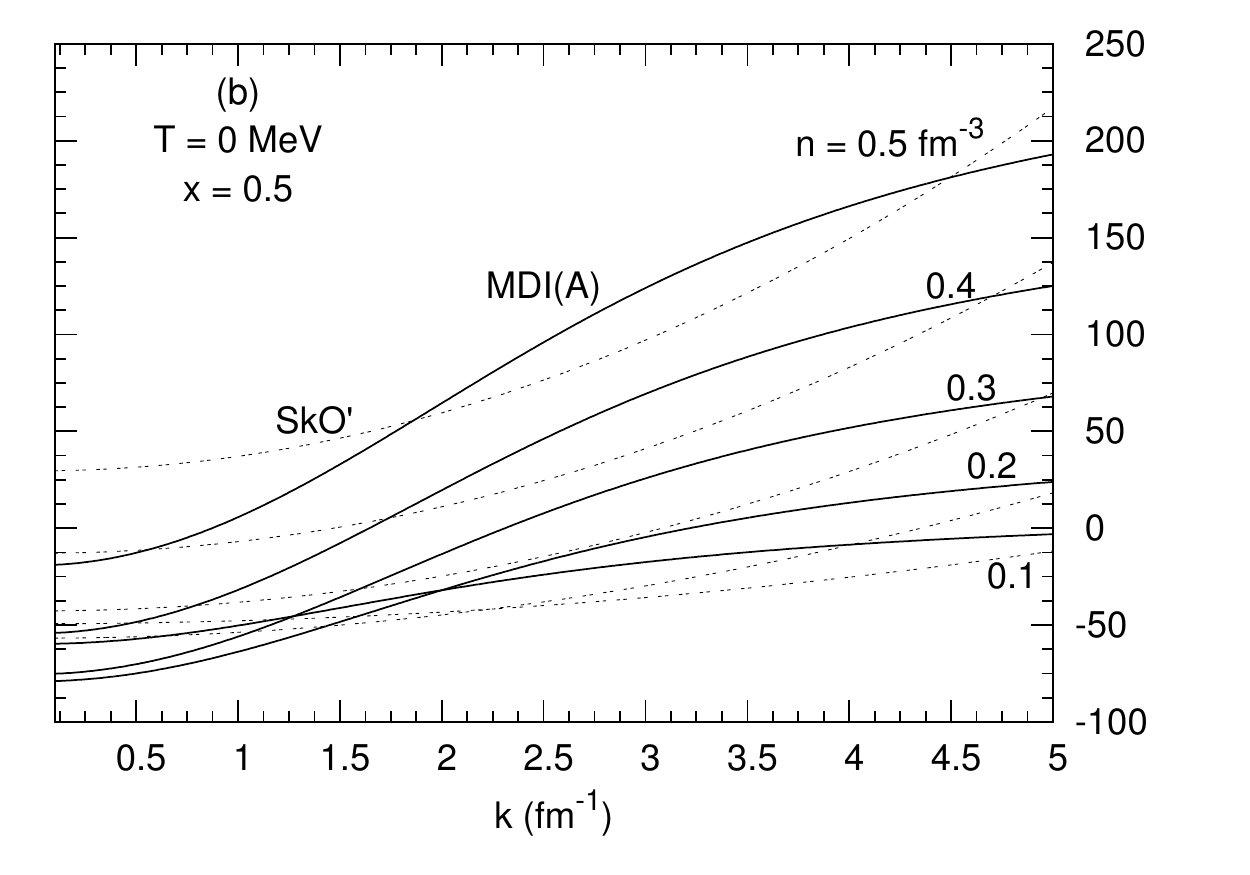}
\end{minipage}
\vskip -0.5cm
\caption{Neutron single particle potentials vs momentum at $T=0$
for the densities $n$ and proton fractions $x=n_p/n$ as marked. Figure from Ref. \cite{cons15a}.}
\label{MDYISk_SPP}
\end{figure}

Fig. \ref{MDYISk_SPP} shows the neutron single particle potentials for the MDI(A) and
SkO$^\prime$ models as functions of momentum for select baryon densities at $T=0$ \cite{cons15a}. 
Results shown are for pure neutron matter ($x=0$)
[Fig. \ref{MDYISk_SPP}(a)] and for isospin asymmetric matter with $x=0.2$ [Fig. \ref{MDYISk_SPP}(b)].  
Owing to the logarithmic structure of the function $R(n,p)$ (see Ref. \cite{cons15a} for its explicit structure), results for MDI(A) tend to saturate at large momenta for both proton fractions.  The quadratic rise with momentum in the
the SkO$^\prime$ model is  common to all Skyrme models.  
For both MDI(A) and SkO$^\prime$ models,  
the effect of a finite $x$ [Fig. 2(b)] is more pronounced at low momenta for which the single particle potential 
becomes more attractive relative to that for pure neutron matter. 

%
%
%

\begin{figure}[h]
\centering
\begin{minipage}[b]{\linewidth}
\centering
\includegraphics[width=12cm]{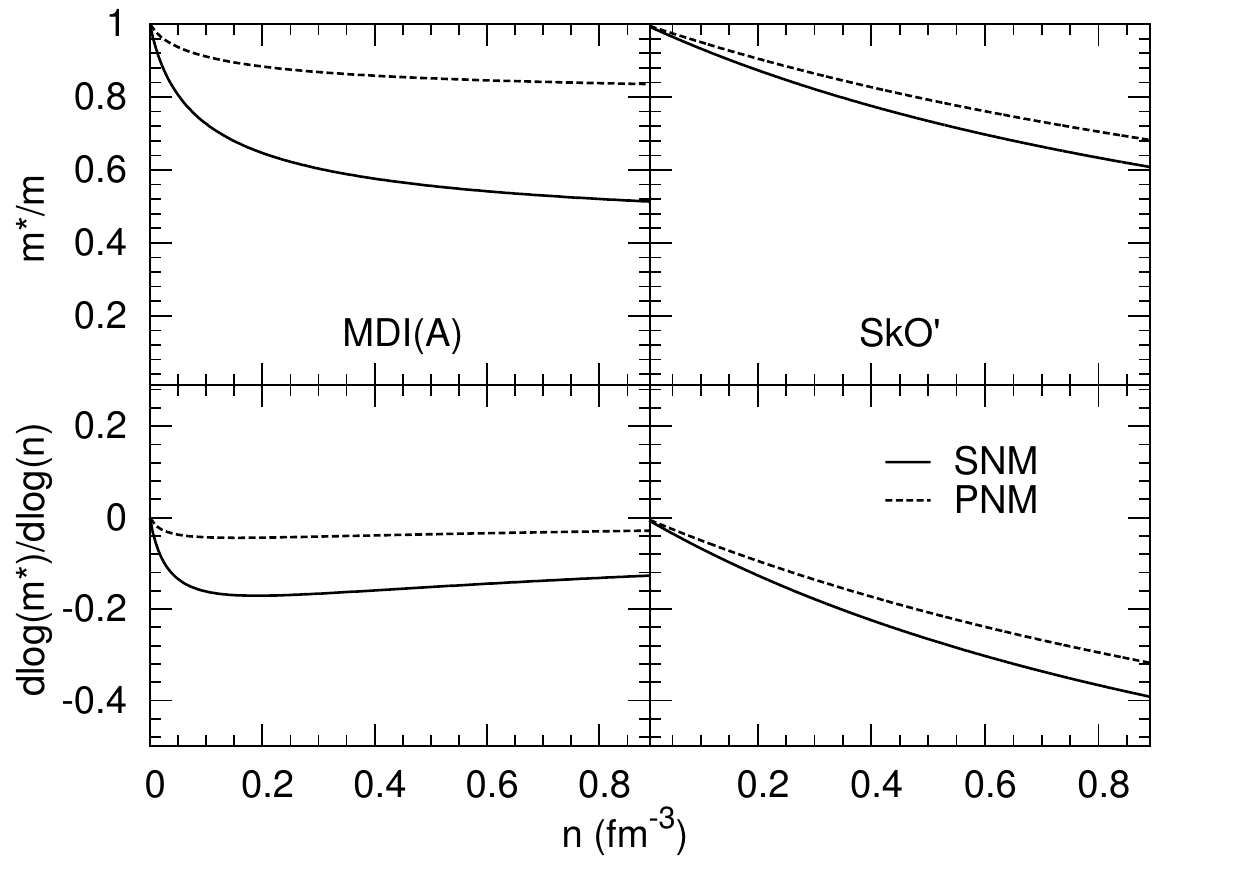}
\end{minipage}
\vskip -0.5cm
\caption{The neutron effective masses (top panels) and their logarithmic derivatives (bottom panels) for non-relativistic potential models (MDI(A) and SkO$^\prime$)  for symmetric nuclear matter (SNM) and pure neutron matter (PNM).   
 Figure extracted from Ref. \cite{cons15b}.}
\label{2mod_Ms_dMs}
\end{figure}

The neutron Landau effective masses $m_n^*$  (scaled with its vacuum value)
and their logarithmic derivatives with respect to density 
are shown as functions of $n$ in  Fig. \ref{2mod_Ms_dMs}. 
Noteworthy points here are: (i)  The isospin splittings are  qualitatively 
similar -- $m_n^*/m_b$ being larger for PNM than for SNM -- although quantitative differences are present, and (ii) except for $n$ up 
to $0.2~{\rm fm}^{-3}$, the decrease with increasing $n$  for the MDI model is relatively slow (logarithmic decline) compared with that for the SkO$^\prime$ model [$(1+ \rm{constant} \times n)^{-1}$ fall off]. 
The logarithmic derivatives $m_n^*$ for MDI(A) show little variation with $n$ at supra-nuclear densities, whereas
results for the SkO$^\prime$ model, which take the simple form $(m^*_n/m_b)-1$, 
show a significant variation with $n$.  The density dependences of the 
effective masses and their logarithmic derivatives determine the behavior of all the thermal properties in FLT. Higher 
order derivatives of the Landau effective mass function in Eq. (\ref{genm}) appear in FLT+NLO. 

%
%

\begin{figure}[h]
\centering
\vskip -1cm
\begin{minipage}[b]{\linewidth}
\centering
\includegraphics[width=12cm]{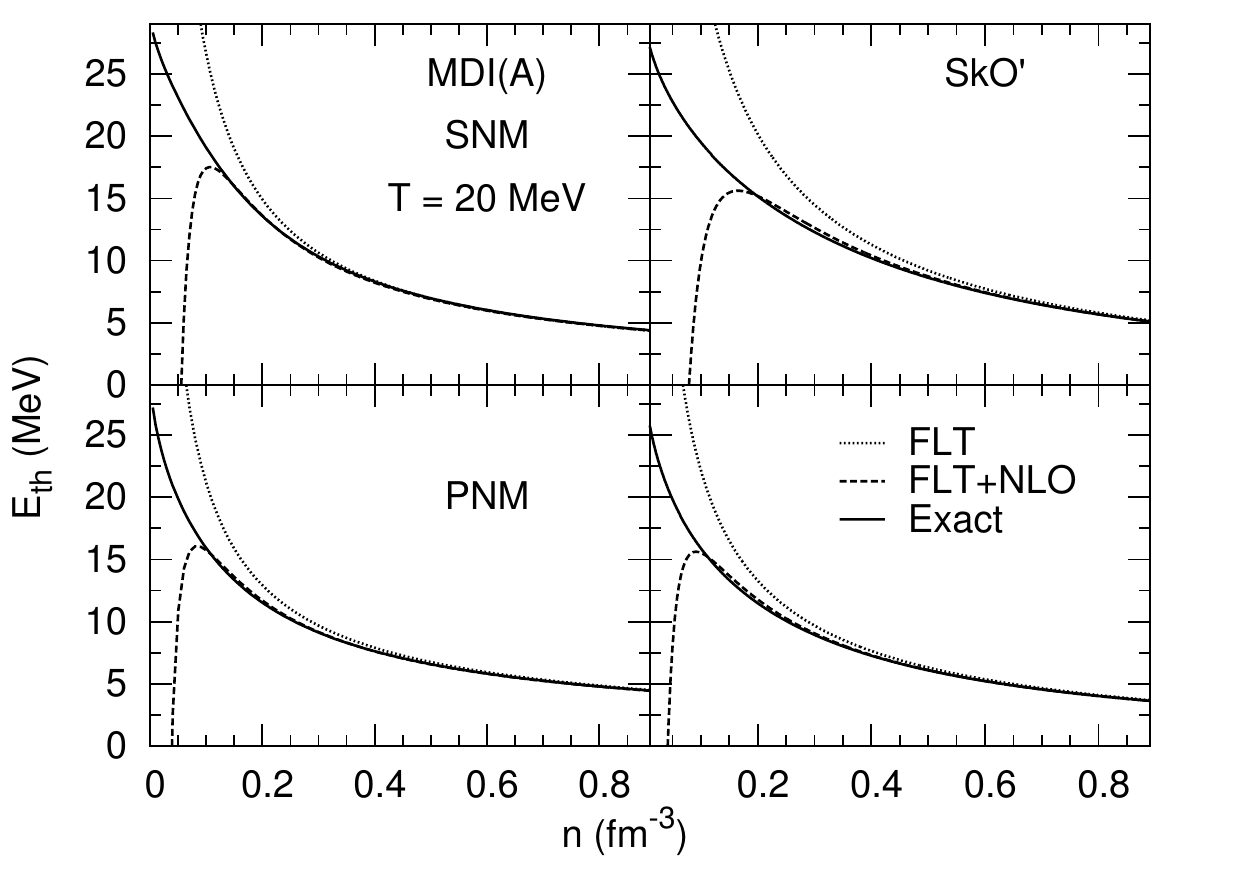}
\end{minipage}
\vskip -0.5cm
\caption{Thermal energy for the non-relativistic models (MDI(A) and SkO$^\prime$) 
at a temperature of  $T=20$ MeV. Results for SNM are 
in the top panels and for PNM in the bottom panels.  Figure extracted from Ref. \cite{cons15b}.}
\label{2mod_Eth}
\end{figure}

\begin{figure}[h]
\centering
\begin{minipage}[b]{\linewidth}
\centering
\includegraphics[width=12cm]{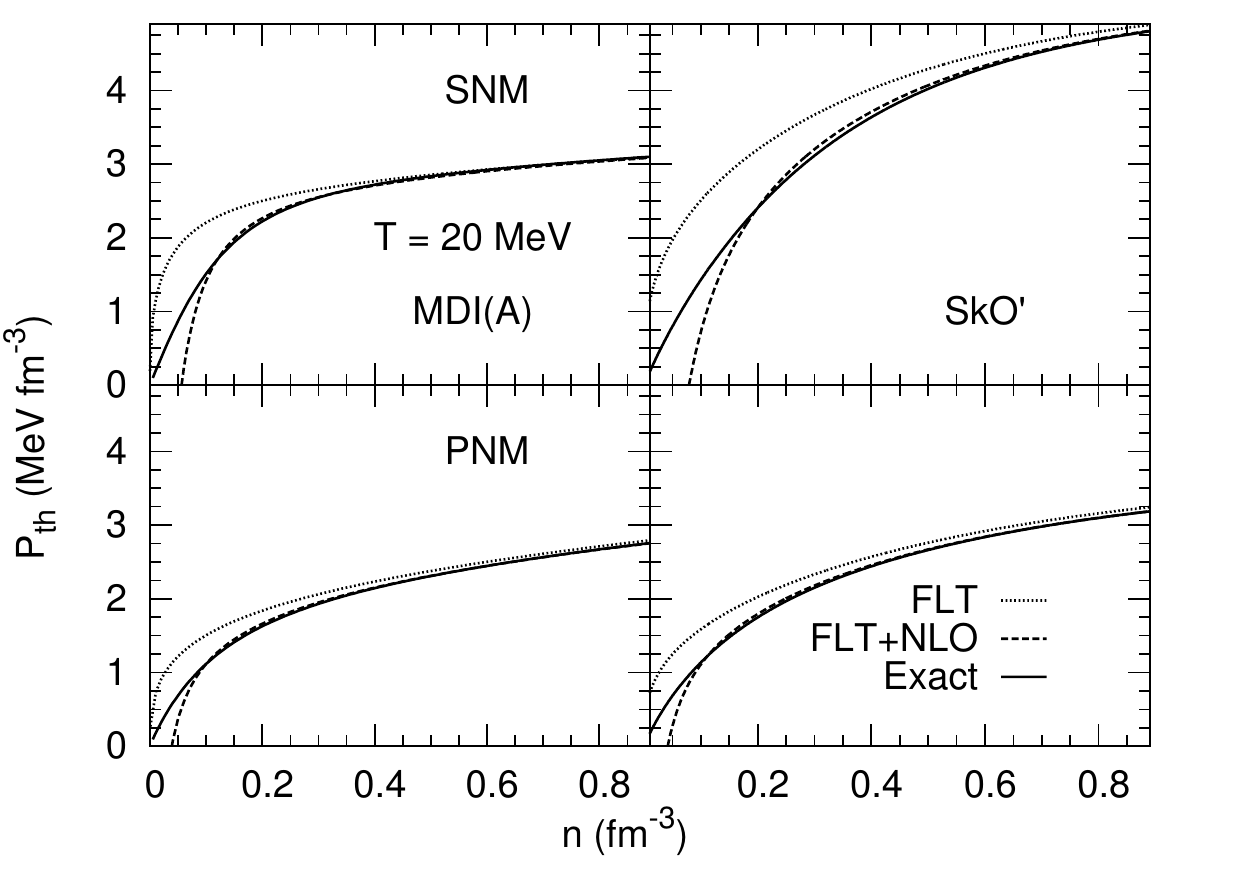}
\end{minipage}
\vskip -0.5cm
\caption{Same as Fig. \ref{2mod_Eth}, but for thermal pressure.  Figure extracted from Ref. \cite{cons15b}.}
\label{2mod_Pth}
\end{figure}

The FLT and FLT+NLO results for the thermal energy, $E_{th}$,  and the thermal pressure, $P_{th}$ are compared with the exact numerical results in Figs. \ref{2mod_Eth} and \ref{2mod_Pth}. The NLO corrections yield  agreement with the exact results down to densities of $(0.5-1) n_s$ compared to $(2-3) n_s$ for FLT. As with FLT, systematically better agreement with FLT+NLO occurs for PNM than for SNM owing to
the neutron density in PNM ($n_n=n$) being twice that  in SNM ($n_n=n/2$); PNM is more degenerate than SNM at the same $n$.  The efficacy of FLT+NLO is such that it reproduces the exact results for  all $(n,T)$ for which the entropy per particle $S \leq 2$, whereas FLT does so only for $S \leq 1$.

%
\vspace*{.75cm}
%

%
%

%
\begin{figure}[h]
\centering
\hspace{-3.cm}
\begin{minipage}[b]{0.4\linewidth}
\centering
\includegraphics[width=7.32cm,height=9cm]{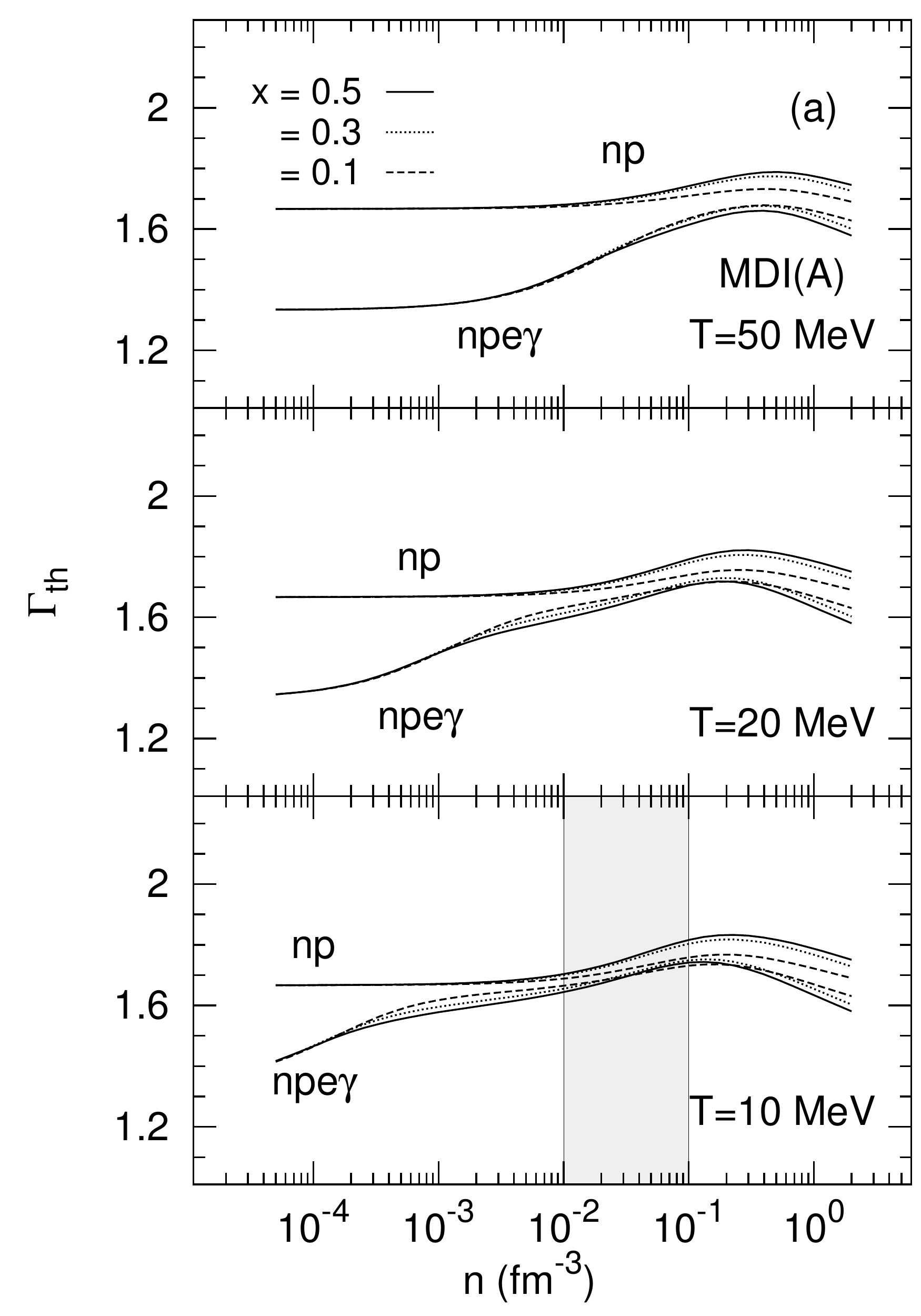}
\end{minipage}
\hspace{0.9cm}
\begin{minipage}[b]{0.4\linewidth}
\centering
\includegraphics[width=7.32cm,height=9cm]{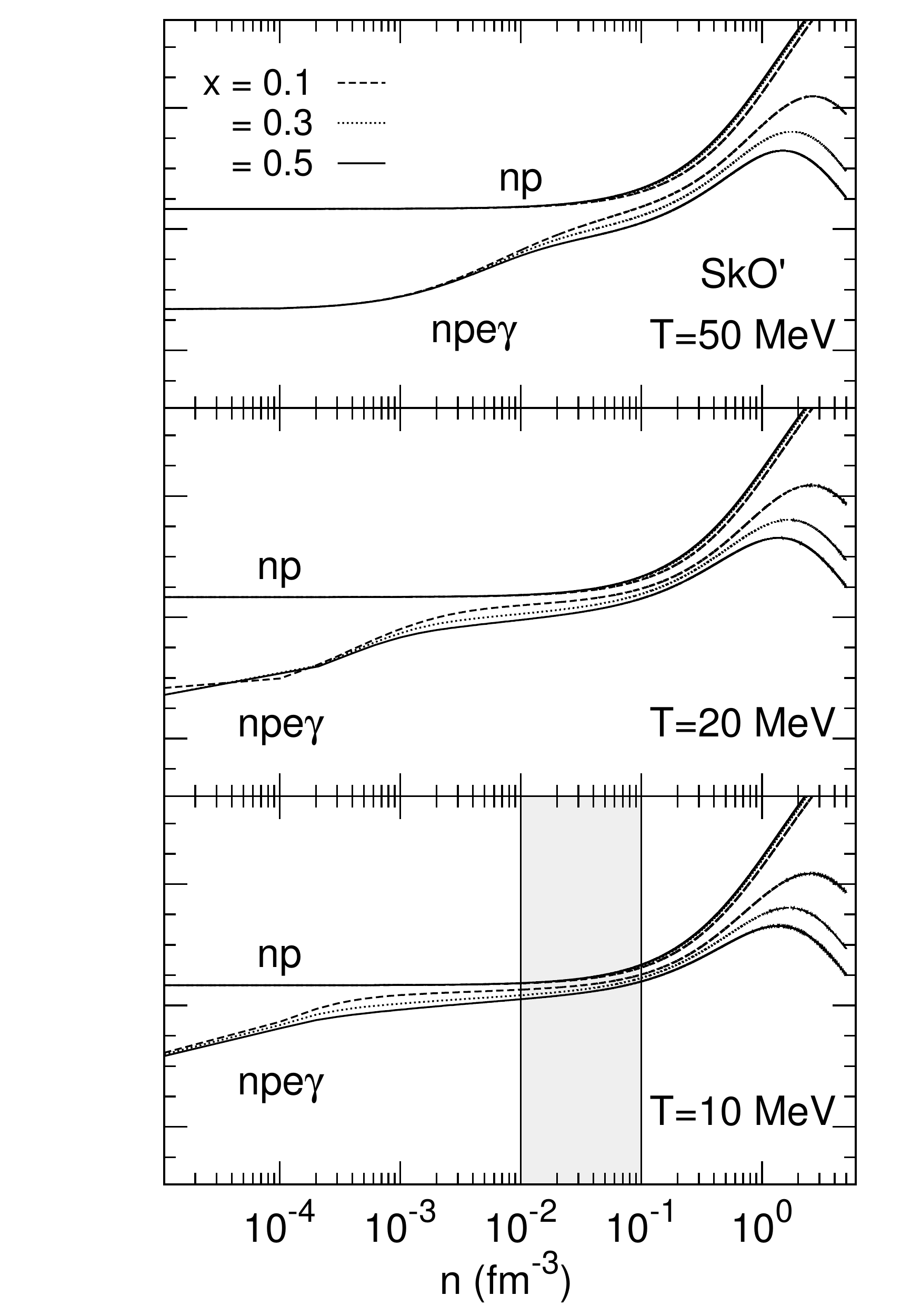}
\end{minipage}
\vskip -0.5cm
\caption{The thermal index $\Gamma_{th}$ for matter with only nucleons (np), and with leptons and photons (npe$\gamma$) for the non-relativistic models MDI(A)and SkO$^\prime$. For $T=10$ MeV, the shaded region shows the range of densities in which nuclei and pasta-like configurations (not considered here) exist. Figure extracted from Ref. \cite{cons15a}.}
\label{Gamma_th}
\end{figure}
%

Currently, many astrophysical simulations incorporate thermal effects through the use of the thermal index  $\Gamma_{th} = 1 + P_{th}/\epsilon_{th}$ \cite{Janka:93,Bauswein:10,Hotokezake:13,Foucart:14,Kaplan14}. 
In Fig. \ref{Gamma_th}, results of $\Gamma_{th}$ vs $n$ in matter with only nucleons $(np)$, and with leptons and photons included $(npe\gamma)$ are contrasted for representative values of $T$ and the proton fraction $x=Y_e$. The results shown are from exact numerical calculations for the non-relativistic models MDI(A) and SkO$^\prime$, which agree with the non-degenerate and degenerate limit expressions discussed above very well in their respective regions of applicability (see Refs. \cite{cons15a,cons15b} for  quantitative analyses).
For $n < 10^{-2}~{\rm fm}^{-3}$, effects of interactions in $np$ matter are small and $\Gamma_{th}\rightarrow 5/3$, the value for non-relativistic ideal fermions.  The presence of relativistic electrons and photons, for which $\Gamma_{th} =  4/3$, causes the total  
$\Gamma_{th}$ to tend toward  $4/3$ in $npe\gamma$ matter as $n$ decreases progressively from $10^{-2}~{\rm fm}^{-3}$. 
In the approximate range  $10^{-2}~{\rm fm}^{-3} < n < 10^{-1}~{\rm fm}^{-3}$ and for  $T<$ 15 MeV, an inhomogeneous phase consisting of nuclei, pasta-like configurations, and $npe\gamma$ (shown by the shaded region for $T=10$ MeV, but not considered in the calculations) is energetically preferred over a uniform phase of $npe\gamma$.  Trace amounts of $\alpha$-particles may be present even at $T=20$ MeV in this density region. For $T\geq 15$ MeV, inhomogeneous phases give way to  a uniform phase pf $npe\gamma$ for all densities. 

The differences in $\Gamma_{th}$  for near-nuclear to supra-nuclear densities  in $np$ matter for the MDI(A) and SkO$^\prime$ models  are caused by the differences in the behaviors of their effective masses (see Fig. \ref{2mod_Ms_dMs}) which control the thermal properties. The mild variation of $\Gamma_{th}$  for the MDI(A) model is because of the relatively flat behavior of 
$m^*$ and $d\ln m^*/d\ln n$ with increasing $n$. In contrast, the rapid rise of $\Gamma_{th}$  for the  SkO$^\prime$ model is a consequence of the monotonic decrease with $n$ of both $m^*$ and $d\ln m^*/d\ln n$. Results of both of these models are, however, modified in 
$npe\gamma$ matter with a tendency to approach the value $4/3$ for very large densities after reaching a peak value. 

\subsubsection{Relativistic field-theoretical models}

Contrasting the above results  with those of relativistic field-theoretical models is taken up below. Results of mean-field theoretical (MFT) calculations from Refs. \cite{cons15a,cons15b} are summarized first, and thereafter the recent results of  two-loop (TL) calculations performed in Ref. \cite{Xilin15}. The TL calculations incorporate exchange-diagram contributions 
arising from scalar ($\sigma$), vector ($\omega$), iso-vector ($\rho$) and $\pi$-meson interactions between nucleons.  In the non-relativistic limit, these contributions are formally the same as in non-relativistic treatments with finite-range interactions.

%
\begin{figure}[h]
\centering
\begin{minipage}[b]{\linewidth}
\centering
\includegraphics[width=12cm]{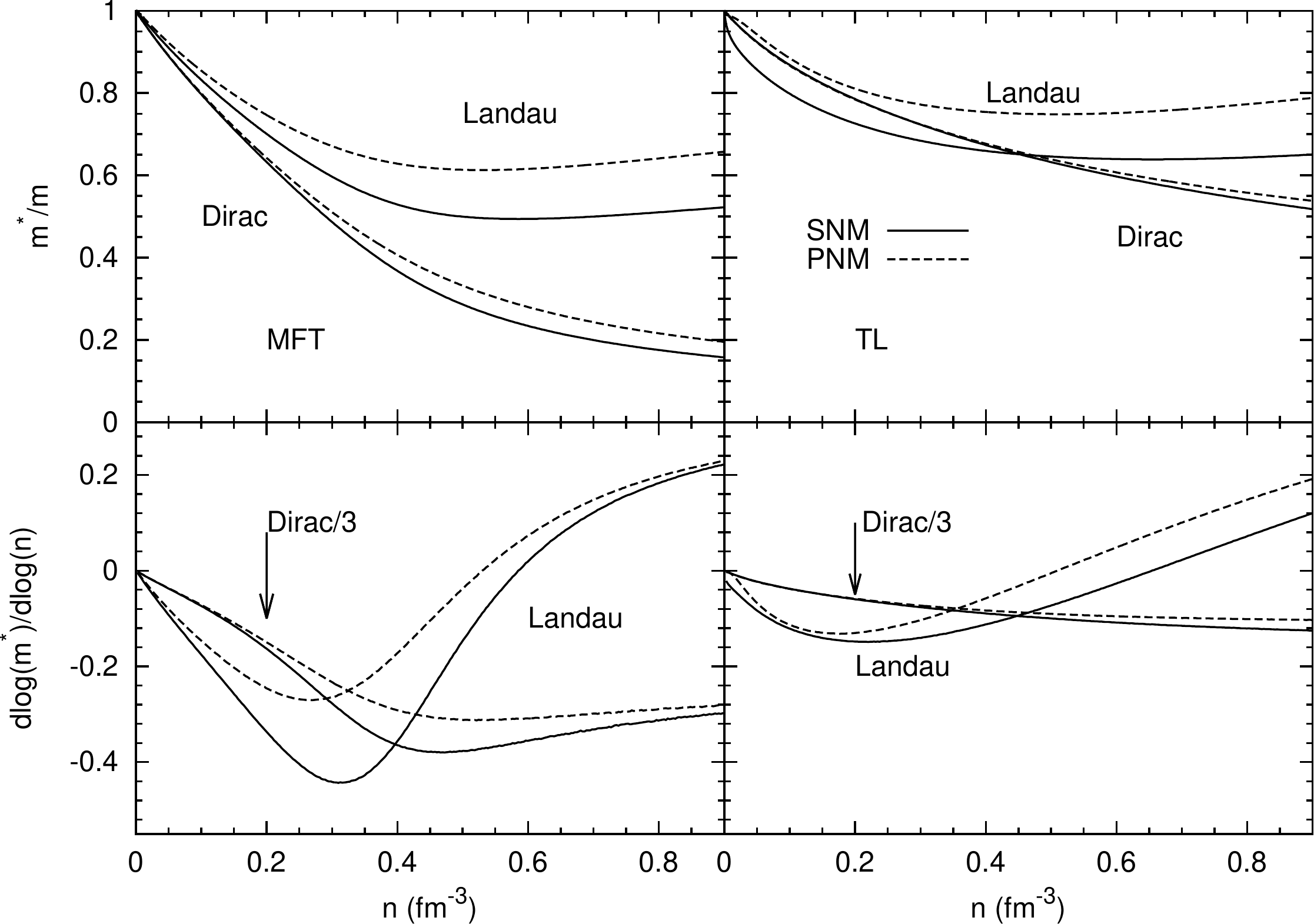}
\end{minipage}
\vskip -0.5cm
\caption{The neutron effective masses (top panels) and their logarithmic derivatives (bottom panels) for field-theoretical models at the MFT and TL levels  for symmetric nuclear matter (SNM) and pure neutron matter (PNM).   
 Figure courtsey Xilin Zhang.}
\label{MFT-TL_Ms_dMs}
\end{figure}
%

In Fig. \ref{MFT-TL_Ms_dMs} are shown the Dirac effective masses $M^*$, the Landau effective masses $m^*$, and their logarithmic derivatives as functions of $n$ for SNM (solid curves) and PNM (dashed curves) \cite{Xilin15}  in MFT. At $T=0$ in MFT, 
\be
m^* = E_F^* = {\sqrt {p_F^2 + M^*{^2} }} \quad {\rm and} \quad 
\frac {d\ln m^*}{d\ln n} = \frac 13 \left[ 1 -  \frac {M^*{^2}}{E_F^*{^2}}  
\left(1 - 3~ \frac {d\ln M^*}{d\ln n} \right) \right] \,.
\ee
The density-dependent  $M^*$ is obtained by minimizing the total energy density (pressure at $T\neq 0$) with respect to $M^*$. 
The asymptotic behaviors for 
$n/n_0 \gg 1 $ are: $m^* \rightarrow p_F$ (as $M^* \rightarrow 0$) and 
$d\ln m^*/d\ln n \rightarrow 1/3 $ .  The approach to these limits is evident from the results in Fig.  \ref{MFT-TL_Ms_dMs}.
The density $n_{min}$ at which $m^*$ is a minimum is obtained from 
\be
\frac {p_F}{M^*} + \frac {dM^*}{dp_F} =  0 \,,
\ee
which for the MFT model considered occurs at $n=0.59~(0.53)~{\rm fm}^{-3}$ for SNM (PNM). The density $n_R$ for which $p_F=M^*$ 
(whence $m^* = \sqrt 2~ p_F$) occurs at
\be
n_R =   \frac {\gamma}{6\pi^2} \left(\frac{m_b}{\sqrt 2 \hbar} \right)^3 \left( \frac {m^*}{m_b} \right)^3~{\rm fm}^{-3} = 0.643~ \gamma \left( \frac {m^*}{m_b} \right)^3~{\rm fm}^{-3} \,,
\ee
where $\gamma = 4~(2)$ for SNM (PNM), and  which is about $(2/3)~n_{min} = 0.39~(35)~{\rm fm}^{-3}$ for SNM (PNM) and marks the transition of nucleons well into the relativistic region.

The inclusion of two-loop effects significantly changes the behavior of $M^*$, $m^*$, and their logarithmic derivatives, their variation 
with $n$ being slower than in MFT (see the right hand panels in Fig. \ref{MFT-TL_Ms_dMs}). Both $M^*$ and $m^*$  acquire substantially larger values than those in MFT for all $n$ in both SNM and PNM.  The densities at which $m^*$'s attain their  minima are similar to those in MFT, but the densities $n_R$ at which $M^*=p_F$ shift to considerably larger densities than in MFT. Consequently, the thermal properties are correspondingly influenced. An analytical analysis of $m^*$'s similar to that for MFT given above is precluded  because of the inherently complicated structure of the TL integrals  which require numerical calculations. Their overall effects are, however, similar to 
those in which exchange contributions from finite-range interactions are included in non-relativistic treatments such as the MDI(A) model discussed before. This similarity is not unexpected as the TL contribution to the total energy density (minimizing which $M^*$ is obtained) reduces formally to its structure in non-relativistic treatments that include exchange interactions \cite{Xilin15}.


\begin{figure}[h]
\centering
\hspace{-0.5cm}
\begin{minipage}[b]{\linewidth}
\centering
\includegraphics[width=11.5cm]{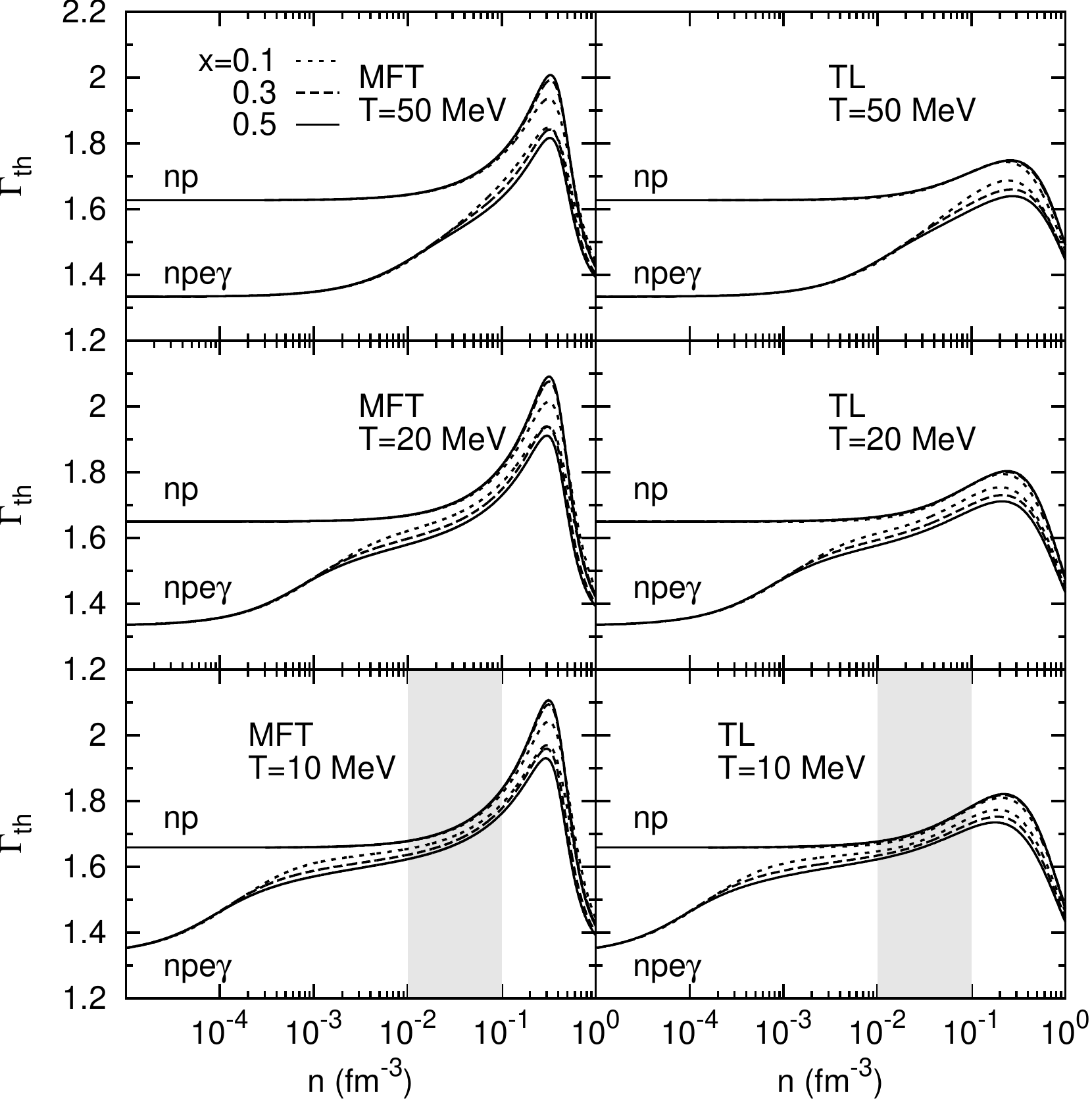}
\end{minipage}
\vskip -0.25cm
\caption{The thermal index $\Gamma_{th}$ from  
relativistic MFT and TL models. Results show contrasts between matter with only nucleons (np), and with  leptons and photons (npe$\gamma$). Figure courtesy Xilin Zhang.}
\label{Gammath_MFT-TL}
\end{figure}
%

The thermal index $\Gamma_{th}$ vs $n$ for MFT and TL models are shown in Fig. \ref{Gammath_MFT-TL}.  Results with and without the inclusion of leptons and photons in this figure are for the same proton fractions $x=Y_e$ and $T$ as in Fig. \ref{Gamma_th} for non-relativistic models. The analysis of $\Gamma_{th} $ vs $n$ for $np$ matter for the MFT model is facilitated by its degenerate limit expression \cite{cons15a,cons15b}
\be
\Gamma_{th} = \frac 43 + \frac 13 \left[ \left( \frac {M^*}{E_F^*}\right)^2 \left (1- 3~  \frac {d\ln M^*}{d\ln n}\right) \right]  
+~{\rm NLO~corrections} \,.
\label{GammathR}
\ee
In the non-relativisitic limit, $M^*/E_F^* \rightarrow 1$ and the logarithmic derivative of $d\ln M^*/d\ln n \rightarrow 0$ leading to $\Gamma_{th} =5/3$. In the ultra-relativistic limit, $M^* \rightarrow 0$ so that  
$\Gamma_{th} =4/3$.  The density at which 
the maximum occurs in $\Gamma_{th}$ (for SNM or PNM) can be determined from
\be
\frac {d\Gamma_{th}}{dp_F} = 0 = \frac {d}{dp_F} \left( 
\frac {p_F^2}{E_F^{*2}} + \frac {p_FM^*}{E_F^{*2}} \frac {dM^*}{dp_F} \right) \,.
\ee
The result is $n\simeq 0.3~{\rm fm^{-3}}$, which also holds for all $x$ in good agreement with the results of exact numerical calculations. The sub-leading NLO corrections do not affect these numbers \cite{cons15b}. 
With increasing values of $T$ (e.g., 50 MeV in the figure), $M^*$ begins to acquire a $T$-dependence, so the values 
of $\Gamma_{th}$ differ slightly from that given by Eq. (\ref{GammathR}), particularly below the peaks.  

Leptons and photons are relativistic components in $npe\gamma$ matter each with   $\Gamma_{th}= 4/3$. 
Except at very high $T$'s, the contribution of photons is small, but that of leptons is significant in charge neutral matter for all $n$. Their presence in a mixture  reduces $\Gamma_{th}$ toward $4/3$ at asymptotically low and high densities as is evident from the results in Fig. \ref{Gammath_MFT-TL} (and Fig. \ref{Gamma_th}).  The peak values of $\Gamma_{th}$ are also reduced from their values in $np$ matter as in the non-relativistic case. The MFT results for $\Gamma_{th}$ differ substantially from 
the non-relativistic cases presented before at near-nuclear and supra-nuclear densities.  
 
The influence of TL effects on $\Gamma_{th}$ is shown in the right hand panels of Fig.  \ref{Gammath_MFT-TL}.  From near-nuclear to supra-nuclear densities and for all $T$'s shown, TL effects reduce $\Gamma_{th}$ from its values in MFT in $np$ matter. This reduction is a consequence of the much larger values of $M^*$'s and $m^*$'s (which determine the thermal properties) in TL calculations compared to those in MFT. The presence of leptons and photons leads to a further reduction of $\Gamma_{th}$ at all $n$ and for all $x$.  These results are semi-quantitatively similar to those of the non-relativistic MDI(A) model (see Fig.  \ref{Gamma_th}) in which exchange contributions from finite-range interactions were considered. 

\subsection{General Comments}
The conclusions that emerge from the results of both non-relativistic
and relativistic field-theoretical models are: (i) the presence of
electrons and photons results in an overall reduction of $\Gamma_{th}$
in $np$ matter for all $n$, (ii) variations of $\Gamma_{th}$ with
$x=Y_e$ and $T$ are relatively small, and (iii) variations with $n$
are significant, but depend sensitively on the density dependence of
the Landau effective mass $m^*$ of the underlying model; the more
rapidly $m^*$ decreases with $n$, the larger is the variation in
$\Gamma_{th}$.

It was mentioned in the introduction the interest Gerry had in
effective masses.  A series of papers published in the 1980's
addressed this issue and brought to light the separate roles of the
momentum- and energy-dependences of effective masses in nuclei and
nuclear matter~\cite{Fantoni81,Negele81,BF81,Prakash83,Mahaux85}.
Relatively few calculations of thermal effects performed so far for
astrophysical applications have considered effects from
long-wavelength fluctuations, single particle-hole excitations,
or collective and pairing correlations near the Fermi surface
\cite{Pethick73,flt}.  These many-body correlations
generally lead to an enhancement of the effective mass at the Fermi
surface and therefore influence the entropy, and, hence, the other state
variables. These correlations are not captured in mean-field theories
even when exchange contributions are included in either
non-relativistic or relativistic models.  Pethick and Carneiro
\cite{Pethick73} have shown that long-wavelength fluctuations lead to
non-analytic behavior of the quasiparticle interaction which in turn
gives rise to $T^3 \ln T$ terms in the specific heat and entropy. Such
terms then become the leading correction to the results of
Fermi-liquid theory at low $T$.  In liquid $^3{\rm He}$, use of only
$T$ and $T^3 \ln T$ terms gives a remarkably good fit to specific heat
data (see Fig. 1.8 in Ref. \cite{flt}).

 Fantoni et al.~\cite{Fantoni83} used correlated basis
functions (CBF) to study the nucleon optical potential in nuclear
matter and found that the enhancement of the effective
mass appears to be much smaller than in liquid $^3{\rm He}$. A comparison of single-particle energies for variational calculations with and without second-order CBF does not show significant differences  (Ref.~\cite{Fantoni83},
Figure 6).  Since 
enhancements of the effective mass and the $T^3 \ln T$ term have the
same origin, it seems likely that contributions from the $T^3 \ln T$
term are not large in dense nuclear matter.  Green's function Monte
Carlo calculations of finite-temperature matter are not yet available
owing to the fermion-sign problem.  Nevertheless, this topic deserves
further scrutiny in light of modern developments in effective field
theories of nuclear matter, at least up to the densities for which they
are reliable. Contributions from these sources should be added to those
reported in this article when deemed appropriate.

\section{NON-THERMAL EFFECTS IN BULK MATTER}
\label{Sec:NTeffects}

Bulk nucleonic matter is also encountered in medium-energy heavy-ion collisions at beam energy per particle in the range 0.5-2 GeV. In these collisions, densities up to 3-4 times the nuclear equilibrium density are accessed albeit in non-equilibrium conditions. The matter, momentum, and energy flow of nucleons in such collisions 
has been characterized by  (i) the mean transverse momentum per nucleon $\langle p_x \rangle /A$ vs rapidity $y/y_{proj}$ \cite{Danielewicz85}, (ii) flow angle from asphericity analysis \cite{Gustafsson84}, (iii) azimuthal distributions \cite{Welke88}, and (iv) radial flow \cite{Siemens79}. In recent years, the analysis of data has been extended to include the collective flow of momentum in terms of  Fourier coefficients (proposed in Ref. \cite{Ollitrault92})  similar to those extracted for relativistic heavy-ion collision experiments at BNL and LHC, but at much higher energies.  

Attempts to account for the experimental findings  in medium-energy heavy-ion collisions either through cascade calculations (sequential scatterings of nucleons with their free-space cross-sections) or through hydrodynamical calculations \cite{SG86} 
(which presume local thermal equilibrium with the EOS as an input) failed insofar as they predicted too little or too much collective flow. The more fundamental kinetic description introduced in Ref. \cite{Bertsch88} employed Monte-Carlo methods to solve for 
 the phase space distribution
function $f({\vec r},{\vec p},t)$ of a nucleon via the Boltzmann-Uhling-Uhlenbeck (BUU) equation incorporating both the mean-field $U$ and a collision term with Pauli blocking of final states:
\def\sst{\scriptscriptstyle}
\begin{eqnarray}
\frac {\partial f}{\partial t} &+& {\vec \nabla}_p U \cdot {\vec \nabla}_r f
- {\vec \nabla}_r U \cdot {\vec \nabla}_p f = \nonumber \\
&-& \frac {1}{(2\pi)^3} \int d^3p_2\,d^3p_{2^{\prime}}\,d\Omega 
\frac {d\sigma_{\sst NN}}{d\Omega} \, v_{12} 
\,\delta^3( {\vec p} + {\vec p}_2 - {\vec p}_{1^{\prime}} 
- {\vec p}_{2^{\prime}} ) \nonumber \\ 
&  \times &
\left[ ff_2 (1-  f_{1^{\prime}}) (1-  f_{2^{\prime}})
- f_{1^{\prime}} f_{2^{\prime}} (1- f) (1- f_2) \right] \,.
\label{BUU}
\end{eqnarray}
The mean field $U$, the functional derivative of the energy density ${\cal H}$ of matter {\em at zero temperature}, i.e., $U(n,{p}) \equiv \delta {\cal H}/\delta n$, depends on both the local density $n$ and the nucleon momentum ${\vec p}$, and serves as an input. The other physical input is the nucleon-nucleon differential cross section, ${d\sigma_{\sst NN}}/{d\Omega}$, which depends on the relative velocity $v_{12}$. The  evolution of  $f({\vec r},{\vec p},t)$ is inherently  {\em off-equilibrium}. 

\begin{figure}[h]
\vspace {-1.cm}
\centering
\hspace{-0.5cm}
\centering
\includegraphics[width=10.cm]{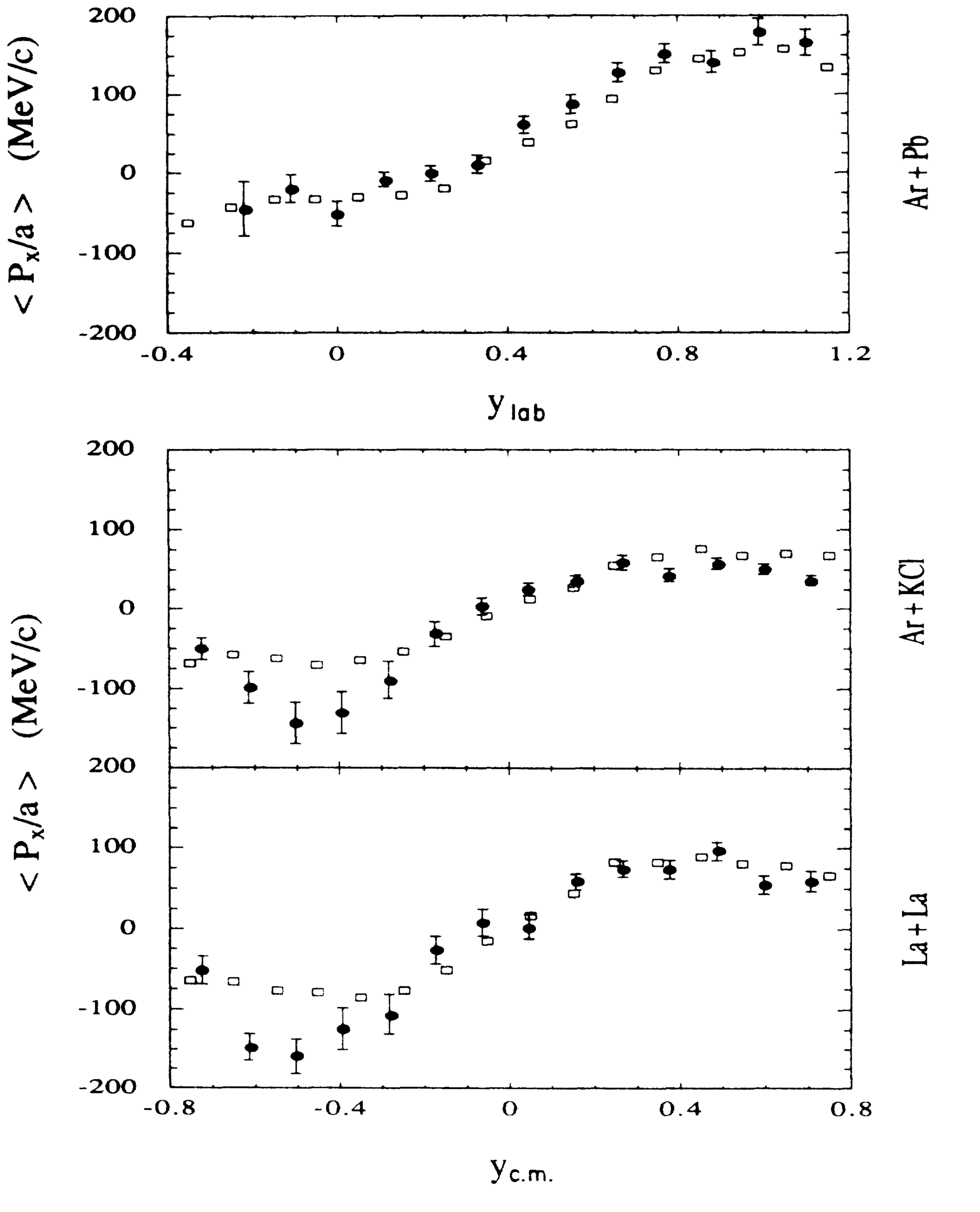}
\vskip -0.5cm
\caption{Transverse momentum per nucleon as a function of rapidity in reactions of 800 MeV per projectile nucleon.  Results of BUU simulations with the MDYI interaction (open squares) are compared with the data of Ref. \cite{Danielewicz88} (solid circles). Results for Ar + Pb are in the lab, for La + La and Ar + KCl in the center of mass. Figure from Ref. \cite{Gale90}.}
\label{HI-th-data}
\end{figure}
Fig. \ref{HI-th-data} shows an example of comparison between  data and theoretical calculations in which isospin averaged cross sections and a momentum-dependent mean field of symmetric nuclear matter were employed \cite{Gale90}.  
 It is worth noting that in these experiments, particle detection inefficiencies caused the experimental  transverse momenta  
 in the backward direction to be  artificially biased towards large   negative values and were   therefore unreliable.  
The interplay of the so-called Vlasov term [second term in Eq. (\ref{BUU}) featuring $U(n, p$)] and collisions [the right hand side in  Eq. (\ref{BUU})] was found to be crucial to account for the data.  Most of the sideways flow gauged through  $\langle p_x \rangle /A$ vs $y/y_{proj}$ is built during the early stages of the collision under non-equilibrium conditions.

Studies performed in Refs.  \cite{Gale87,Prakash88b,Welke88,Gale90,Danielewicz:00,Danielewicz:02} have shed valuable insight into  the nature of the momentum-dependence in the mean field $U(n,p)$ required to account for the collective flow observed in heavy-ion data. For example, a quadratic momentum-dependence in $U(n,p)$, characteristic of zero-range Skyrme models, gives too much sideways flow \cite{Prakash88b}. Relativistic MFT models for which the Schrodinger-equivalent potential is linear in energy  \cite{Jaminon:81,ABBCP} also give too much sideways flow.  
A mean field that saturates at high momenta, such as those in finite-range models, adequately accounts for the data. A similar mean field is also required in optical model fits  of nucleon-nucleus scattering data  \cite{Hama:90,Cooper:93,Danielewicz:00,Danielewicz:02}. Examples of microscopic calculations of $U(n,p)$  that show a saturating behavior at high momenta in 
non-relativistic approaches can be found in Refs.  \cite{w88,Zuo14}. A similar behavior is observed both in MFT models that employ non-linear derivatives  \cite{Gaitanos13} and in the two-loop calculations of Ref. \cite{Xilin15}.

The conclusion that has emerged 
is that as long $U(n,p$) saturates at high momenta as required in optical model fits to nuclear data,
a symmetric matter incompressibility parameter $K_s\approx 240$ MeV fits the heavy-ion data \cite{Danielewicz02}.   This is comforting in view of the fact that the value $K_s\simeq240\pm20$ MeV is suggested by the analysis of the giant monopole resonance data \cite{Youngblood99,Garg04,Colo04}.
Additionally, it has  been confirmed that an EOS based on such a mean field can yield a neutron star with $M_{max}\simge2M_\odot$ \cite{cons15a}.

Rare-isotope accelerators (RIAs) that can collide
highly neutron-rich nuclei has encouraged work to study  collisions featuring  high neutron excess
\cite{Das03,Li04,Li04b}. Generalizing Eq.~(\ref{BUU}) to a mixture, the kinetic equation for neutrons is
\begin{equation}
\frac {\partial f_n}{\partial t} + {\vec \nabla}_p U \cdot {\vec \nabla}_r f_n
- {\vec \nabla}_r U \cdot {\vec \nabla}_p f_n = J_n = \sum_{i=n,p} J_{ni}\,, 
\end{equation}
where $J_n$ describes collisions of a neutron with all other neutrons
and protons.  A similar equation applies for protons with
appropriate modifications.  The mean field $U\equiv U(n_n,n_p;{\vec p}\,)$ now depends
explicitly on the neutron-proton asymmetry and establishes a connection to the
symmetry energy in that $U$ is now obtained from a
functional differentiation of the Hamiltonian density ${\cal H}(n_n,n_p)$ of isospin asymmetric matter.
Examples of such mean fields may be found in
Refs.~\cite{Prakash97,Li04,Li04b}.  Isospin asymmetry influences the neutron-proton differential flow and the ratio of neutron to
proton multiplicity as a function of transverse momentum at mid-rapidity.  
Future investigations of these signatures will shed additional light on the EOS at supra-nuclear densities.  
Details of on-going studies along these lines can be found in many articles in the proceedings of the meeting on ``Topical Issue on Nuclear Symmetry Energy'', published in 2014 in volume 50 of the {\it European Physics Journal A}.

\section{CONCLUSIONS}
\label{Sec:Conclusions}

The advances in dense matter theory during the time that Gerry Brown was interested in the topic were enormous, and the pace of advancement has not slowed.  Astronomical observations have provided the important evidence that the neutron-rich EOS above $n_s$ is quite stiff, being able to support neutron star masses in excess of $2M_\odot$.  At the same time, the EOS near $n_s$ seems to be relatively soft, judging from experimental evidence and neutron matter studies indicating 40 MeV $\simle L\simle70$ MeV.  Assuming that neutron stars have crusts, neutron matter calculations are reasonable correct near $n_s$, causality is valid, and GR is the correct theory of gravity at the required densities, typical neutron star radii are confined to the range 9 km $<R_{1.4}<14$ km.  If strong phase transitions do not exist in the range $n_s<n<2n_s$, this range is even smaller: 10.7 km $\simle R_{1.4}\simle13.1$ km.   Many upcoming astrophysical observations should be able to confirm these predictions and reduce these ranges further, allowing for a tighter description of neutron star matter and its composition.  

 Further evidence concerning the internal composition of neutron stars is becoming available from observations of neutron star cooling, which consists of the 
body of observed surface temperatures and ages of a few dozen 
neutron stars to date. In one spectacular case, the neutron star 
in the supernova remnant Cassiopeia A has, over the period from 2000--2010, 
been observed~\cite{Heinke10} to be cooling much faster than 
expected for a star it's age ($\sim330$ years). While it has been
suggested from examining observations from a variety of detectors 
that this cooling represents an anomalous instrumental
effect \cite{Posselt13}, continued observations for an additional five years
\cite{Ho15}, employing one carefully calibrated detector 
({\it Chandra}'s ACIS-S), have
reinforced the original report of rapid cooling. This rapid 
cooling has been interpreted~\cite{Page11,Shternin11} to be 
caused by a temporary phase of enhanced neutrino emission due to 
Cooper pair breaking and formation during the onset of neutron 
superfluidity in the star's core.  Furthermore, the rapidity of 
the cooling suggests that not only did it begin within a few 
decades of the present time, but that protons must be superconducting 
in the interior as well.  It is possible to determine the critical 
temperature for the neutron superfluid with some accuracy 
($T_c=5-10\times10^8$ K) since it depends mostly on the well-determined 
age of this neutron star.\footnote{The more recent analysis of Ref.
\cite{Ho15} suggests $T_c\simeq6\times10^8$ K.} 
Continued observations of this star, and other cooling neutron stars, 
will reveal further information about their interiors and the 
thermal properties of dense matter.   

The physics and astrophysics of core-collapse supernovae, the birth of neutron stars and their evolution to old age, and mergers of binary stars involving neutron stars and black holes, also involve thermal effects to varying degrees. In addition to influencing the hydrodynamic evolutions of these objects, thermal effects play significant roles in the photon, neutrino, and gravitational radiation emissivities at different stages of  their evolutions.  With vastly improved capabilities in astronomical observations and laboratory experiments involving rare-isotope accelerators that can access highly neutron-rich matter, we are now well poised to explore and establish the properties of dense hadronic matter under conditions of extreme density, isospin content, temperature, and magnetic fields.

\section*{ACKNOWLEDGEMENTS}
This work was supported by the U.S. DOE under Grants No. DE-FG02-87ER-40317 and DE-FG02-93ER-40756. The authors are grateful for the generous help of collaborators Constantinos Constantinou, Yeunhwan Lim, Brian Muccioli, Andrew Steiner, and Xilin Zhang.  We are also grateful to collaborators Christian Drischler and Achim Schwenk for providing us with prepublication results from theoretical neutron and nuclear matter studies of arbitrary proton fractions, and to Xilin Zhang for prepublication results concerning thermal properties of matter.

%
%

\section*{REFERENCES}

\bibliography{gerry15}

\end{document}